\documentclass[a4paper,fleqn,usenatbib]{mnras}

\usepackage[T1]{fontenc}
\usepackage{ae,aecompl}

\usepackage{graphicx}	
\usepackage{amsmath}	
\usepackage{amssymb}	
\usepackage{epsfig,ragged2e}
\usepackage{dcolumn}
\newcolumntype{.}{D{.}{.}{-1}}
\newcolumntype{u}{D{,}{\;\pm\;}{-1}}

\usepackage{caption}
\usepackage{xcolor}
\usepackage{pdfpages}
\usepackage{rotating}
\usepackage{booktabs}	

\newcommand{\degrees}{$^\circ$}
\newcommand{\per}{$^{-1}$}

\title[610 MHz GMRT follow-up of ACT clusters]{GMRT 610 MHz observations of galaxy clusters in the ACT equatorial sample}

\author[K. Knowles et al.]{Kenda Knowles,$^{1}$\thanks{E-mail: kendaknowles.astro@gmail.com (KK)}
Andrew J. Baker,$^{2}$
J. Richard Bond,$^{3}$
Patricio A. Gallardo,$^{4}$
\newauthor Neeraj Gupta,$^{5}$
Matt Hilton,$^{1}$
John P. Hughes,$^{2}$
Huib Intema,$^{6}$
Carlos H. L\'{o}pez-\newauthor Caraballo,$^{7,8}$
Kavilan Moodley,$^{1}$
Benjamin L. Schmitt,$^{9}$
Jonathan Sievers,$^{10}$
\newauthor Crist\'{o}bal Sif\'{o}n,$^{4,11}$
Edward Wollack,$^{12}$
\\
$^{1}$Astrophysics \& Cosmology Research Unit, School of Mathematics, Statistics \& Computer Science, University of KwaZulu-Natal,\\ \; Durban, 3690, South Africa\\
$^{2}$Department of Physics and Astronomy, Rutgers, The State University of New Jersey, 136 Frelinghuysen Road, Piscataway,\\ NJ 08854-8019, USA\\
$^{3}$Canadian Institute for Theoretical Astrophysics, 60 St. George Street, University of Toronto, Toronto, ON, M5S 3H8, Canada\\
$^{4}$Department of Physics, Cornell University, Ithaca, NY USA\\
$^{5}$IUCAA, Post Bag 4, Ganeshkhind, Pune 411007, India\\
$^{6}$Leiden Observatory, Leiden University, PO Box 9513, NL2300 RA Leiden, Netherlands\\
$^{7}$Instituto de Astrof\'{i}sica and Centro de Astro-Ingenier\'{i}a, Facultad de F\'{i}sica, Pontificia Universidad Cat\'{o}lica de Chile, Av. Vicu\~{n}a \\ \;Mackenna 4860, 7820436 Macul, Santiago, Chile\\
$^{8}$Departamento de Matem\'{a}ticas, Universidad de La Serena, Av. Juan Cisternas 1200, La Serena, Chile\\
$^{9}$Department of Physics and Astronomy, University of Pennsylvania, 209 South 33rd Street, Philadelphia, PA 19104, USA\\
$^{10}$Astrophysics \& Cosmology Research Unit, School of Chemistry \& Physics, University of KwaZulu-Natal, Durban, 3690, South Africa\\
$^{11}$Department of Astrophysical Sciences, Peyton Hall, Princeton University, Princeton, NJ 08544, USA\\
$^{12}$NASA/Goddard Space Flight Center, 8800 Greenbelt Rd, Greenbelt, MD 20771, USA
}

\date{Accepted XXX. Received YYY; in original form ZZZ}

\pubyear{2018}

\begin{document}
\label{firstpage}
\pagerange{\pageref{firstpage}--\pageref{lastpage}}
\maketitle

\begin{abstract}

We present Giant Metrewave Radio Telescope 610 MHz observations of 14 Atacama Cosmology Telescope (ACT) clusters, including new data for nine. The sample includes 73\% of ACT equatorial clusters with $M_{500} > 5 \times 10^{14}\;M_\odot$. We detect diffuse emission in three of these (27$^{+20}_{-14}$\%): we detect a radio mini-halo in ACT-CL J0022.2$-$0036 at $z=0.8$, making it the highest-redshift mini-halo known; we detect potential radio relic emission in ACT-CL J0014.9$-$0057 ($z=0.533$); and we confirm the presence of a radio halo in low-mass cluster ACT-CL J0256.5+0006, with flux density $S_{610} = 6.3\;\pm\;0.4$ mJy. We also detect residual diffuse emission in ACT-CL J0045.9$-$0152 ($z=0.545$), which we cannot conclusively classify. For systems lacking diffuse radio emission, we determine radio halo upper limits in two ways and find via survival analysis that these limits do not significantly affect radio power scaling relations. Several clusters with no diffuse emission detection are known or suspected mergers, based on archival X-ray and/or optical measures; given the limited sensitivity of our observations, deeper observations of these disturbed systems are required in order to rule out the presence of diffuse emission consistent with known scaling relations. In parallel with our diffuse emission results, we present catalogs of individual radio sources, including a few interesting extended sources. Our study represents the first step towards probing the occurrence of diffuse emission in high-redshift ($z\gtrsim0.5$) clusters, and serves as a pilot for statistical studies of larger cluster samples with the new radio telescopes available in the pre-SKA era.
\end{abstract}



\section{Introduction} \label{sec:intro}

Over the past two decades, low frequency radio follow-up of galaxy clusters has provided insight into the non-thermal physics of the intracluster medium through observations of diffuse, cluster-scale synchrotron emission \citep{BrunettiJones.2014.Review}. There are historically three main classes of cluster diffuse radio emission, namely halos, relics, and mini-halos, characterised by their position relative to the cluster core, size, polarization, and their general morphology. It is well established that the presence of these diffuse structures is correlated with the dynamical state of the host cluster {\citep[e.g.,][]{Buote.2001.GRH, Cassano.2010.GRHMergerConn}}, with halos and relics found in merging systems, while mini-halos are preferentially found around the brightest cluster galaxy (BCG) in cool-core clusters.

There are however several open questions relating to the formation mechanisms of these sources, and as more sensitive cluster observations are carried out, more diffuse structures are being discovered that do not conform to the standard classifications \citep{vanWeeren.2017.Nature}. A particular problem in trying to understand the origin of diffuse cluster radio emission is a lack of homogeneous cluster samples with radio follow-up: although over 50 radio halos have been discovered to date, they form a heterogeneous cluster sample \citep{Yuan.2015.DEscalrel}, making it difficult to disentangle selection effects from the scatter in the various scaling relations. The first cluster samples to be studied with respect to diffuse radio emission were taken from X-ray selected samples, with the Extended GMRT Radio Halo Survey \citep{Kale.2013.EGRHSUL} being the largest homogeneous X-ray selected sample studied to date, with 67 clusters above a 0.1 - 2.4 keV band X-ray luminosity of $L_X > 5 \times 10^{44}$ erg s\per, and within a redshift range of $0.2 < z < 0.4$. As X-ray luminosity selection may bias samples towards relaxed clusters \citep{Poole.2007.MergerSims}, more recent studies have used Sunyaev-Zel'dovich effect \citep[SZ;][]{SunyaevZeldovich.1972.SZE} properties to select the cluster samples. The largest SZ-selected sample studied to date consists of 56 mass-selected Planck PSZ2 clusters \citep{Cuciti.2015} within a mass and redshift range of $M_{\rm 500,SZ} \gtrsim 6 \times 10^{14} M_\odot$ and $0.2 \lesssim z \lesssim 0.4$, respectively. Both samples are restricted to low to intermediate redshift and high mass systems. Diffuse radio emission has been detected in higher redshift and lower mass systems \citep{Lindner.2014.ElGordo, Knowles.2016.J0256}, but no systematic survey of a larger statistical sample of such systems has been undertaken. {As inverse Compton (IC) losses are expected to dominate at high redshift \citep{Marscher.1983.IClosses}, detecting diffuse radio emission above $z\;=\;0.5$ has 
been predicted to be difficult in clusters below a mass of $M_{\rm 500,SZ} = 10^{15}\; M_\odot$ \citep{Cassano.2010.GRHLOFAR}.}


Here we present the results of a pilot radio study {of a homogeneously selected sample of clusters} from the Atacama Cosmology Telescope's equatorial cluster catalog \citep{Hasselfield.2013.ACTE}. {In this study we search for diffuse cluster emission in a sample where, for the first time, the cluster selection criteria has been extended to lower mass and higher redshift. Although we have small number statistics, with this sample we can make the first step towards probing the occurence of diffuse radio emission in clusters above $z\;\gtrsim\;0.5$. }

The paper is structured as follows. {The cluster sample and selection is discussed in Section \ref{sec:sample}, with the observations and data reduction strategy specified in Section \ref{sec:obs+reduction}. Section \ref{sec:clusterDE} reports new diffuse radio emission detections in the sample, with non-detections in dynamically disturbed sytems reported in Section \ref{sec:nondets}. Non-detections in relaxed clusters are discussed in Appendix \ref{app:nondets}. Upper limits for all non-detections, as well as sample scaling relations, are discussed in Section \ref{sec:UL}. Notes on interesting radio sources throughout the sample are given in Section \ref{sec:fovsrcs}, with full details provided in Appendix \ref{app:fovsrcs}. A summary of our results and concluding remarks are given in Section \ref{sec:summary}. All appendices are available online only.}
 
In this paper we adopt a $\Lambda$CDM flat cosmology with $H_0 = 70$ km s\per Mpc\per, $\Omega_m$ = 0.3 and $\Omega_\Lambda$ = 0.7. We assume $S_\nu \propto \nu^{\alpha}$ throughout the paper, where $S_\nu$ is the flux density at frequency $\nu$ and $\alpha$ is the spectral index. $R_{500}$ denotes the radius within which the average density is 500 times the critical density of the Universe. Colour versions of all figures are available in the online journal. The full catalog of source flux densities is available online.

\section{The cluster sample}
\label{sec:sample}

The Atacama Cosmology Telescope \citep[ACT;][]{Swetz.2011.ACT} is a 6 m telescope providing arcminute resolution observations of the millimetre sky. The ACT Equatorial sample \citep[hereafter ACT-E;][]{Hasselfield.2013.ACTE} was compiled from three years (2008-2011) of 148 GHz observations of a 504 deg$^2$, zero-declination strip overlapping the Sloan Digital Sky Survey \citep[SDSS;][]{York.2000.SDSS} Stripe 82 \citep{Adelman-McCarthy.2007, Annis.2014.S82}. The ACT-E catalog consists of 68 galaxy clusters detected via the SZ effect, with a mass and redshift range of $1.4 \times 10^{14} < M_{\rm 500c,SZ}^{\rm UPP} [M_\odot] < 9.4 \times 10^{14}$ and $0.1 \lesssim z \lesssim 1.4$, respectively. Here, UPP refers to the Universal Pressure Profile and its associated mass--scaling relation \citep{Arnaud.2010.UPP}, which was used to model the cluster SZ signal \citep[for details, refer to Section 3.2 of][]{Hasselfield.2013.ACTE}. The UPP masses quoted in this paper are with respect to the critical density at the cluster redshift. 

\begin{table*}
 \centering
 \caption{The sub-sample of ACT-E clusters. (1) ACT cluster catalog name. (2) J2000 RA of the SZ peak. (3) J2000 Dec of the SZ peak. (4) Redshift. (5) Integrated Compton-$y$ parameter within $R_{500}$. (6) SZ-derived mass within $R_{500}$. (7) Alternate cluster designation. }
 \label{tab:sample}
 \begin{tabular}{l..cuuuc}
  \toprule
  Cluster name & \multicolumn{1}{c}{RA$_{J2000}$} & \multicolumn{1}{c}{Dec$_{J2000}$} & {z} & \multicolumn{1}{c}{$\tilde{y}_{0}$} & \multicolumn{1}{c}{$Y_{500}$} & \multicolumn{1}{c}{$M_{\rm 500c}^{\rm UPP}$} & Alternate Name\\
  (ACT-CL...) & \multicolumn{1}{c}{(deg)} & \multicolumn{1}{c}{(deg)} &  & \multicolumn{1}{c}{($10^{-4}$)} & \multicolumn{1}{c}{($10^{-4}$ arcmin$^2$)} & \multicolumn{1}{c}{($10^{14} h_{70}^{-1} M_\odot$)}\\
  \midrule
  J0014.9$-$0057 & 3.7276 & -0.9502 & 0.533 & 1.34 , 0.18 & 5.0 , 0.9 & 5.7 , 1.1 & GMB11 J003.71362-00.94838 \\
  J0022.2$-$0036 & 5.5553 & -0.6050 & 0.805 & 1.35 , 0.16 & 3.8 , 0.6 & 5.5 , 0.9 & WHL J002213.0-003634 \\
  J0045.2$-$0152 & 11.3051 & -1.8827 & 0.545 & 1.31 , 0.18 & 4.8 , 0.9 & 5.6 , 1.1 & WHL J004512.5-015232 \\
  J0059.1$-$0049 & 14.7855 & -0.8326 & 0.786 & 1.24 , 0.15 & 3.5 , 0.6 & 5.2 , 0.9 & -\\
  J0152.7+0100 & 28.1764 & 1.0059 & 0.230 & 1.30 , 0.15 & 13.0 , 2.3 & 5.7 , 1.1 & Abell 267 \\
  J0239.8$-$0134 & 39.9718 & -1.5758 & 0.375 & 1.61 , 0.18 & 9.4 , 1.6 & 6.7 , 1.3 & Abell 370 \\ 
  J0256.5+0006 & 44.1354 & 0.1049 & 0.363 & 0.82 , 0.15 & 3.4 , 1.0 & 3.8 , 0.9 & RXC J0256.5+0006 \\
  J2051.1+0215 & 312.7885 & 2.2628 & 0.321 & 1.36 , 0.26 & 7.6 , 2.5 & 5.3 , 1.4 & RXC J2051.1+0216 \\
  J2129.6+0005 & 322.4186 & 0.0891 & 0.234 & 1.23 , 0.17 & 11.4 , 2.4 & 5.3 , 1.1 & RXC J2129.6+0005\\
  J2135.2+0125 & 323.8151 & 1.4247 & 0.231 & 1.47 , 0.16 & 14.2 , 2.4 & 6.3 , 1.2 & Abell 2355 \\
  J2135.7+0009 & 323.9310 & 0.1568 & 0.118 & 0.68 , 0.17 & 10.8 , 6.3 & 2.6 , 1.1 & Abell 2356  \\
  J2154.5$-$0049 & 328.6319 & -0.8197 & 0.488 & 0.95 , 0.17 & 3.4 , 0.9 & 4.3 , 0.9 & WHL J215432.2-004905\\
  J2327.4$-$0204 & 351.8660 & -2.0777 & 0.705 & 2.65 , 0.21 & 10.1 , 1.0 & 9.4 , 1.5 & RCS2 J2327.4-0204\\
  J2337.6+0016 & 354.4156 & 0.2690 & 0.275 & 1.43 , 0.18 & 11.5 , 2.2 & 6.1 , 1.2 & Abell 2631 \\
  \bottomrule
 \end{tabular}
 \vspace*{0.2cm}
\end{table*}

For our deep GMRT radio follow-up of the ACT-E clusters, we selected all clusters above a mass limit of $M_{\rm 500c,SZ}^{\rm UPP} \geqslant 5.0 \times 10^{14} M_\odot$, excluding those with existing GMRT data, and observed two sub-samples: a pilot sample with a wide redshift range, and a high-redshift sample. As the pilot sample was proposed for and observed before the ACT-E cluster masses were published, the selection was done using the preliminary masses. Based on the published masses, three of these clusters (ACT-CL J2135+0009, ACT-CL J2154$-$0049, and ACT-CL J0256.5+0006) are outside of our mass cut, however we include them here for completeness.

The full list of 14 ACT-E clusters discussed in this paper is given in Table \ref{tab:sample}. This list includes the pilot and high-redshift samples (eight clusters) as well as five ACT-E clusters with archival 610 MHz GMRT data. One further cluster, ACT-CL J0239.8$-$0134, was observed by us as part of an ACTPol \citep[polarization-sensitive upgrade to ACT;][]{Thornton.2016.ACTPol, Hilton.2017.ACTPol} observing programme. We now have deep radio follow-up of 11 of 15 (i.e., 73\% of) ACT-E clusters with $M_{\rm 500c,SZ}^{\rm UPP} \geqslant 5.0 \times 10^{14} M_\odot$.

\citet{Sifon.2016} performed spectroscopic observations of 44 ACT clusters, 10 of which are part of our radio follow-up sample. They identified a minimum of 44 cluster members for each system in our sample. For each cluster, they used the member galaxy identifications to perform a DS test analysis \citep{DresslerShectman.1988.DSTest}, obtaining a measure of the significance of substructure in the cluster, given by the parameter $S_\Delta$. A value of $S_\Delta < 0.05$ is indicative of substructure. All but one of the clusters in our sample have X-ray imaging from either the Chandra or XMM-Newton telescopes, although there is a wide range of exposure times. A full multiwavelength analysis of the sample will be presented in a forthcoming paper (Knowles et al., in prep), but a subset of this data is used for specific clusters in this paper.

\section{Radio Observations and Data Reduction}
\label{sec:obs+reduction}
\subsection{New Radio Observations}
\label{subsec:newobs}
For our observations of the pilot and high-redshift samples and ACT-CL J0239.8$-$0134 (PI: Knowles; Project ID 22\_044, 26\_031, and 30\_012, respectively) we used the GMRT's default continuum mode at 610 MHz, with the 33 MHz bandwidth split into 256 channels. Data were acquired in the RR and LL polarizations. The on-source and integration time for each observation is given in columns five and six of Table \ref{tab:obs}. For each cluster, a flux and bandpass calibrator was observed {for $\sim$ 15 min} at the beginning and end of the observation block. This source was also used to estimate the instrumental antenna gains{, providing initial phase and amplitude corrections for the science target. Additional details of the reduction strategy are given in section \ref{subsec:datared}.} 

\begin{table*}
 \centering
 \caption{GMRT Observation details. (1) ACT cluster designation. (2) GMRT proposal ID. (3) Observation date. (4) Central observing frequency. (5) On-source time. (6) Integration time. (7) Central RMS noise. (8) Synthesised beam. }
 \label{tab:obs}
 \begin{tabular}{lccccccc}
  \toprule
  Cluster name & GMRT ID & Obs. Date & Freq. & $t_{\rm src}$ & $t_{\rm int}$ & rms noise & Beam \\
   & & (DD-MM-YY) & (MHz) & (hrs) & (sec) & ($\mu$Jy beam\per) & \arcsec$\times$\arcsec, PA (\degrees) \\
  \midrule
  ACT-CL J0014.9$-$0056 & 22\_044 & 27-08-12 & 608.0 & 6.3 & 16.1 & 35 & 6.2 $\times$ 4.2, 72.3 \\
  ACT-CL J0022.2$-$0036 & 26\_031 & 11-08-14 & 607.9 & 7.4 &  8.0 & 37 & 6.8 $\times$ 5.2, -56.9 \\
  ACT-CL J0045.2$-$0152 & 26\_031 & 11-08-14 & 607.9 & 7.3 &  8.0 & 35 & 5.9 $\times$ 4.6, -68.0 \\
  ACT-CL J0059.1$-$0049 & 26\_031 & 10-08-14 & 608.1 & 6.7 &  8.0 & 30 & 5.2 $\times$ 4.3, 76.0 \\
  ACT-CL J0152.7+0100 	& 16\_117 & 23-08-09 & 613.4 & 4.4 & 16.1 & 98 & 5.8 $\times$ 4.2, 70.3 \\
  ACT-CL J0239.8$-$0134 & 30\_012 & 17-06-16 & 608.1 & 4.8 &  8.0 & 37 & 5.9 $\times$ 3.9, 32.0 \\
  ACT-CL J0256.5+0006 	& 22\_044 & 26-08-12 & 608.0 & 6.9 & 16.1 & 22 & 5.7 $\times$ 4.1, 72.2 \\
  ACT-CL J2051.1+0215 	& 26\_021 & 20-07-14 & 608.0 & 3.3 & 16.1 & 42 & 5.1 $\times$ 4.1, 89.3 \\
  ACT-CL J2129.6+0005 	& 26\_050 & 31-08-12 & 612.4 & 6.0 & 16.1 & 50 & 5.2 $\times$ 3.9, 67.5 \\
  ACT-CL J2135.2+0125 	& 26\_021 & 04-08-14 & 608.1 & 3.4 & 18.6 & 110 & 5.7 $\times$ 4.2, 64.7 \\
  ACT-CL J2135.7+0009 	& 22\_044 & 28-08-12 & 607.1 & 2.8 & 16.1 & 90 & 5.1 $\times$ 4.5, 59.4 \\
  ACT-CL J2154.5$-$0049 & 22\_044 & 26-08-12 & 607.2 & 5.9 & 16.1 & 46 & 6.1 $\times$ 4.0, 68.3 \\
  ACT-CL J2327.4$-$0204 & 26\_031 & 06-08-14 & 607.3 & 7.5 &  8.0 & 58$^\dagger$ & 5.4 $\times$ 5.0, 63.1 \\
  ACT-CL J2337.6+0016 	& 18\_078 & 08-05-10 & 610.8 & 6.3 & 16.1 & 170 & 5.5 $\times$ 4.4, 64.5 \\
  \bottomrule
 \end{tabular}
 
 \vspace{-0.1cm}
 \justify
 $^\dagger$ Due to a bright interfering source to the West of the cluster region, the central map noise is as high as 0.1 mJy beam\per.\\
 
\end{table*}

\subsection{Archival Observations}
\label{subsec:archivalobs}
Five clusters in our sample have deep 610 MHz GMRT data. Two of the archival observations {, ACT-CL J0152.7+0100 \citep{Kale.2013.EGRHSUL} and ACT-CL J2337.6+0016 \citep{Venturi.2007.GRHS1}, made use of the GMRT's Hardware Backend (GHB), while ACT-CL J2129.6+0005 \citep{Kale.2015.EGRHS} was observed using the dual 610/235 MHz Software Backend (GSB). The remaining two archival clusters, ACT-CL J2135.2+0125 \citep{Cassano.2016} and ACT-CL J2051.1+0215, were observed with the default GSB continuum mode at 610 MHz.} The on-source and integration time for each observation is given in columns five and six of Table \ref{tab:obs}.

\subsection{Data Reduction}
\label{subsec:datared}
We processed all datasets using the SPAM \citep[Source Peeling and Atmospheric Modelling;][]{Intema.2009.SPAM} software, which makes use of AIPS \citep[NRAO Astronomical Image Processing System;][]{Wells.1985} and Obit \citep{Cotton.2008.Obit} tools. A full description of the SPAM software is given in \citet{Intema.2014.SPAM}. Here we outline the main calibration steps. First, strong RFI is removed via statistical outlier methods. Before phase calibration, the data is averaged down to 24 channels as a compromise between imaging speed and loss of spectral resolution due to bandwidth smearing. Using models derived from the VLA Low-Frequency Sky Survey \citep[VLSS;][]{Cohen.2007.VLSS} and the NRAO VLA Sky Survey \citep[NVSS;][]{Condon.1998.NVSS}, initial phase calibration solutions are determined, followed by several self-calibration (selfcal) loops. Imaging uses AIPS's polyhedron facet-based wide-field imaging to compensate for non-coplanarity of the GMRT array. After performing several rounds of imaging and selfcal, Obit is used to remove low-level RFI after inspecting the residual visibilities. Ionospheric effects over the full field of view are corrected for using a time-variable phase screen during imaging. The phase screen is fit over the array using direction-dependent gains determined for strong sources in the field of view.

\subsection{Imaging}
\label{subsec:imaging}

For each cluster, we produced a primary beam-corrected, full-resolution image using Briggs $robust = -1$ weighting. The central rms noise and synthesised beam parameters for these images are given in columns seven and eight of Table \ref{tab:obs}, respectively. The full field-of-view primary beam-corrected images are provided in Appendix \ref{app:fovimgs}. Due to the near-zero declination of the clusters, several of the fields are contaminated by residual North-South sidelobes around bright radio sources which could not be improved through SPAM's source peeling. Despite this, the central noise values of the images are similar to those achieved by others with similar on-source time at this frequency \citep[see e.g.][]{Venturi.2008.GRHS2, Kale.2015.EGRHS}.

{For those clusters which showed no evidence for diffuse emission in the full resolution image,} to investigate the presence of low-level diffuse emission we performed a process of point source subtraction and re-imaging. First, the point sources were modelled using the clean components of a high resolution image, produced using a $uv$-cut of {$> 8\; \rm k\lambda$, corresponding to $\sim$ 55 kpc and 190 kpc at the lowest and highest redshifts in our sample, respectively.} This model was then { Fourier transformed and} subtracted from the $uv$-data. The source-subtracted data was then imaged at full resolution to provide a residual image. Finally, the source-subtracted data was imaged at lower resolution, using a $uv$-cut of $< 8\; \rm k\lambda$ and an outer taper of 5 k$\lambda$. {These cuts ensure a fixed angular scale; setting a fixed physical scale for the sample, and adjusting the $uv$-cuts per dataset accordingly is a more optimal approach, to be incorporated in future work.} We discuss the results of this imaging in the following sections.

\subsection{Point source contamination}
\label{subsec:ptsrccontam}

When determining flux densities for emission revealed in the source-subtracted, low-resolution images, we need to consider the effect of subtracting compact emission from the $uv$-data and to what extent the source subtraction contaminates the low resolution image. To this end we follow a similar procedure as in previous work \citep[see Section 4.2 of][]{Knowles.2016.J0256} by performing a statistical analysis of the low resolution, source-subtracted images, using compact source and source-free positions. We summarise the main steps here.

For a given dataset, we first use the source model to create a catalog of compact source positions and $\sim$ 100 random off-source positions. Using the low resolution, source-subtracted image, we then calculate the flux density within a low-resolution beam-sized area centred on each position in the catalog. Using these flux densities, we can determine the mean, $\mu$, and standard deviation, $\sigma$, of the off-source and compact source flux density populations. The bias in subtraction of compact source emission is quantified by the mean of the on-source population, $\mu_{\rm srcs}$. The systematic uncertainty introduced by the subtraction process, $\sigma_{\rm syst}$, is contained within the standard deviation of the on-source population, i.e., $\sigma_{\rm srcs}^2 = \sigma_{\rm off-src}^2 + \sigma_{\rm syst}^2$, where $\sigma_{\rm off-src}$ is effectively the map uncertainty, or rms noise.

We use these statistical measurements to correct the measured flux densities of low resolution sources, and incorporate the systematic uncertainties introduced by the point source removal into the flux density uncertainties. The flux density uncertainties for all 610 MHz source measurements include a $\sim$ 5\% absolute flux calibration and residual amplitude error \citep{Chandra.2004.GMRTfluxerr}. The flux density, $S$, of low-resolution sources and the corresponding uncertainty, $\Delta{S}$, are therefore calculated as follows: \begin{align}
 S = S_{\rm meas} - \left(\mu_{\rm srcs}\times N_S\right)\indent \label{eqn:snu}\\
 \Delta{S}^2 = \left(0.05 S\right)^2 + N_S \left(\sigma^2_{\rm rms} + \sigma^2_{\rm syst}\right) \label{eqn:deltasnu}
\end{align}

\noindent where $\sigma_{\rm rms}$ is the central map noise, and $N_S$ is the number of independent beams within the flux aperture.

\section{New cluster diffuse emission detections}
\label{sec:clusterDE}
The observations presented in this paper are summarised in Table \ref{tab:obs}. The imaging process revealed diffuse radio emission in the cluster region in the form of a mini-halo and a radio halo in ACT-CL J0022.2$-$0036 and ACT-CL J0256.5+0006, respectively. Residual low-resolution emission is found in clusters ACT-CL J0014.9$-$0056 and ACT-CL J0045.2$-$0152. Here we present and briefly discuss the imaging results for these clusters.

\begin{figure}
 \centering
 \includegraphics[width=0.48\textwidth,clip=True,trim=70 10 100 10]{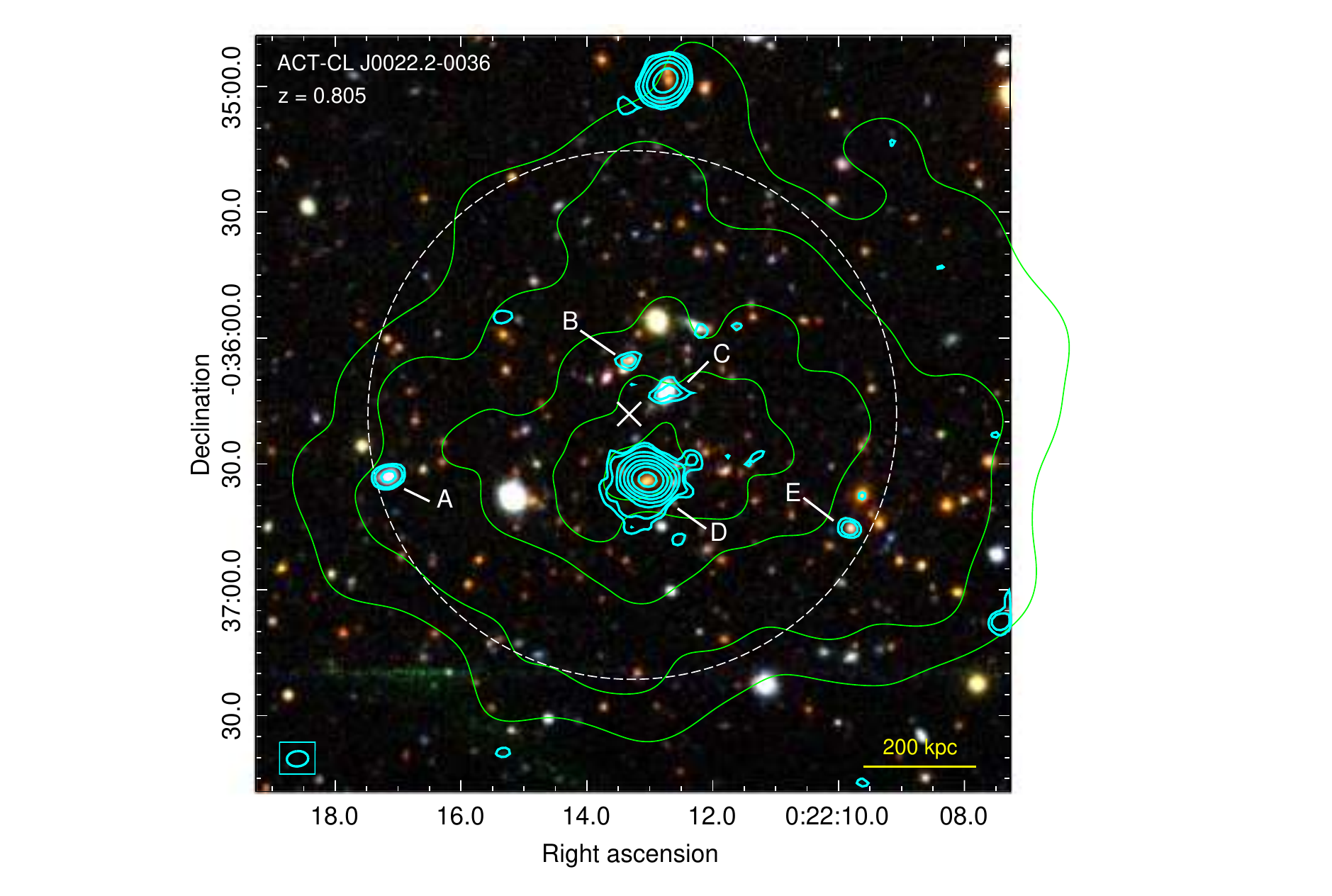}
 \caption{SDSS DR12 composite $gri$-image of ACT-CL J0022.2$-$0036 with the primary beam-corrected 610 MHz contours overlaid in cyan, with contour levels [3,5,10,20,50,100,200,500]$\sigma$, where $1\sigma = 37\; \mu$Jy/beam. The synthesised beam is 5.2{\arcsec} $\times$ 3.6{\arcsec}, p.a. 84.2\degrees, and is shown by the boxed ellipse. There is no $-3\sigma$ flux in the image. \textit{Chandra} contours, as in Figure \ref{fig:j0022_spec}, are shown in green. The $R_{500}$ cluster region and SZ peak are indicated by the dashed circle and X, respectively. The yellow bar at the bottom right shows the physical scale at the cluster redshift. {Flux densities} for sources A-E are given in Table \ref{tab:j0022}.}
 \label{fig:j0022}
\end{figure}

\subsection{ACT-CL J0022.2$-$0036: A radio mini-halo}

ACT-CL J0022.2$-$0036 (hereafter J0022) is the highest redshift cluster in our sample at $z = 0.805$. Our full resolution image, shown by cyan contours in the right panel of Figure \ref{fig:j0022}, reveals the presence of {extended emission around source D, identified as the BCG, which we characterise as a radio mini-halo.} This is the highest redshift mini-halo discovered to date, with the previously most distant detection being in the Phoenix cluster which lies at a redshift of $z = 0.596$ \citep{vanWeeren.2014.phoenixMH}. 

To confirm the presence of the extended emission, we model the compact BCG emission with a single 2D Gaussian and fit it to source D, allowing the central position, size ($\sigma_x$ and $\sigma_y$), angle $\theta$, and amplitude $I_{0_{\rm BCG}}$ to vary. After subtracting the best fit model, residual emission at the 6$\sigma$ level remains, confirming the presence of low level extended emission attributed to a mini-halo. The residual has an angular extent of 19.4\arcsec, corresponding to a physical size of 146 kpc at the redshift of the cluster. To attempt to disentangle the BCG and mini-halo emission, we model the mini-halo with an exponential profile \citet{Murgia.2009.mh}, adding a 2D exponential component of the following form to the above Gaussian model:
\begin{equation}
 I_{\rm MH}(x,y) = I_{0_{\rm MH}} \exp \left(\sqrt{(x - x_0)^2 + (y - y_0)^2} / r_e\right),
\end{equation}
where $I_{0_{\rm MH}}$ is the central radio surface brightness, and $r_e$ is the radius at which the brightness falls to $I_0/e$. We convolve the exponential term with the 610 MHz synthesised beam before adding the Gaussian component, and fit for this combined model in the 2D plane. The mini-halo fit parameters obtained are therefore deconvolved parameters.

With the two-component model we achieve a significant $\Delta\chi^2$ improvement over the Gaussian-only model, with a $p$-value $<$ 0.001 indicating a strong preference for the mini-halo component. After subtracting the best fit model, residual emission remains at the 2$\sigma$ level in the region of the mini-halo. A more complex model is therefore required to account for the full structure, however this is beyond the scope of this paper.

We obtained the following best fit values for parameters of interest in our two-component model: $I_{0_{\rm BCG}} = 26.5 \pm 0.1$ mJy beam\per, $I_{0_{\rm MH}} = 0.7 \pm 0.2$ mJy beam\per, and $r_e = 5.5 \pm 3.1$\arcsec. By integrating the fitted Gaussian out to the full width half maximum, we calculate a BCG flux density of $S_{\rm BCG} = 46.8 \pm 0.2$ mJy. As the mini-halo effective radius is poorly constrained, we estimate a mini-halo flux density by assuming the mean fit value for $r_e$ and integrating the best fit exponential model out to $3r_e$, as per \citet{Murgia.2009.mh}, such that $S_{\rm  MH} = 2\pi f r_e^2 I_{0_{\rm MH}}$, where $f = (1 - 4e^{-3})$. This gives a mini-halo flux density of $S_{\rm MH} = 6.4 \pm 1.8$ mJy. We note that IC losses are expected to be dominant at the cluster redshift, and could affect the synchrotron power by more than an order of magnitude. Higher resolution observations of this system would improve our ability to better constrain the physical properties of the mini-halo.

In addition to the mini-halo around the BCG, our full resolution image shows four compact radio sources in the cluster region, labelled A-C, and E. Through spectroscopic matching, we determine that sources A and B belong to the cluster, with redshifts of 0.8016 and 0.8001, respectively. Sources C and E also have spectroscopic redshifts, identifying them as emanating from foreground ($z = 0.1593$) and background ($z = 0.8236$) galaxies, respectively. The SDSS Data Release 12 \citep[DR12;][]{Alam.2015} image of the cluster, with our full resolution 610 MHz contours overlaid, is shown in Figure \ref{fig:j0022}. The 610 MHz {flux densities} for sources A-E are listed in Table \ref{tab:j0022}. Only the BCG (source D) is detected in FIRST, with a 1.435 GHz {flux density} of $S_{\rm D,FIRST} = 19.02 \pm 0.93$ mJy. No extended mini-halo emission is visible in the FIRST map. { Using this FIRST flux density and the 610 MHz fitted value, we determine an integrated spectral index of $\
alpha_{608}^{1435} = -0.9 \pm 0.1$ for the BCG. }

\begin{table}
 \centering
 \caption{610 MHz {flux densities} for the sources in the ACT-CL J0022.2$-$0036 $R_{500}$ cluster region. Source labels are indicated in Figure \ref{fig:j0022}. Identifications are made spectroscopically, unless indicated by $^*$ when visual colour identification was made: M - cluster member, F - foreground galaxy, B - background galaxy. The BCG is denoted by $^\bullet$. $^\dagger$ The {flux density} for source D is measured for the entire structure, including the mini-halo emission. }
 \label{tab:j0022}
 \begin{tabular}{cccccc}
  \toprule
  ID & R.A. & Dec. & $S_{\rm 610 MHz}$ & Notes\\
   & (deg) & (deg) & (mJy) & \\
  \midrule
  A & 5.571470 & -0.608780 & 0.53 $\pm$ 0.07 & M  \\
  B & 5.555534 & -0.601103 & 0.25 $\pm$ 0.06 & M  \\
  C & 5.552970 & -0.603190 & 0.75 $\pm$ 0.05 & F  \\
  D & 5.554318 & -0.609001 & 36.22 $\pm$ 1.81$^\dagger$ & M$^\bullet$  \\
  E & 5.540945 & -0.612209 & 0.25 $\pm$ 0.06 & B\\
  \bottomrule
 \end{tabular}
\end{table}

\begin{figure*}
 \centering
 \includegraphics[width=0.45\textwidth,clip=True,trim=95 25 115 25]{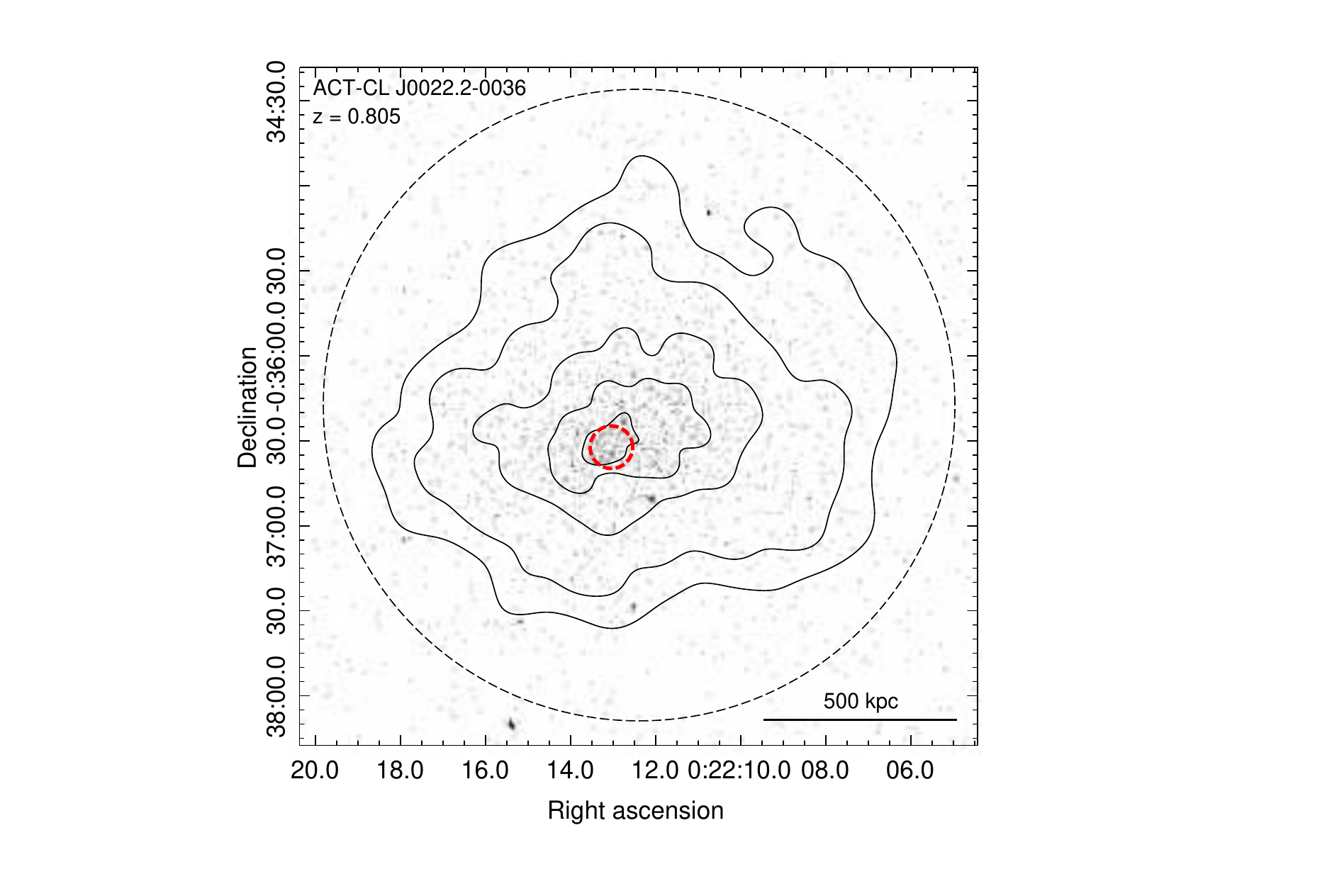}
 \includegraphics[width=0.45\textwidth,clip=True,trim=0 100 10 110]{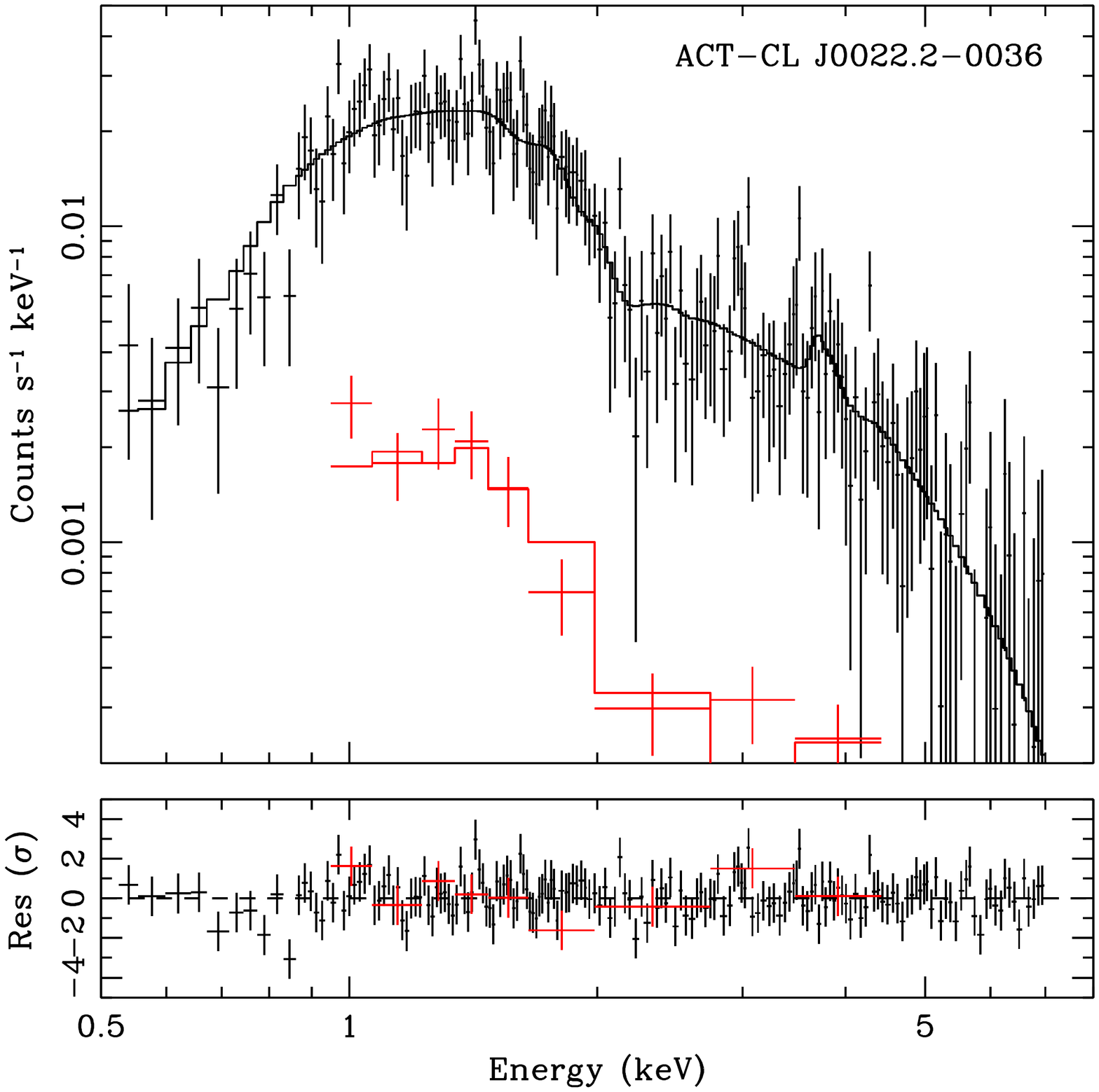}
 \caption{\textit{Left:} Raw \textit{Chandra} counts image superimposed with adaptively smoothed contours from the exposure-corrected, background-subtracted image. { Contours levels are [0.75, 1.6, 3.5, 7.5, 17.5]$\;\times\; 10^{-9}$ cts s{\per} cm$^{-2}$ arcsec$^{-2}$.} Dashed thin black and dashed thick red circles indicate the regions from which spectra were obtained, namely the total cluster and the mini-halo/BCG region, respectively. See text for details. \textit{Right:} X-ray spectra obtained from the exposure-corrected, background-subtracted, adaptively smoothed \textit{Chandra} image of J0022.2$-$0036. The red spectrum was obtained from the mini-halo/BCG region (red dashed circle in the left panel of Figure \ref{fig:j0022}). The black spectrum was obtained from the total cluster region (black solid circle in the left panel of Figure \ref{fig:j0022}) minus the mini-halo/BCG region. Table \ref{tab:j0022_spec} shows the spectral fits. }
 \label{fig:j0022_spec}
\end{figure*}

Both optical spectroscopy and \textit{Chandra} X-ray imaging are available to investigate the dynamical state of J0022. Spectroscopic observations of 55 cluster galaxies by \citet{Sifon.2016} show tentative evidence of dynamical disturbance in the cluster. Their DS test results indicate marginal evidence for substructure, with a significance value\footnote{A value of $\Delta S < 0.05$ is understood to indicate substructure.} of $S_\Delta = 0.07^{+0.18}_{-0.03}$. \citet{Sifon.2016} identify the BCG at a redshift of 0.8096 $\pm$ 0.0003, coinciding with source D. 

\begin{table}
 \centering
 \caption{X-ray Spectral fits to the total cluster and the mini-halo/BCG regions shown in the left panel of Figure \ref{fig:j0022_spec}. The redshift and column density were fixed at $z = 0.805$ and $N_H = 2.76 \times 10^{20}$ atoms/cm$^2$, respectively.}
 \label{tab:j0022_spec}
 \begin{tabular}{llcccc}
  \toprule
   & &  $\chi^2$ & d.o.f. & $kT$ & Fe abundance\\
  \midrule  
  \multicolumn{5}{l}{\underline{Mini-halo}}    \\
  & fixed Fe & 10.1 & 7 & 4.2$^{+1.9}_{-1.1}$ & 0.3 \\
  & free Fe & 8.6 & 6 & 3.2$^{+1.4}_{-0.7}$ & 1.4$^{+1.3}_{-0.9}$\\
  \multicolumn{5}{l}{\underline{Total cluster minus mini-halo}} \\
  & free Fe & 154 & 180 & 7.9$^{+1.0}_{-0.8}$ & 0.19$\pm$ 0.12 \\
  \bottomrule
 \end{tabular}

\end{table}

J0022 was observed for 64 ks with \textit{Chandra} ACIS-I (ObsID: 16226, PI: Hughes) in 2014. The exposure time after cleaning is $\sim$61 ks. A basic data reduction was followed using CIAO version 4.6 (CALDB 4.6.3), with filtering carried out on grade, status, VFAINT mode, and time. Finally, astrometry was corrected for based on cross matching between X-ray point sources and optical stars from SDSS/S82 images. 

X-ray image analysis was carried out over the (0.5--2.0)\,keV band. Point sources were removed and their pixel values replaced with Poisson-distributed random variables, based on the estimated local {flux density}. The image was then exposure-corrected, background-subtracted using blank-sky-background files, and adaptively smoothed with a Gaussian kernel whose sigma increases with decreasing intensity. A raw counts image of J0022 with minimal processing is shown in the left panel of Figure \ref{fig:j0022_spec}. There is evidence of asymmetry within the cluster core, however we do not attempt to estimate its significance. The green contours overlaid on the counts image are from the fully processed, adaptively smoothed \textit{Chandra} image.

\begin{figure*}
 \centering
 \includegraphics[width=0.45\textwidth, clip=True, trim=30 20 40 50]{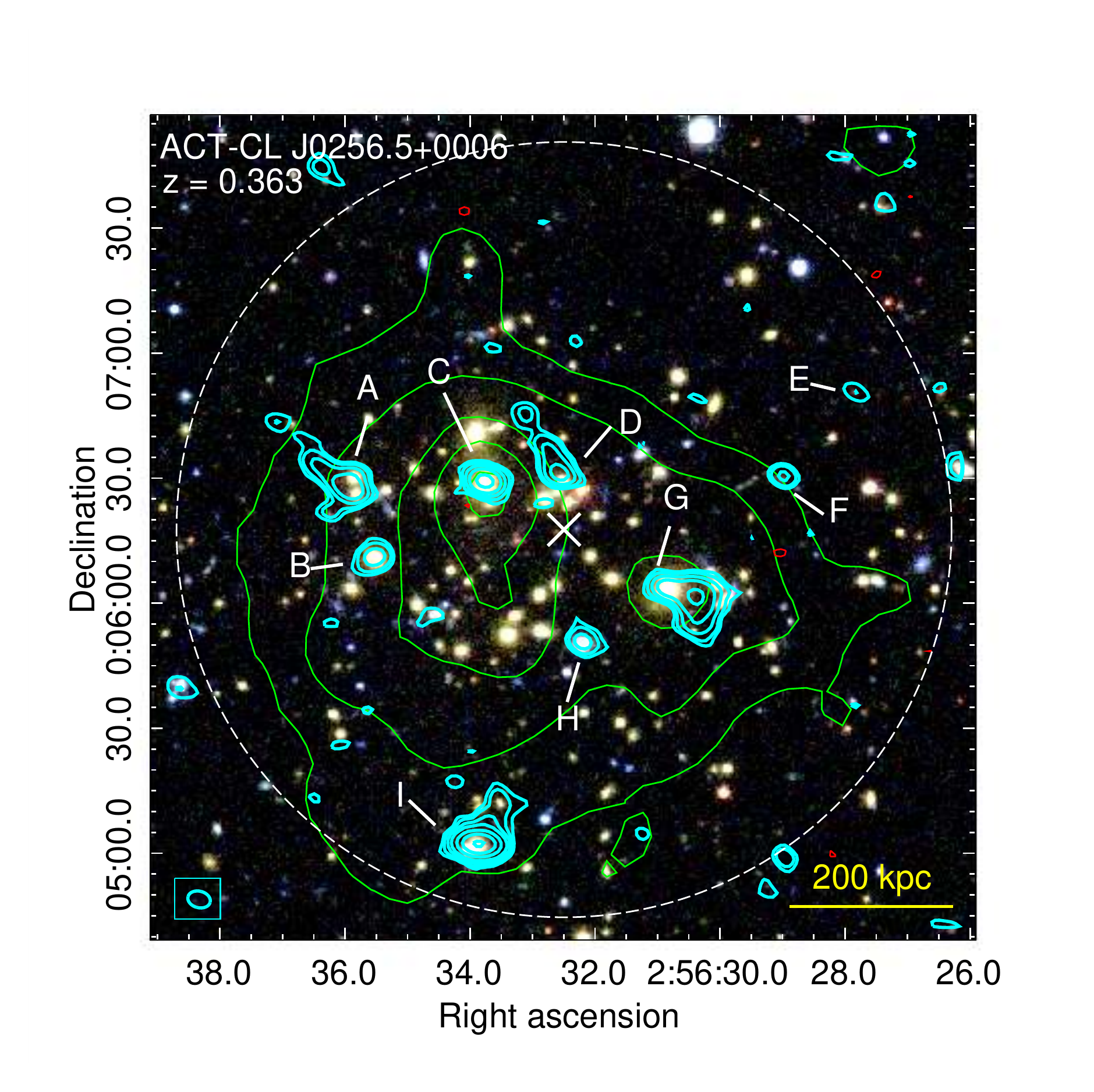}
 \includegraphics[width=0.49\textwidth, clip=True, trim=30 10 60 0]{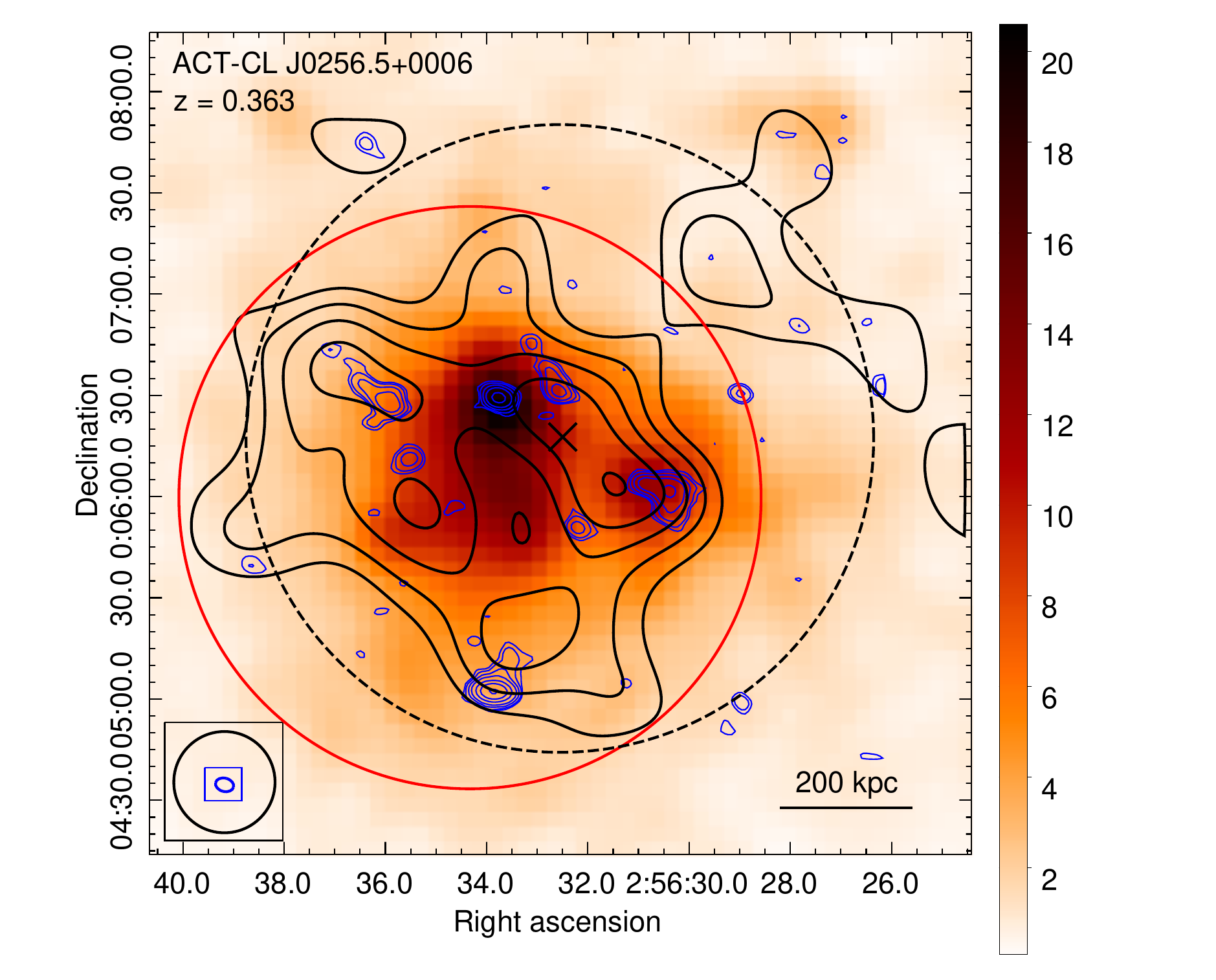}
 \caption{ACT-CL J0256.5+0006. \textit{Left:} SDSS DR12 $gri$-image with full resolution 610 MHz GMRT contours overlaid (+ve: thick, cyan, -ve: thin, red). Contour levels are [$\pm$3, 5, 10, 20, 50, 100]$\sigma$, where 1$\sigma = 22\; \mu$Jy beam{\per}. Thin, green contours {([2.6, 5, 10, 14, 19] counts)} indicate the X-ray emission shown in the right pane. \textit{Right:} Exposure corrected, smoothed \textit{XMM-Newton} EPIC image with 610 MHz GMRT full-resolution (thin, blue; same as the cyan in left panel) and low resolution source subtracted (thick, black; [3,4,5,6]$\sigma$, where 1$\sigma = 80\; \mu$Jy beam\per) contours overlaid. {The colourbar is in units of counts, and the image is smoothed with a 3px Gaussian kernel.} The halo {flux density} was measured within the region indicated by the solid red circle. In both panels, beams are shown by the boxed ellipses, and the X and dashed circle are as in Figure \ref{fig:j0022}. The bar in the bottom right of each image shows the physical scale at the cluster redshift. {Flux densities} for sources A-I and the radio halo are provided in Table \ref{tab:j0256}.}
 \label{fig:j0256}
\end{figure*}

Finally, a spectral analysis was carried out in two regions of the cluster after point source subtraction: the mini-halo/BCG only, and the total cluster minus the mini-halo region. These are indicated by the dashed red and black circles, respectively, superimposed on the counts image shown in the left panel of Figure \ref{fig:j0022_spec}. The right panel of Figure \ref{fig:j0022_spec} shows both spectra (red: mini-halo/BCG; black: total minus mini-halo). Using fixed values of $z = 0.805$ and $N_H = 2.76 \times 10^{20}$ atoms/cm$^2$, we obtain good spectral fits, assuming two cases for the mini-halo region: a fixed iron abundance of 0.3, and a free iron abundance. The fit parameters for both spectra are given in Table \ref{tab:j0022_spec}. 

As mini-halos are typically found in cool-core clusters, we use these results to determine how different the X-ray temperature of the mini-halo is compared to the rest of the cluster. For the fixed iron case ($kT_{\rm MH} = 4.2^{+1.9}_{-1.1}$ keV), we find the temperature difference to be significant at approximately 1.5$\sigma$ (86.0\% C.L., $\Delta\chi^2 = 2.18$). In the case of a free iron abundance ($kT_{\rm MH} = 3.2^{+1.4}_{-0.7}$ keV), the temperature difference is significant at $\sim 2\sigma$ (94.4\% C.L., $\Delta\chi^2 = 3.65$). Therefore, given the current \textit{Chandra} data, there is an indication that the X-ray gas in the region of the radio mini-halo is somewhat cooler than the rest of the cluster. However, this result is not statistically significant.

Additional X-ray and spectroscopic data may provide a clearer picture of the dynamical state of this cluster. The irregularities seen in the raw X-ray counts image may indicate gas sloshing in the cluster core{, however, this is currently speculative given the current data.}

\subsection{ACT-CL J0256+0006: A radio halo}
\label{subsubsec:rh}

ACT-CL J0256+0006 ($z = 0.363$) is a known merging cluster based on \textit{XMM-Newton} observations \citep{Majerowicz.2004.J0256} and confirmed by a DS test significance value of $S_\Delta = 0.003^{+0.027}_{-0.001}$. Our data revealed a radio halo in this cluster after removing the several compact radio sources from the cluster region. The detection is presented in detail in \citet{Knowles.2016.J0256}, where we measured a halo {flux density} of $S_{\rm 610 MHz} = 5.6 \pm 1.4$ mJy. Since publishing the detection, where the halo was detected only in heavily smoothed images, we have reprocessed the radio data with updated software, leading to an improvement in the central noise of the full-resolution image of 15\%. Although the halo is still not visible in the full resolution image, several $3\sigma$ residuals can be seen in the cluster region before source subtraction. 

The radio halo, shown by the thick black contours in the left panel of Figure \ref{fig:j0256}, is detected in the smoothed source-subtracted image, which has a resolution of beam$_{\rm LR}\; =\;$ 30{\arcsec} $\times$ 30{\arcsec}, p.a. 0\degrees. The subtracted sources are shown by the thin blue contours from the full-resolution image. These sources are discussed below. The solid circle indicates the region in which the halo {flux density} was measured: accounting for a source-subtraction bias of $\mu_{\rm srcs} = -32 \pm 35 \;\mu$Jy beam$_{\rm LR}$\per and a systematic subtraction uncertainty of $\sigma_{\rm syst} = 0.1$ mJy beam$_{\rm LR}$\per, we measure a 610 MHz {flux density} of 6.9 $\pm$ 0.7 mJy, in agreement with the {flux density} measured from the previous data reduction, albeit more constrained. This converts to a 1.4 GHz k-corrected radio power of $P_{\rm 1.4 GHz} = (1.2 \pm 0.4) \times 10^{24}$ W Hz{\per} in the cluster rest frame, using a fiducial spectral index of -1.2 $\pm$ 0.2. Based on the UPP $M_{500}$ mass from \citet{Hasselfield.2013.ACTE}, this is still the lowest mass cluster found to date to host radio halo emission.

\begin{table}
 \centering
 \caption{610 MHz {flux densities} for the sources in the ACT-CL J0256.5+0006 $R_{500}$ cluster region. Source labels are indicated in Figure \ref{fig:j0256}. Notes are as in Table \ref{tab:j0022}.}
 \label{tab:j0256}
 \begin{tabular}{cccccc}
  \toprule
  ID & R.A. & Dec. & $S_{\rm 610 MHz}$ & Notes\\
   & (deg) & (deg) & (mJy) & \\
  \midrule
  A & 44.150178 & 0.108155 & 2.28 $\pm$ 0.15 & F/M$^\dagger$  \\
  B & 44.148053 & 0.103056 & 0.52 $\pm$ 0.05 & M  \\
  C & 44.140665 & 0.108134 & 1.75 $\pm$ 0.10 & M$^\bullet$  \\
  D & 44.135829 & 0.108676 & 0.48 $\pm$ 0.05 & M$^*$ \\
  E & 44.115960 & 0.114084 & 0.16 $\pm$ 0.04 & B$^*$ \\
  F & 44.120759 & 0.108472 & 0.27 $\pm$ 0.04 & - \\
  G & 44.128574 & 0.101065 & 4.07 $\pm$ 0.22 & M \\
  H & 44.134114 & 0.097459 & 0.42 $\pm$ 0.05 & F$^*$ \\
  I & 44.141004 & 0.084114 & 7.84 $\pm$ 0.40 & M \\
  \midrule
  RH & 44.143033 & 0.099908 & 6.9 $\pm$ 0.7 & - \\
  \bottomrule
 \end{tabular}
 \justify
 $^\dagger$ Source A encompasses two galaxies. See text for details.
\end{table}

There are nine discrete sources in the cluster $R_{500}$ region detected above $5\sigma$, several of them exhibiting tailed emission, common in merging clusters \citep{Bliton.1998.NATs}. These sources, labelled A-I, are shown in Figure \ref{fig:j0256} and their {flux densities} are given in Table \ref{tab:j0256}. Four sources (B, C, G, I) are cluster members, identified spectroscopically using the \citet{Sifon.2016} data; source C is the BCG. Source G is the probable BCG of the subcluster, based on its spatial match with the subcluster core in the X-ray image shown in Figure \ref{fig:j0256}. This source also exhibits tailed emission, opposite in direction to the subcluster infall. Although the resolution of our 610 MHz image is not fine enough to discern structure within this tail, it is likely that this is a bent tailed radio source experiencing ram pressure stripping due to the merger \citep{Bliton.1998.NATs}. The other three tailed radio sources, A, D, and I, are also confirmed cluster members or suspected members based on colour identification in the SDSS 3-colour $gri$ image shown in Figure \ref{fig:j0256}. We note, however, that source A encompasses two galaxies, one of which is spectroscopically confirmed to be a foreground galaxy at $z = 0.3511$. The directions of the radio tails on these sources naively fit the merger scenario outlined in \citet{Knowles.2016.J0256}, with the exception of source I, the tail of which is directed in projection \textit{towards} the merger centre. Higher resolution VLBI and/or higher frequency GMRT imaging is required to fully investigate the tailed structures in this cluster and understand the geometry of the ongoing merger. Source I is also the only radio source detected in FIRST with a {flux density} of 3.66 {$\pm$} 0.27 mJy. From this we determine an integrated spectral index of $\alpha_I = -0.89 \pm 0.20$. 

\begin{figure*}
 \centering
 \includegraphics[width=0.44\textwidth, clip=True, trim=30 20 40 40]{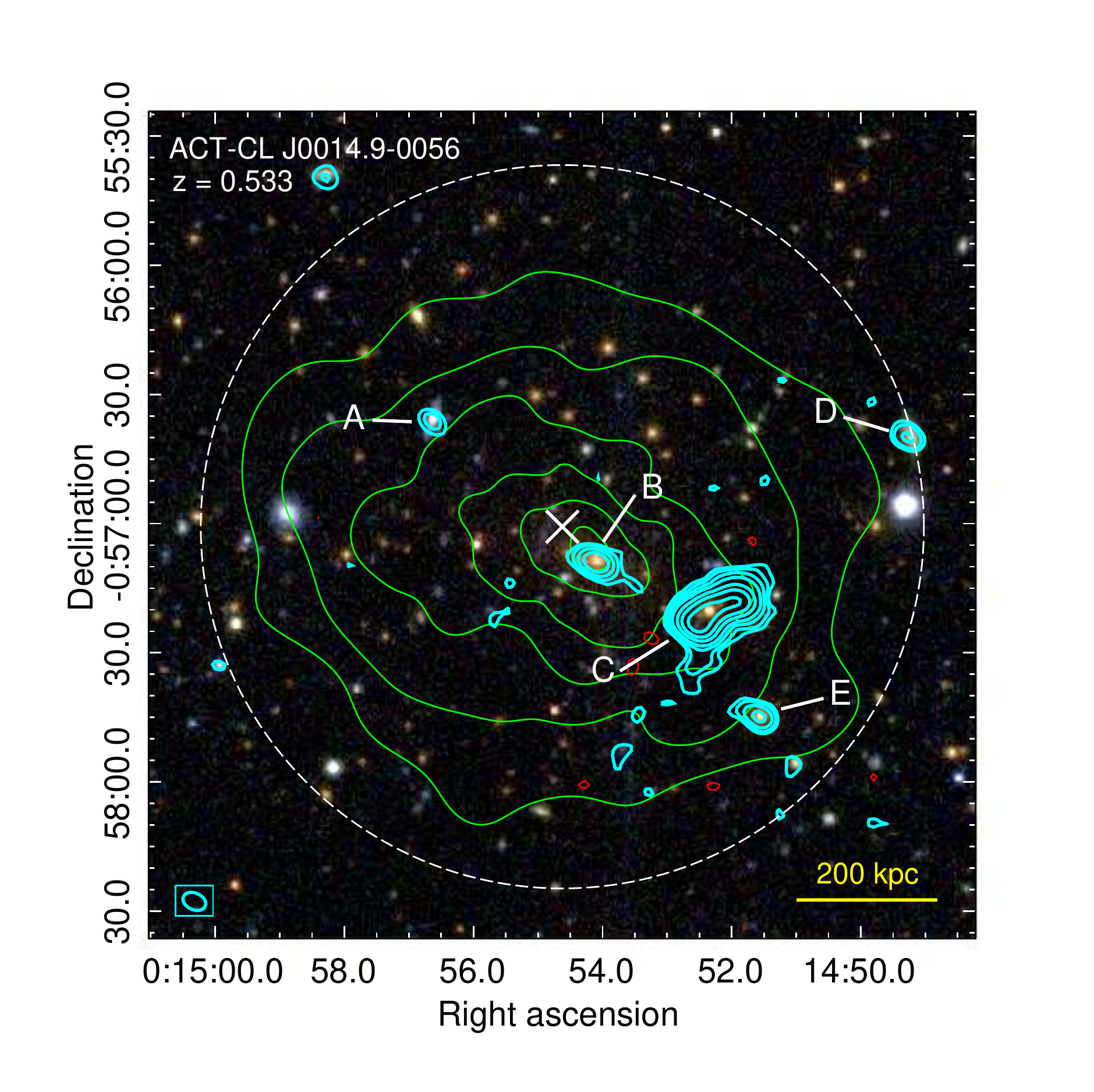}
 \includegraphics[width=0.5\textwidth, clip=True, trim=25 10 45 10]{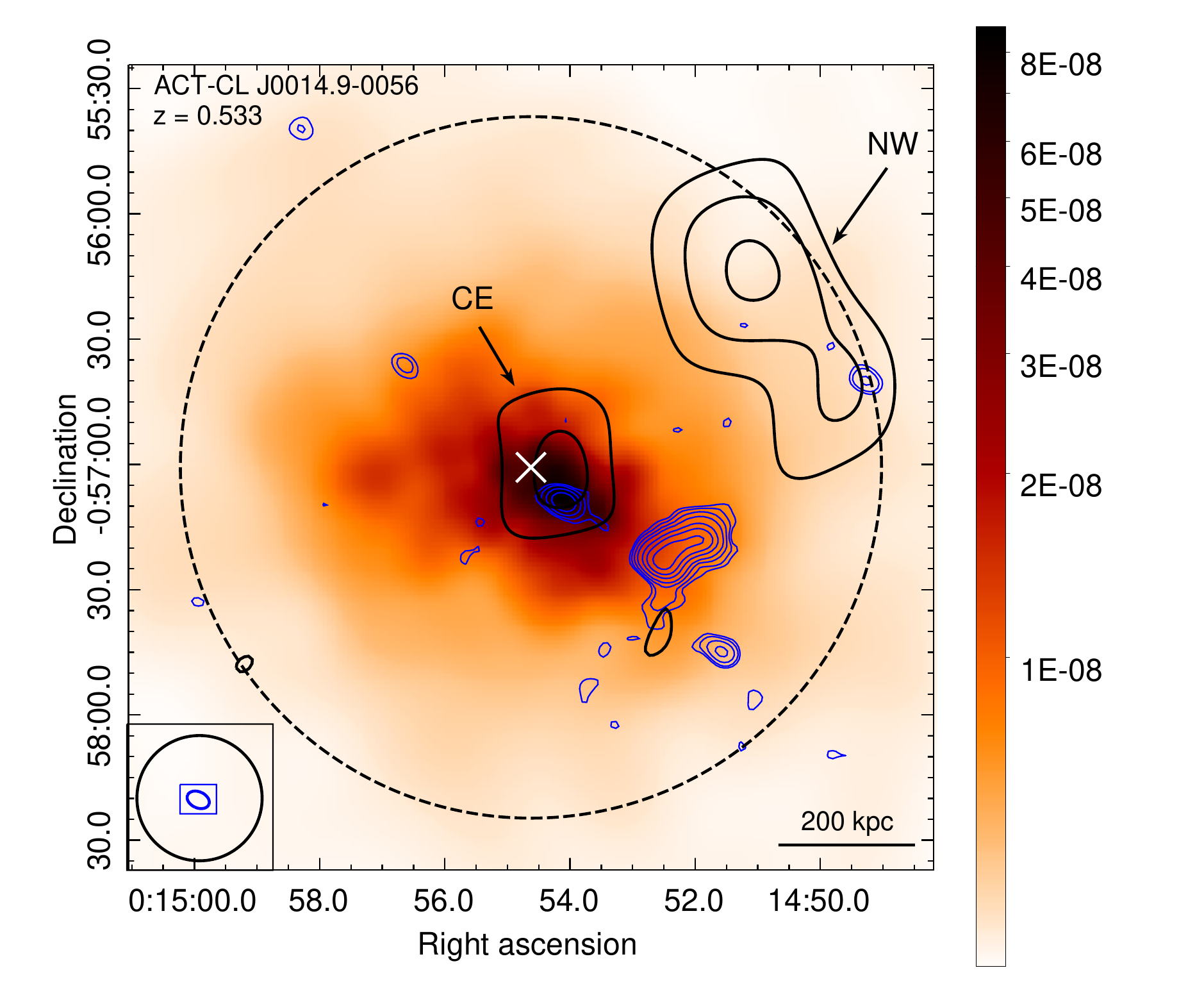}
 \caption{ACT-CL J0014.9$-$0057. \textit{Left:} SDSS DR12 $gri$-image of the cluster region with full resolution 610 MHz GMRT contours (+ve: thick, cyan, -ve: thin, red) overlaid. Contour levels are [$\pm$3,5,10,20,50,100,200]$\sigma$, where 1$\sigma = 35\; \mu$Jy beam\per. The X-ray emission, imaged in the right panel, is indicated by thin, green contours { with levels [2, 4, 8, 16, 32, 64]$\;\times\;10^{-9}$ cts s\per cm$^{-2}$ arcsec$^{-2}$}. \textit{Right:} { Background subtracted, adaptively smoothed, \textit{Chandra} image, overlaid} with contours from the full resolution (thin, blue; as cyan in the left panel) and smoothed, source-subtracted (thick, black; [3,4,5]$\sigma$, where 1$\sigma = 110\; \mu$Jy beam\per) 610 MHz GMRT images. {The colourbar is in units of cts s{\per} cm$^{-2}$ arcsec$^{-2}$.} In both panels, beams are shown by the boxed ellipses, and the X and dashed circle are as in Figure \ref{fig:j0022}. The bar in the bottom right of each image shows the physical scale at the cluster redshift. {Flux densities} for sources A-E and low-resolution sources NW and CE are given in Table \ref{tab:j0014}. }
 \label{fig:j0014}
\end{figure*}

\begin{table}
 \centering
 \caption{610 MHz {flux densities} for the sources in the ACT-CL J0014.9$-$0056 $R_{500}$ cluster region. Source labels are indicated in the left panel of Figure \ref{fig:j0014}. Notes are as in Table \ref{tab:j0022}.}
 \label{tab:j0014}
 \begin{tabular}{cccccc}
  \toprule
  ID & R.A. & Dec. & $S_{\rm 610 MHz}$ & Notes\\
   & (deg) & (deg) & (mJy) & \\
  \midrule
  A & 3.735974 & -0.943412 & 0.34 $\pm$ 0.05 & F  \\
  B & 3.725413 & -0.952494 & 1.37 $\pm$ 0.08 & M$^\bullet$  \\
  C & 3.717799 & -0.955695 & 36.64 $\pm$ 1.84 & M  \\
  D & 3.705313 & -0.944409 & 0.51 $\pm$ 0.06 & F$^*$ \\
  E & 3.714956 & -0.962428 & 1.01 $\pm$ 0.07 & M \\
  \midrule
  NW & 3.712831 & -0.937167 & 1.7 $\pm$ 0.4 & -\\
  CE & 3.725577 & -0.950470 & 0.7 $\pm$ 0.3 & -\\
  \bottomrule
 \end{tabular}
\end{table}

\subsection{ACT-CL J0014.9$-$0057}

ACT-CL J0014.9$-$0056 ($z = 0.533$) was first detected in the Gaussian Mixture Brightest Cluster Galaxy (GMBCG) cluster catalog based on SDSS Data Release 7 \citep[GMBCGJ003.72543-00.95236;][]{Hao.2010.GMBCG}. It is also identified by the Southern Astrophysical Research Gravitational Arc Survey (SOGRAS) as a strong gravitational lensing system with a gravitational arc to the West of the cluster BCG \citep[SOGRAS0014-0057;][]{Furlanetto.2013.SOAR}. 

The dynamical state of this system is unclear, with X-ray imaging from \textit{Chandra} showing a slightly disturbed morphology, while a spectroscopic optical analysis by \citet{Sifon.2016} of 62 cluster members shows no evidence of substructure in the galaxy velocity distribution ($S_\Delta = 0.331^{+0.282}_{-0.132}$). If there is merger activity in this system, these results may indicate that it is occurring in the plane of the sky. Left and right panels of Figure \ref{fig:j0014} show the SDSS DR12 3-colour $gri$-image and \textit{Chandra} (ObsID: 16228; 29 ks) image of the cluster, respectively. Green contours on the SDSS image indicate the \textit{Chandra} emission.

\begin{figure*}
 \centering
 \includegraphics[width=0.44\textwidth, clip=True, trim=30 20 40 30]{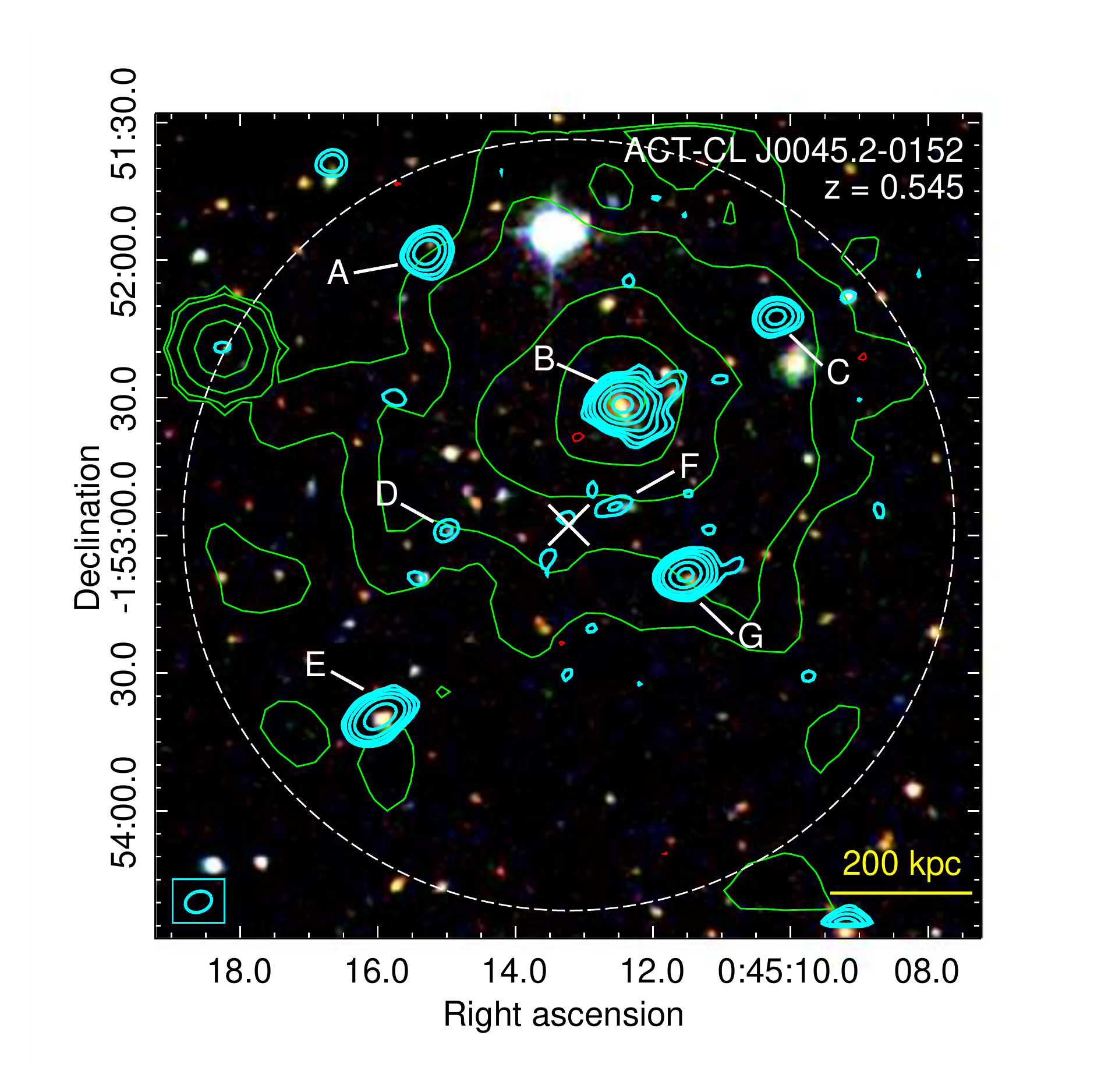}
 \includegraphics[width=0.49\textwidth, clip=True, trim=20 10 65 10]{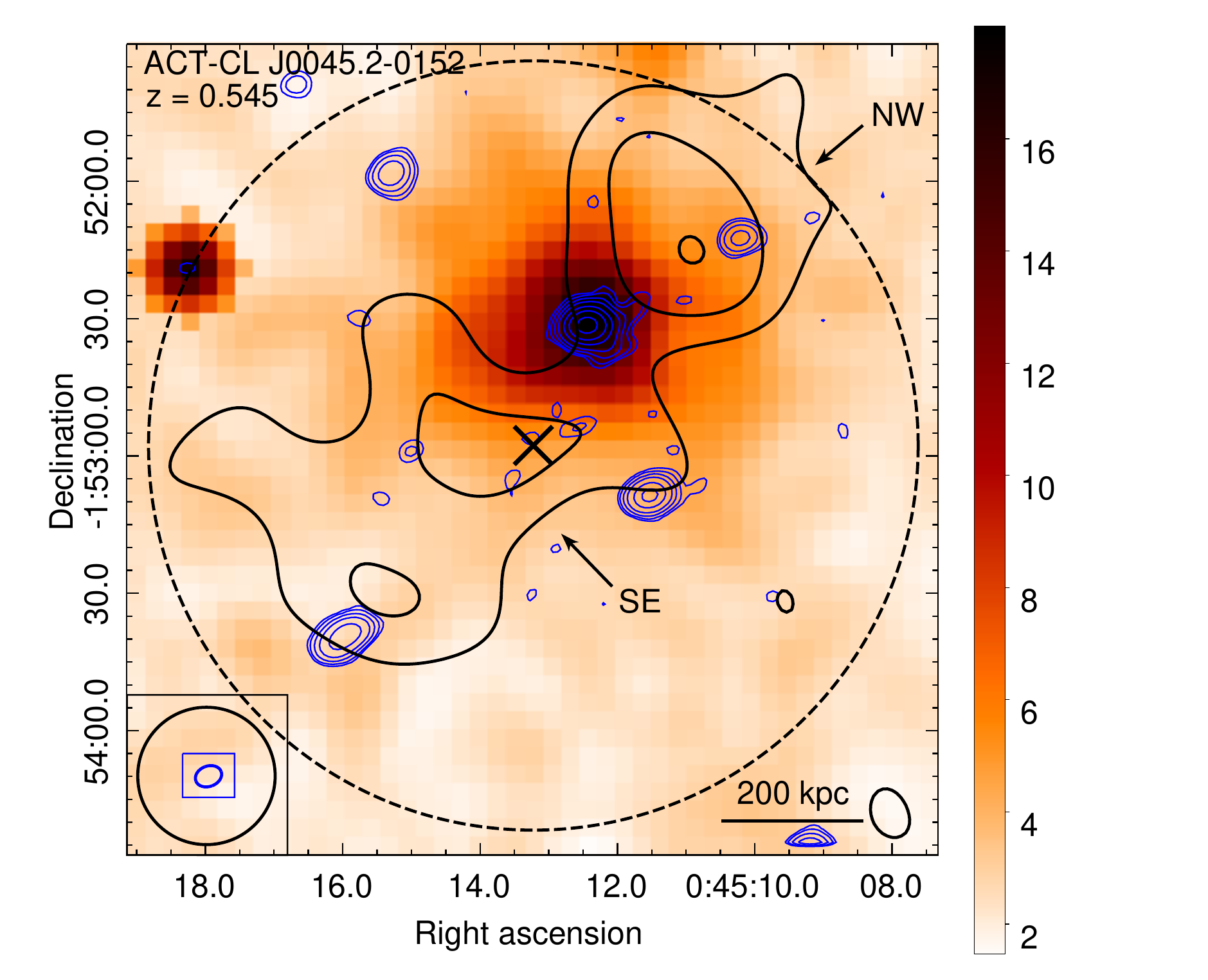}
 \caption{ACT-CL J0045.2$-$0152. \textit{Left:} SDSS DR12 $gri$-image with full resolution 610 MHz contours (+ve: thick, cyan, -ve: thin, red) overlaid, with levels of [$\pm$3,5,10,20,50,100,200]$\sigma$, where 1$\sigma = 35\; \mu$Jy beam\per. The X-ray emission detected by \textit{Chandra} is indicated by thin, green contours { (levels are [3, 3.875, 6.5, 10.875, 17] counts)}. \textit{Right:} \textit{Chandra} image with 610 MHz full resolution (thin, blue; as cyan in left panel) and smoothed, source-subtracted (thick, black; [3,4,5]$\sigma$, where 1$\sigma = 100\; \mu$Jy beam\per) contours overlaid. {The colourbar is in units of counts and the image has been smoothed with a 3px Gaussian kernel.} In both panels, beams are shown by the boxed ellipses, and the X and dashed circle are as in Figure \ref{fig:j0022}. The bar in the bottom right of each image shows the physical scale at the cluster redshift. {Flux densities} for sources A-G and low-resolution sources NW and SE are given in Table \ref{tab:j0045}.}
 \label{fig:j0045}
\end{figure*}

Our source-subtracted, smoothed 610 MHz GMRT image of the cluster region, indicated by black contours in the right panel of Figure \ref{fig:j0014}, shows an extended source North-West of the cluster SZ peak, indicated by `NW' in the figure, which we classify as a candidate relic. Incorporating the uncertainty and bias due to the source subtraction, we measure a 610 MHz {flux density} for this source of $S_{\rm NW} = 1.7 \pm 0.4$ mJy. Similarly, the unresolved residual emission in the centre of the cluster, indicated by `CE' (central emission), has a measured 610 MHz {flux density} of $S_{\rm CE} = 0.7 \pm 0.3$ mJy. It is marginally offset from the BCG by $\sim$ 5.6{\arcsec} and could be the peak of a low power radio halo which is below the noise of the current data. It could also be a low power radio mini-halo, however the spatial offset from the BCG is peculiar if this is the case. An alternative explanation is that this residual emission could be an old radio lobe, possibly related to the compact radio source B. {Using the noise in the FIRST image of this cluster, we deterime a spectral index upper limit of $\alpha_{\rm CE}\;=\;-3.1$, which makes the old radio lobe scenario more likely.} Additional radio data would be required to confirm the presence of these faint sources, with improved estimates of the cluster dynamical state necessary for accurate classification. {We note that IC losses may be retarding the synchrotron signals: at the cluster redshift of $z=0.533$, the observed synchrotron power could be a factor of $\sim$ 0.2 lower than expected.}

The discrete compact sources subtracted from the data are shown by the cyan (blue) contours in the left (right) panel of Figure \ref{fig:j0014}. There are five sources detected above $5\sigma$, labelled A-E, and their 610 MHz {flux densities} are given in Table \ref{tab:j0014}. All but source D coincide with spectroscopically identified galaxies from \citet{Sifon.2016}, with the redshifts indicating that sources B (the BCG), C, and E are cluster members at redshifts of 0.5344, 0.5354, and 0.5347 respectively, and source A is a foreground source at $z = 0.1094$. Based on colour matching in the SDSS $gri$-image, source D is likely a foreground galaxy. Source C is extended but the resolution of our 610 MHz images is insufficient to determine its structure. Two possible classifications are a head-tail galaxy, or a FR-II source with the Eastern lobe angled towards us. Based on the position of the associated cluster member, in the centre of the extended source, the latter classification is more likely. Source C is also the only radio source detected in FIRST, with a {flux density} of 15.11 $\pm$ 1.52 mJy. We {determine an integrated} spectral index of $\alpha_C = -1.03 \pm 0.27$. 

\subsection{ACT-CL J0045.2$-$0152}

ACT-CL J0045.2$-$0152 ($z = 0.545$), otherwise known as WHL J004512.5-015232, was discovered in SDSS-III by \citet{Wen.2012}. \citet{Sifon.2016} performed an optical spectroscopic analysis of 56 cluster members and calculated a DS test value of $S_\Delta = 0.024^{+0.024}_{-0.002}$, indicating a disturbed system with significant substructure. \textit{Chandra} X-ray imaging of this cluster (ObsID: 19588; 50 ks) shows a fairly regular morphology, with the X-ray peak coincident with the cluster BCG, as shown by the green contours overlaid on the 3-colour SDSS DR12 image of the cluster in the left panel of Figure \ref{fig:j0045}. From the \textit{Chandra} data, we measure an X-ray temperature for the cluster of $T_{\rm X} = 8.5 \pm 0.9$ keV. {This system may therefore be undergoing a line of sight merger.}

Our 610 MHz full resolution, primary beam corrected image reveals seven compact radio sources in the cluster region, labelled A-G, along with positive residuals at the 3$\sigma$ level, as shown by the cyan (blue) contours in the left (right) panel of Figure \ref{fig:j0045}. The 610 MHz {flux densities} for these sources are given in Table \ref{tab:j0045}. Three of the sources, B, E, and G, are spectroscopically identified as cluster members at redshifts of 0.5486, 0.5535, and 0.5399, respectively, with source B identified as the cluster BCG. All three sources are detected in FIRST, with 1.4 GHz {flux densities} of $S_{\rm 1.4 GHz, B} = 4.57 \pm 0.49$ mJy, $S_{\rm 1.4 GHz, E} = 1.94 \pm 0.43$ mJy, and $S_{\rm 1.4 GHz, G} = 2.09 \pm 0.38$ mJy, allowing for integrated spectral index measurements of $\alpha_B = -1.19 \pm 0.29$, $\alpha_E = -0.96 \pm 0.60$, and $\alpha_G = -0.85 \pm 0.49$. Source C has no optical counterpart in the SDSS DR12 $gri$-image shown in Figure \ref{fig:j0045}. Based on colour matching from the SDSS data, sources A and D are probable background sources, and source F is a foreground source.

After subtracting compact sources A-G, the smoothed source-subtracted 610 MHz image reveals a bridge of faint emission spanning almost the entire $R_{500}$ region of the cluster, shown by the black contours in the right panel of Figure \ref{fig:j0045}. We label this source in two parts, namely `NW' and `SE' as indicated in the figure. We measure 610 MHz {flux densities} of $S_{\rm NW} = 1.5 \pm 0.2$ mJy and $S_{\rm SE} = 3.0 \pm 0.4$ mJy, respectively, after taking into account the uncertainty and bias due to the source subtraction. Although the emission is detected with low significance, we suggest possibilities for its classification. Based on the emission position relative to the cluster SZ $R_{500}$ region, it is possible that SE source is a candidate radio halo, whereas the NW source may be a candidate radio relic. However, the X-ray emission is significantly offset from the SE source which is uncommon for radio halos. Another possibility is that the SE and NW sources are lobes from AGN jets emanating from the cluster BCG, based on their symmetric morphology around source B, however it is uncommon for AGN jets to extend out to $R_{500}$. {With IC losses potentially reducing the synchrotron signal by a factor of $\sim$ 0.2,} deeper and/or multi-frequency radio data are required to unveil the origin of this emission.

\begin{table}
 \centering
 \caption{610 MHz {flux densities} for the sources in the ACT-CL J0045.2$-$0152 $R_{500}$ cluster region. Source labels are indicated in Figure \ref{fig:j0045}. Notes are as in Table \ref{tab:j0022}.}
 \label{tab:j0045}
 \begin{tabular}{cccccc}
  \toprule
  ID & R.A. & Dec. & $S_{\rm 610 MHz}$ & Notes\\
   & (deg) & (deg) & (mJy) & \\
  \midrule
  A & 11.313699 & -1.866189 & 1.56 $\pm$ 0.09 & B$^*$  \\
  B & 11.301696 & -1.875454 & 12.71 $\pm$ 0.64 & M$\bullet$  \\
  C & 11.292533 & -1.870118 & 0.98 $\pm$ 0.07 & -  \\
  D & 11.312520 & -1.883044 & 0.19 $\pm$ 0.05 & B$^*$ \\
  E & 11.316560 & -1.894312 & 4.43 $\pm$ 0.23 & M \\
  F & 11.302300 & -1.881644 & 0.19 $\pm$ 0.05 & F$^*$ \\
  G & 11.298039 & -1.885726 & 4.34 $\pm$ 0.22 & M \\
  \midrule
  NW & 11.295482 & -1.870882 & 1.5 $\pm$ 0.2 & -\\
  SE & 11.309153 & -1.883260 & 3.0 $\pm$ 0.4 & -\\
  \bottomrule
 \end{tabular}
\end{table}

\section{Disturbed clusters with non-detections}
\label{sec:nondets}
The remaining ten clusters in our sample show no evidence of diffuse radio emission at the level of the noise in our images. From X-ray imaging and optical spectroscopy, we estimate that six of these systems are experiencing ongoing or recent merger activity. Here we briefly discuss these six clusters, which include two known massive mergers, ACT-CL J0239.8$-$0.134 (Abell 370) and ACT-CL J2135.2+0125 (Abell 2355), and catalog the 610 MHz radio sources within the cluster $R_{500}$ regions. Discussion of the remaining four clusters with no diffuse emission is provided in Appendix \ref{app:nondets}. 

\subsection{ACT-CL J0059.1$-$0049}

\begin{figure}
 \centering
 \includegraphics[width=0.46\textwidth,clip=True, trim=20 20 20 40]{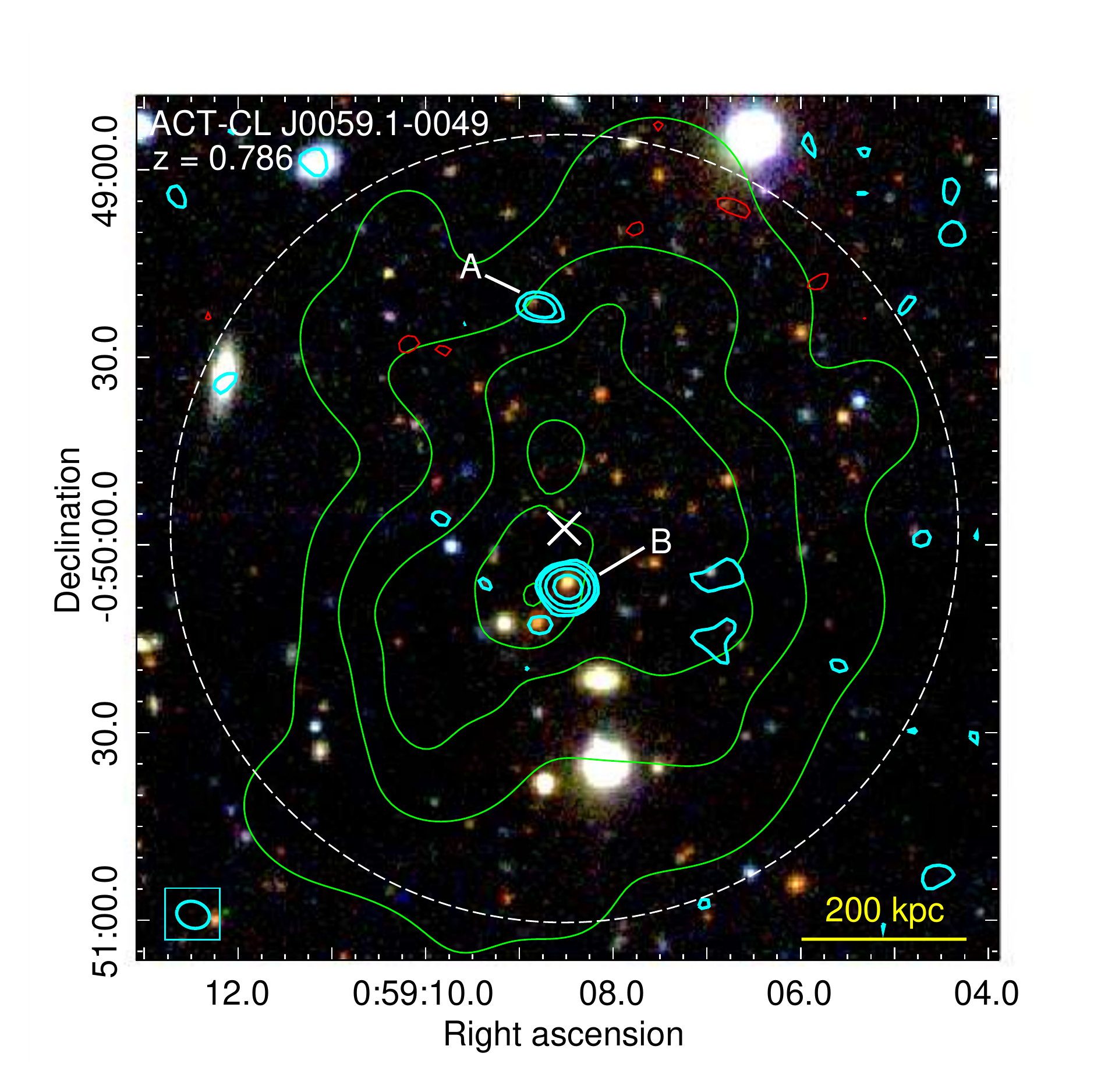}
 \caption{SDSS DR12 3-colour \textit{gri}-image of ACT-CL J0059$-$0049 with 610 MHz GMRT contours overlaid (+ve: thick, cyan, -ve: thin, red). Contour levels are $[\pm3,5,10,20]\sigma$, where $1\sigma = 30\; \mu$Jy beam\per, with the beam shown by the boxed ellipse. Thin, green contours ({levels: [1.65, 3.4, 6.2, 11.7, 23.0]$\;\times\;10^{-9}$ cts s\per cm$^{-2}$ arcsec$^{-2}$}) indicate \textit{Chandra} X-ray emission. The cross and dashed circle are the SZ peak and $R_{500}$ region for the cluster, respectively. The physical scale at the cluster redshift is shown by the bar in the bottom right corner. {Flux densities} for sources A and B are given in Table \ref{tab:j0059}.}
 \label{fig:j0059}
\end{figure}

ACT-CL J0059$-$0049 ($z = 0.786$) is one of the new cluster detections in the ACT-E sample and there is little information about this cluster in the literature beyond ACT-led follow-up programmes. \textit{Chandra} data (PI: Hughes - Obs ID: 16227; 40 ks), shown by the green contours in Figure \ref{fig:j0059}, reveals an elongated morphology in the NW-SE direction, indicating a disturbed system. However, the \citet{Sifon.2016} spectroscopic analysis of 44 cluster members produced a DS test significance value of $S_\Delta = 0.574^{+0.125}_{-0.208}$, indicating no substructure along the line of sight. This system may therefore be experiencing a plane of the sky merger.

Only two radio sources, labelled A and B, are detected more than $5\sigma$ above the noise within the cluster $R_{500}$, as shown by the cyan contours in Figure \ref{fig:j0059}. Neither source is detected in FIRST. Both sources spatially coincide with spectroscopically confirmed cluster members with redshifts of $z_A = 0.7886$ and $z_B = 0.7874$. The 610 MHz {flux densities} for these sources are given in Table \ref{tab:j0059}, with the brighter of the two sources (source B) belonging to the cluster BCG. The SDSS $gri$-image presented in Figure \ref{fig:j0059} shows that 3$\sigma$ 610 MHz emission coincides with another spectroscopically confirmed cluster member at $z = 0.7864$ South-East of the BCG, but due to the low significance we do not determine a 610 MHz {flux density}. Based on the $M_{\rm 500,SZ}$ mass of this cluster, the \citet{Cassano.2013.GRHScalRel} scaling relations predict a radio halo power of $P_{\rm 1.4 GHz} = 0.7 \times 10^{24}$ W Hz\per, substantially lower than the 
upper limits determined in Section \ref{sec:UL} and shown in Table \ref{tab:UL}. {IC losses are expected to be significant at the redshift of this cluster, affecting the synchrotron emisson by an order of magnitude. As such,} the presence of a radio halo in this cluster cannot be ruled out without more sensitive radio data.

\begin{table}
 \centering
 \caption{610 MHz {flux densities} for the sources in the ACT-CL J0059$-$0049 $R_{500}$ cluster region. Source labels are indicated in Figure \ref{fig:j0059}. Notes are as in Table \ref{tab:j0022}.}
 \label{tab:j0059}
 \begin{tabular}{cccccc}
  \toprule
  ID & R.A. & Dec. & $S_{\rm 610 MHz}$ & Notes\\
   & (deg) & (deg) & (mJy) & \\
  \midrule
  A & 14.786593 & -0.822782 & 0.25 $\pm$ 0.06 & M  \\
  B & 14.785346 & -0.835209 & 1.00 $\pm$ 0.08 & M$^\bullet$  \\
  \bottomrule
 \end{tabular}
\end{table}

\subsection{ACT-CL J0239.8$-$0134}

\begin{figure}
 \centering
 \includegraphics[width=0.47\textwidth,clip=True,trim=45 20 45 40]{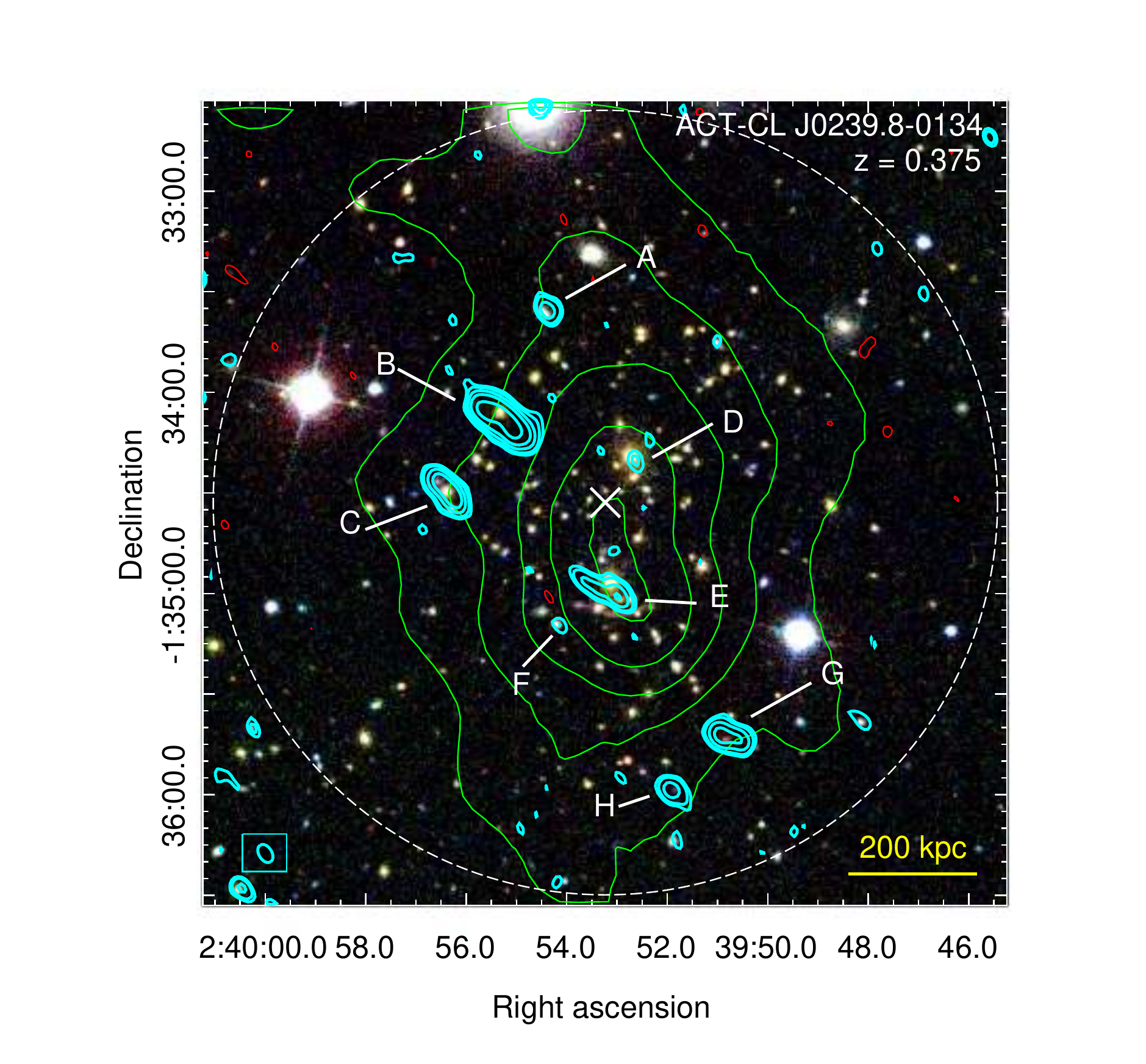}
 \caption{SDSS $gri$-image of ACT-CL J0239.8$-$0134 with 610 MHz GMRT (+ve: thick, cyan, -ve: thin, red) and \textit{XMM-Newton} (thin, green{; levels: [0.50, 0.93, 1.35, 1.78, 2.20] counts}) contours overlaid. The X and dashed circle indicate cluster SZ peak and the $R_{500}$ region, respectively. GMRT contour levels are $[\pm3, 5, 10, 20, 50, 100]\sigma$, where $1\sigma = 37\; \mu$Jy beam\per, with the beam shown by the boxed ellipse. The physical scale at the cluster redshift is shown by the bar in the bottom right corner. {Flux densities} for sources A-H are given in Table \ref{tab:j0239}.}
 \label{fig:j0239}
\end{figure}

More commonly referred to as Abell 370, ACT-CL J0239.8$-$0134 ($z = 0.375$; hereafter A370) is a merging cluster \citep{Ota.1998} with two equal mass subclusters which are merging along the line of sight \citep{Richard.2010.A370}, although recent strong lensing analysis may indicate the presences of more substructures \citep{Lagattuta.2017.A370SL}. This cluster was the first found to host an Einstein ring \citep{Soucail.1987.A370arc, Paczynski.1987.GravArcs} and has since been extensively studied as a gravitational lensing system, being chosen as one of the Hubble Frontier Fields (HFF) targets \citep{Koekemoer.2017}. The new HFF data has found A370 to host a Type Ia supernova \citep{Graham.2016}. \textit{Chandra} X-ray imaging (ObsID: 7715, 515 - 95 ks) identified two substructures, each centred on a BCG \citep{Shan.2010}, with a bolometric $R_{500}$ X-ray luminosity of $L_{X,500} = 1.89 \pm 0.05 \times 10^{45}$ erg s{\per} and power ratio of $P_3/P_0 = (0.62 \pm 0.49) \times 10^{-7}$ \citep{Mahdavi.2013}. 
The SDSS 3-
colour $gri$-image of the A370 cluster region is shown in Figure \ref{fig:j0239}, with arbitrary \textit{XMM-Newton} contours (ObsID: 0782150101; 133 ks) overlaid in green.

A370 has been targeted at radio wavelengths above 20 GHz and falls within the sky coverage of lower frequency surveys such as FIRST and NVSS. Most recently, \citet{Wold.2012} performed deep VLA 1.4 GHz observations of the cluster region, imaging at a resolution of 1.7\arcsec. We use their source catalog to measure spectral indices for any detected sources in our map. Here we present targeted low-frequency radio observations of A370 for the first time. Our full resolution, primary beam-corrected image of the $R_{500}$ cluster region is presented by cyan contours in Figure \ref{fig:j0239}. Eight discrete radio sources, labelled A-H, are detected above a noise threshold of $5\sigma$. All eight sources are detected in the \citeauthor{Wold.2012} data and all have optical counterparts; all but source F have spectroscopic redshifts along with the 1.4 GHz detections. The 610 MHz {flux densities} and 1.4$-$0.61 GHz spectral index measurements are given in Table \ref{tab:j0239}. Sources A, D, F, and H are 
compact or unresolved in the 610 MHz images. The remaining four sources (B, C, E, G) all show evidence for tailed or extended structure in the 610 MHz map. Source E is resolved at 1.4 GHz into two sources, one a compact source associated with the subcluster BCG, and the other a narrow angle tail radio galaxy. The other three extended sources are all resolved into double-lobed FR-II structures at 1.4 GHz. 


No diffuse cluster emission is found in the 1.4 GHz \citeauthor{Wold.2012} data, and we do not detect any diffuse structures in our full resolution 610 MHz image either. After modelling and subtracting the compact sources from the $uv$-data, re-imaging at several resolutions reveals residual emission at several of the subtracted source positions, but no reliable large-scale emission. {Scaling relations predict a 1.4 GHz radio halo power of $1.7^{+0.5}_{-0.4} \;\times\;10^{24}$ W Hz\per, with our highest upper limit estimate being 1.5$\sigma$ below this value (see Table \ref{tab:UL}). Given this is a massive merger, the lack of diffuse emission is unexpected according to theoretical models.} One possibility is that the diffuse emission has an ultra-steep spectrum ($\alpha < -1.5$) and it may only be observable in sensitive imaging at frequencies lower than 610 MHz. Another possibility is that the line of sight merger is prohibitive to observing a potential radio relic due to its optical depth, as 
in this case the relic would have a narrower column along the line sight. A370 falls within the sky coverage of the low frequency LOFAR surveys and if diffuse emission in this cluster exists, we would expect it to be observed in those data.

\begin{table}
 \centering
 \caption{610 MHz {flux densities} for the sources in the ACT-CL J0239.8$-$0134 $R_{500}$ cluster region. Source labels are indicated in Figure \ref{fig:j0239}. The last column is the spectroscopic or photometric ($^*$) redshift from \citet{Wold.2012}. The BCG is denoted by $^\bullet$.}
 \label{tab:j0239}
 \begin{tabular}{lccuuc}
  \toprule
  ID & R.A. & Dec. & \multicolumn{1}{c}{$S_{\rm 610 MHz}$} & \multicolumn{1}{c}{$\alpha^{1400}_{610}$} & $z$ \\
   & (deg) & (deg) & \multicolumn{1}{c}{(mJy)} & & \\
  \midrule 
  A & 39.97658 & -1.55984 & 0.79 , 0.07 & -1.3 , 0.3 & 0.3870  \\ 
  B & 39.98042 & -1.56889 & 12.08 , 0.62 & -0.7 , 0.1 & 0.3690  \\ 
  C & 39.98488 & -1.57482 & 3.62 , 0.20 & -0.9 , 0.1 & 0.4210  \\ 
  D & 39.96929 & -1.57234 & 0.28 , 0.06 & -1.5 , 0.9 & 0.3750$^\bullet$ \\ 
  E & 39.97068 & -1.58351 & 2.04 , 0.15 & -0.8 , 0.2 & 0.3729$^{\dagger}$ \\ 
  F & 39.97562 & -1.58596 & 0.15 , 0.06 & 0.5 , 1.3  & 0.3818$^*$ \\ 
  G & 39.96159 & -1.59520 & 2.56 , 0.16 & -0.8 , 0.1 & 0.3596 \\ 
  H & 39.96625 & -1.59966 & 0.97 , 0.07 & -1.3 , 0.2 & 1.0340 \\ 
  \bottomrule
 \end{tabular}
 \justify
 $^\dagger$ The optical match is either a subcluster BCG at this redshift, or a galaxy at $z_{\rm spec} = 0.3822$. 
\end{table}

\subsection{ACT-CL J2051.1+0215}

\begin{figure}
 \centering
 \includegraphics[width=0.47\textwidth,clip=True,trim=40 20 40 41]{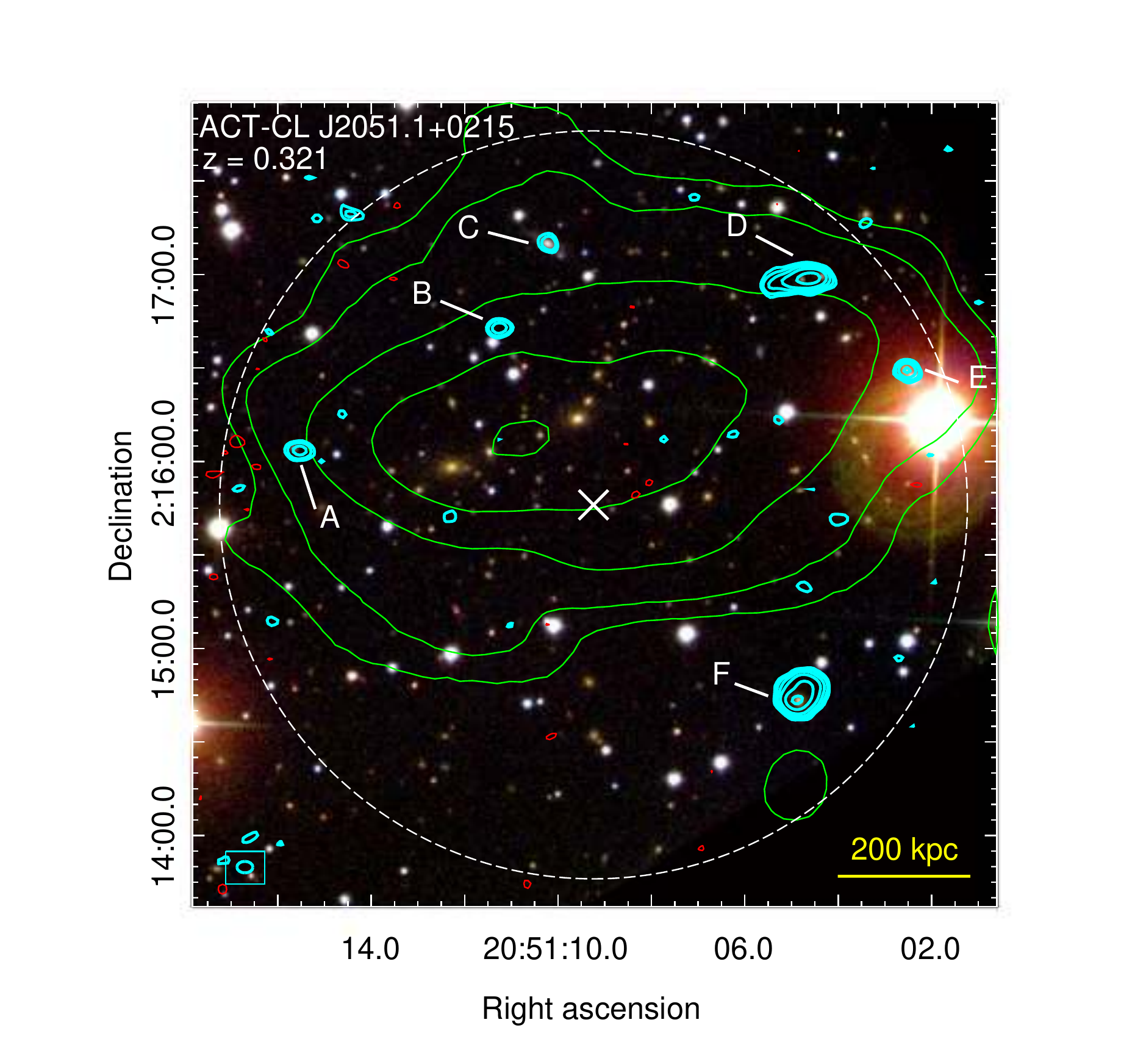}
 \caption{SDSS $gri$-image of ACT-CL J2051.1+0215 with 610 MHz GMRT (+ve: thick, cyan, -ve: thin, red) and \textit{XMM-Newton} (thin, green{; levels: [1.5, 1.9, 2.9, 4.7, 7.1] counts}) contours overlaid. The cross and dashed circle indicate cluster SZ peak and the $R_{500}$ region, respectively. GMRT contour levels are $[3,5,10,20,50,100]\sigma$, where $1\sigma = 42\; \mu$Jy beam\per, with the beam shown by the boxed ellipse. The physical scale at the cluster redshift is shown by the bar in the bottom right corner. {Flux densities} for sources A-F are given in Table \ref{tab:j2051}.}
 \label{fig:j2051}
\end{figure}

ACT-CL J2051.1+0215 ($z = 0.321$) was first detected in the ROSAT All-Sky Survey \citep[RASS;][]{Bade.1998} and was given the designation RXC J2051.1+0216. \citet{Piffaretti.2011} include this cluster in their meta-catalogue of X-ray detected clusters, listing a (0.1-2.4\,keV) X-ray luminosity of $L_{X,500} = 4.54 \times 10^{44}$ erg s\per within $R_{500}$, and mass of $M_{\rm X,500} = 4.06 \times 10^{14} M_\odot$, consistent with the SZ-derived mass presented in Table \ref{tab:sample}. Visual inspection of the \textit{XMM-Newton} image (ObsID: 0650383701) reveals an elongated morphology in the E-W direction and a somewhat asymmetric structure, with a separation between the X-ray peak and the cluster BCG of $\sim$ 22\arcsec, leading to the conclusion that this cluster is undergoing or has recently undergone a cluster merger. The X-ray morphology is indicated by the green contours (arbitrary levels) overlaid on the SDSS $gri$-image shown in Figure \ref{fig:j2051}.

Our full resolution, primary beam-corrected image of the ACT-CL J2051.1+0215 reveals six discrete sources in the cluster region. These sources, labelled A-F, are indicated by the cyan contours in Figure \ref{fig:j2051} and their 610 MHz {flux densities} are listed in Table \ref{tab:j2051}. Sources D and F have extended emission with a peak to one side of the source. Although our 610 MHz resolution is insufficient to discern structure within the emission, it is likely that these are tailed radio galaxies, with source F possibly a wide angle tail due to the breadth of the emission. As there is no spectroscopic follow-up of this cluster, all galaxy source matching is done by eye in the SDSS DR12 image shown in Figure \ref{fig:j2051}. All sources except source B have a likely optical counterpart, all of which are potential cluster members based on $gri$ colour.

There is no evidence for diffuse emission in the cluster region above the noise threshold of our images, even after subtraction of compact sources. The predicted radio power, given the cluster mass, is $P_{\rm 1.4 GHz} = 0.7 \times 10^{24}$ W Hz\per. As our upper limit determined in Section \ref{sec:UL} and shown in Table \ref{tab:UL} is slightly above this theoretical value, deeper radio data are required. In addition, a full morphological analysis of the cluster would be helpful in confirming the dynamical cluster state.

\begin{table}
 \centering
 \caption{610 MHz {flux densities} for the sources in the ACT-CL J2051.1+0215 $R_{500}$ cluster region. Source labels are indicated in Figure \ref{fig:j2051}. Notes are as in Table \ref{tab:j0022}. Here all galaxy identification has been done based on colour matching.}
 \label{tab:j2051}
 \begin{tabular}{cccccc}
  \toprule
  ID & R.A. & Dec. & $S_{\rm 610 MHz}$ & Notes\\
   & (deg) & (deg) & (mJy) & \\
  \midrule
  A & 312.815208 & 2.268342 & 0.76 $\pm$ 0.08 & M  \\
  B & 312.797377 & 2.279267 & 0.72 $\pm$ 0.08 & -  \\
  C & 312.792957 & 2.286826 & 0.46 $\pm$ 0.07 & M  \\
  D & 312.770273 & 2.283620 & 7.41 $\pm$ 0.39 & M \\
  E & 312.768854 & 2.385923 & 0.23 $\pm$ 0.07 & M \\
  F & 312.770502 & 2.246574 & 13.37 $\pm$ 0.68 & M \\
  \bottomrule
 \end{tabular}
\end{table}

\subsection{ACT-CL J2135.2+0125}

\begin{figure}
 \centering
 \includegraphics[height=0.47\textwidth, clip=True, trim=40 20 40 38]{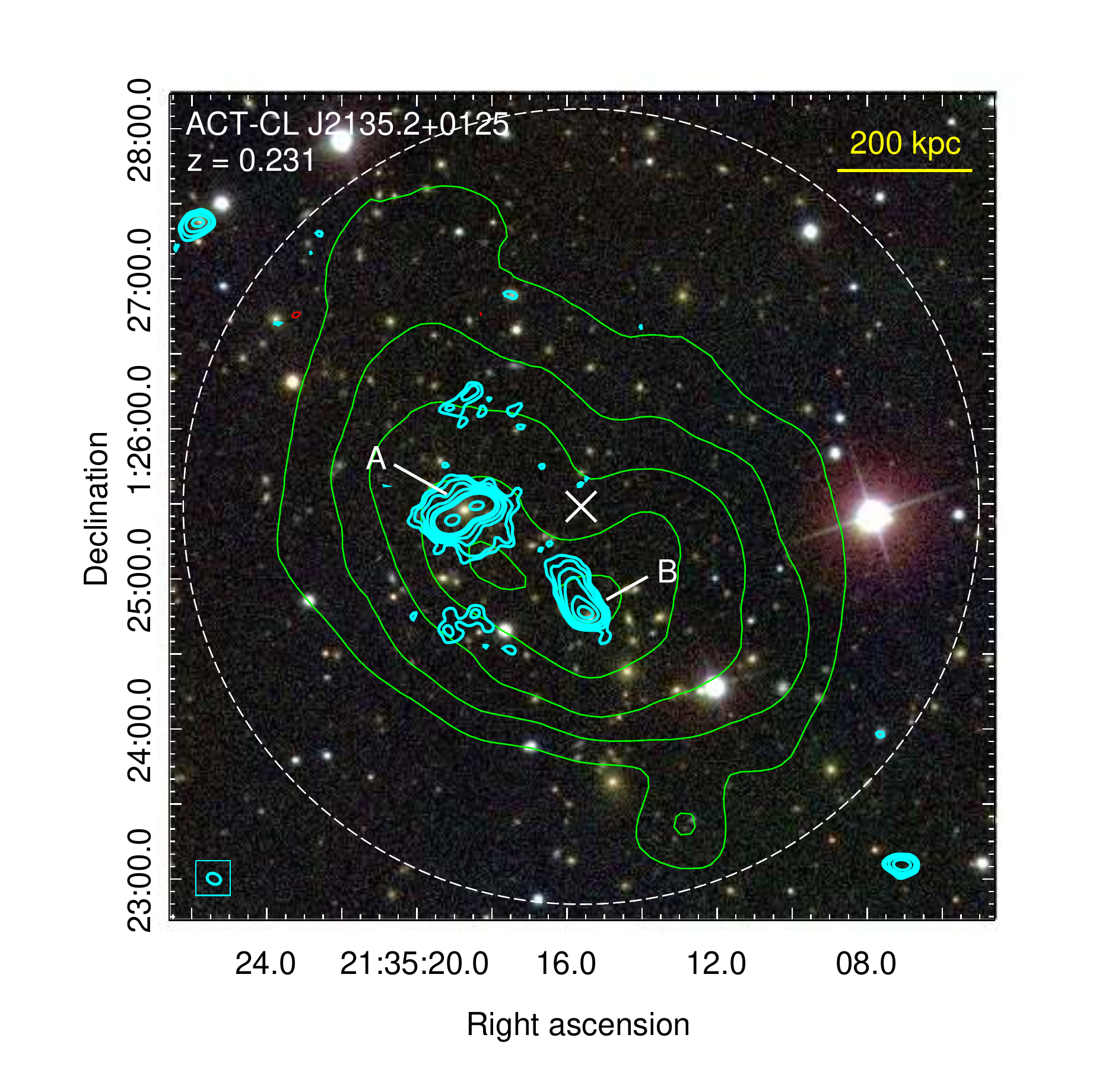}
 \caption{SDSS DR12 3-colour \textit{gri}-image of ACT-CL J2135.2+0125 with 610 MHz GMRT (+ve: thick, cyan, -ve: thin, red) and \textit{XMM-Newton} (thin, green{; levels: [7.0, 9.5, 12.0, 14.5, 17.0] counts}) contours overlaid. Radio contour levels are $[\pm3,5,10,20,50,100,200,500]\sigma$, where $1\sigma = 110\; \mu$Jy beam\per, with the beam shown by the boxed ellipse. The X and dashed circle are the SZ peak and $R_{500}$ region for the cluster, respectively. The physical scale at the cluster redshift is shown by the bar in the bottom right corner. {Flux densities} for sources A and B are given in Table \ref{tab:j2135p0125}.}
 \label{fig:j2135p0125}
\end{figure}

ACT-CL J2135.2+0125 is an Abell cluster \citep[A2355; ][]{Abell.1958} at a redshift of $z = 0.231$. We detect no extended diffuse emission in our reprocessing of the 610 MHz GMRT data, however our full resolution, primary beam-corrected 610 MHz GMRT image of the cluster, shown as thick, cyan (+ve) and thin, red (-ve) contours in Figure \ref{fig:j2135p0125}, reveals two extended sources, labelled A and B, above 10$\sigma$ of the noise. The {flux density} for each source is given in Table \ref{tab:j2135p0125}. Source A is a bright FR-II radio galaxy associated with the cluster BCG. Source B appears to be a tailed radio galaxy associated with a probable cluster member, and coincides spatially with one of the X-ray peaks. As this region is not covered by FIRST, we do not determine spectral indices for either source. The positive emission to the North and South of source A are imaging artifacts due to the bright source.

ACT-CL J2135.2+0125 has been observed by both the \textit{XMM-Newton} (Obs ID: 0692931301; 22 ks) and \textit{Chandra} (Obs ID: 15097; 20 ks) X-ray telescopes. The X-ray analysis shows a definitive disturbed morphology with two X-ray peaks, as shown by the green XMM contours overlaid on the SDSS DR12 3-colour $gri$-image in Figure \ref{fig:j2135p0125}. There is a spatial offset of $\sim$ 25{\arcsec} between the BCG and the Eastern X-ray peak, providing more evidence for significant substructure in the cluster. Indeed, \citet{Cassano.2016} quote X-ray morphological parameters, based on the \textit{Chandra} data, which are consistent with a merging system ($P_3/P_0 = 7.5 ^{+4.8}_{-3.1} \times 10^{-7}$; $w = 0.0049^{+0.0018}_{-0.0042}$; $c = 0.075^{+0.005}_{-0.004}$). 

{Given current theoretical models, the lack of observable diffuse radio emission in this massive, plane of the sky merger is unexpected.} The \citet{Cassano.2013.GRHScalRel} scaling relations predict a 1.4 GHz radio halo power of $1.4^{+0.5}_{-0.4} \times 10^{24}$ W Hz{\per} for this cluster, based on the $M_{\rm 500,SZ}$ mass. Our measured radio halo upper limit, discussed in Section \ref{sec:UL}, of $< 0.5 \times 10^{24}$ W Hz\per, is well below the theoretical prediction. However, as for Abell 2146 \citep{Russell.2011.A2146, HlavacekLarrondo.2018.A2146}, low power radio emission may still exist in this system, requiring very sensitive imaging to detect.

\begin{table}
 \centering
 \caption{610 MHz {flux densities} for the sources in the ACT-CL J2135.2+0125 $R_{500}$ cluster region. Source labels are indicated in Figure \ref{fig:j2135p0125}. Notes are as in Table \ref{tab:j0022}.}
 \label{tab:j2135p0125}
 \begin{tabular}{cccccc}
  \toprule
  ID & R.A. & Dec. & $S_{\rm 610 MHz}$ & Notes\\
   & (deg) & (deg) & (mJy) & \\
  \midrule
  A & 323.828161 & 1.423957 & 331.31 $\pm$ 16.57 & M$^\bullet$  \\
  B & 323.815081 & 1.413930 & 51.93 $\pm$ 2.61 & -  \\

  \bottomrule
 \end{tabular}
\end{table}

\subsection{ACT-CL J2154.5$-$0049}

\begin{figure}
 \centering
 \includegraphics[height=0.47\textwidth, clip=True, trim=30 20 10 45]{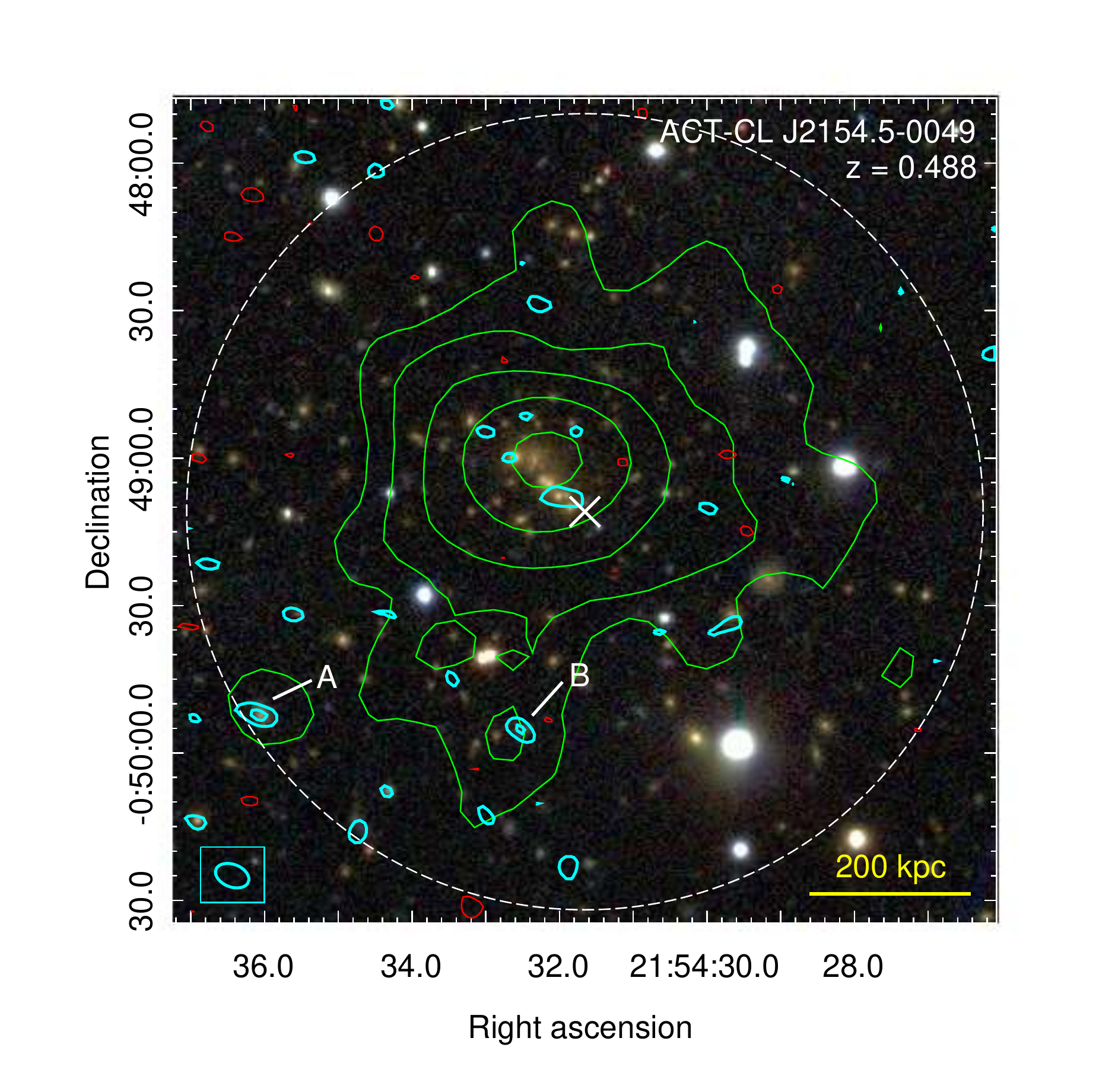}
 \caption{SDSS DR12 3-colour \textit{gri}-image of ACT-CL J2154.5$-$0049 with 610 MHz GMRT contours overlaid (+ve: thick, cyan, -ve: thin, red). Contour levels are $[3,5]\sigma$, where $1\sigma = 46\; \mu$Jy beam\per, with the beam shown by the boxed ellipse. \textit{Chandra} X-ray contours are overlaid in green (thin){ with levels of 3.5, 5, 8, 14, and 26 counts}. The X and dashed circle are the SZ peak and $R_{500}$ region for the cluster, respectively. The physical scale at the cluster redshift is shown by the bar in the bottom right corner. {Flux densities} for sources A and B are provided in Table \ref{tab:j2154}.}
 \label{fig:j2154}
\end{figure}

ACT-CL J2154.5$-$0049 ($z = 0.488$) is a poorly studied system with very little information in the literature. ACT-led follow-up in the optical and X-ray bands hint at potential plane of the sky merger activity. \citet{Sifon.2016} identify 52 cluster members with which they measure a DS test significance value of $S_\Delta = 0.092^{+0.070}_{-0.019}$, indicating no substructure along the line of sight. However, \textit{Chandra} data (Obs ID: 16230; 64 ks) show a region of hot gas extending outwards from the cluster core, which may indicate remnants of merger activity. This X-ray emission is shown by thin green contours overlaid on the SDSS DR12 3-colour $gri$-image in Figure \ref{fig:j2154}.

\begin{table}
 \centering
 \caption{610 MHz {flux densities} for the sources in the ACT-CL J2154.5$-$0049 $R_{500}$ cluster region. Source labels are indicated in Figure \ref{fig:j2154}. }
 \label{tab:j2154}
 \begin{tabular}{ccccc}
  \toprule
  ID & R.A. & Dec. & $S_{\rm 610 MHz}$ \\
   & (deg) & (deg) & (mJy) & \\
  \midrule
  A & 328.650348 & -0.831180 & 0.31 $\pm$ 0.09 & F \\
  B & 328.635551 & -0.831950 & 0.15 $\pm$ 0.06 & - \\

  \bottomrule
 \end{tabular}
\end{table}

At the noise level of our full resolution primary beam-corrected 610 MHz GMRT image, this cluster appears to be {quite radio quiet, with only} two sources detected above 5$\sigma$, labelled A and B in Figure \ref{fig:j2154}. The 610 MHz {flux densities} are given in Table \ref{tab:j2154}. Source A is coincident with a foreground galaxy with a spectroscopic redshift of 0.5012. Source B has no visible optical counterpart.

There are also several 3$\sigma$ radio residuals within the cluster region. After removing sources A and B, smoothing reveals two sources, but we do not attempt to classify them as they are unresolved and have low significance. Further radio data are required to confirm their existence and probe their origin. As the measured upper limit for J2154 of $P_{\rm 1.4 GHz} < 4.0 \times 10^{24}$ W Hz{\per} (see Section \ref{sec:UL} and Table \ref{tab:UL} for details) is well above the predcited scaling relation power of $P_{\rm 1.4 GHz} = 0.3 \times 10^{24}$ W Hz\per, the additional, deeper, radio data will be necessary to determine the presence of a radio halo. {IC losses may be affecting the synchrotron signal: at a redshift of $z=0.488$, 20\% of the signal is potentially upscattered to higher frequencies.}

\subsection{ACT-CL J2337.6+0016}
\begin{figure}
 \centering
 \includegraphics[height=0.47\textwidth, clip=True, trim=35 20 7 50]{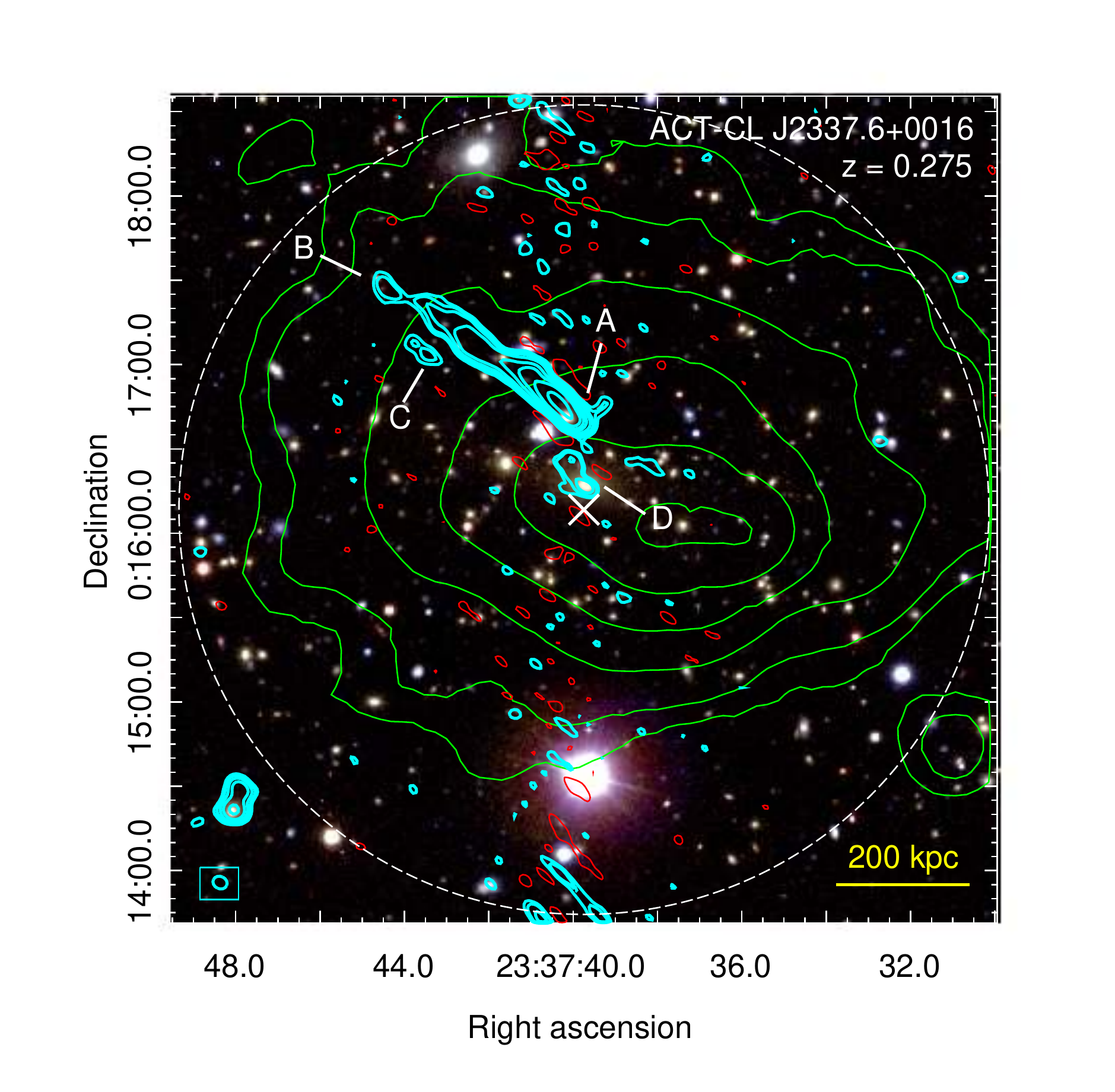}
 \caption{SDSS DR12 3-colour \textit{gri}-image of ACT-CL J2337.6+0016 with 610 MHz GMRT contours overlaid (+ve: thick, cyan, -ve: thin, red; levels are $[3,5,10,20,50,100,200]\sigma$, where $1\sigma = 170\; \mu$Jy beam\per), with the beam shown by the boxed ellipse. {\textit{Chandra} X-ray contours are overlaid in green (thin) with levels of 1.5, 1.8, 2.6, 4.0, 6.0, and 8.5 counts.} The X and dashed circle are the SZ peak and $R_{500}$ region for the cluster, respectively. The physical scale at the cluster redshift is shown by the bar in the bottom right corner. {Flux densities} for sources A-D are provided in Table \ref{tab:j2337}.}
 \label{fig:j2337}
\end{figure}

ACT-CL J2337.6+0016, hereafter J2337, is a rich Abell cluster \citep[A2631;][]{Abell.1958} at $z = 0.275$. It is also part of the  REFLEX sample \citep[RXC J2337.6+0016][]{Bohringer.2004} and has a 0.1-2.4 keV band luminosity of $L_{\rm X,[0.1-2.4 keV]} = 7.571 \times 10^{44}$ erg s\per. Archival ROSAT images show a complex morphology, with follow-up \textit{XMM-Newton} observations (Obs ID: 0042341301) revealing a gas temperature of $kT = 9.6 \pm 0.3$ kev \citep{Zhang.2006}. Using this XMM observation, \citet{Finoguenov.2005} characterise J2337 as experiencing a late stage core disruption. J2337 has also been observed twice by the \textit{Chandra} X-ray Telescope for a total of 26 ks (Obs IDs: 3248, 11728) - the emission, shown by green contours overlaid on the SDSS DR12 $gri$ composite image in Figure \ref{fig:j2337}, is slightly elongated in the E-W direction, with a significant offset of 30.7{\arcsec} between the X-ray peak and the cluster BCG. In their spectroscopic analysis \citet{Sifon.2016} use 154 cluster members to measure a DS statistic of $S_\Delta = 0.008^{+0.020}_{-0.006}$, indicative of significant substructure in the cluster. 


In the radio, J2337 has been studied at 1.4 GHz with the VLA in A-band \citep{Rizza.2003} as well as at 610 MHz with the GMRT as part of the GRHS \citep{Venturi.2007.GRHS1}. Our reprocessing of the 610 MHz data has a slightly higher central noise level compared to that quoted by \citet{Venturi.2007.GRHS1} due to contamination of the cluster region by sidelobes of a bright source South of the cluster. The full resolution, primary beam-corrected image is shown by the cyan contours overlaid on Figure \ref{fig:j2337}. We detect four sources above 5$\sigma$ of the central rms noise of 84.1 $\mu$Jy beam\per, labelled A-D. Table \ref{tab:j2337} reports the 610 MHz {flux densities} for these sources.

Source A is a bright head-tail galaxy which dominates the $R_{500}$ region. It has a largest angular size of 80.2{\arcsec}, corresponding to a physical length of 338 kpc at the redshift of the cluster. It is the only source identified in FIRST, providing a 1.4 GHz/610 MHz spectral index of $\alpha = -1.2 \pm 0.1$. All but source C have optical counterparts: sources A and C are coincident with cluster members at spectroscopic redshifts of 0.2717 and 0.2777, respectively, and source B may be related to a high redshift galaxy based on colour matching in the SDSS image.

No diffuse emission is detected in J2337. Based on the cluster mass, the \citet{Cassano.2013.GRHScalRel} scaling relations predict a radio halo power of $P_{\rm 1.4 GHz} = 1.2^{+0.5}_{-0.3} \times 10^{24}$ W Hz\per, slightly above the upper limit of $0.8 \times 10^{24}$ W Hz{\per} determined in Section \ref{sec:UL}. Given the multiwavelength evidence for substructure in this cluster, more sensitive radio data will be necessary to determine whether this massive merger hosts a radio halo.

\begin{table}
 \centering
 \caption{610 MHz {flux densities} for the sources in the ACT-CL J2337.6+0016 $R_{500}$ cluster region. Source labels are indicated in Figure \ref{fig:j2337}. Notes are as in Table \ref{tab:j0022}.}
 \label{tab:j2337}
 \begin{tabular}{cccccc}
  \toprule
  ID & R.A. & Dec. & $S_{\rm 610 MHz}$ & Notes\\
   & (deg) & (deg) & (mJy) & \\
  \midrule
  A & 354.419405 & 0.280519 & 250.93 $\pm$ 12.56 & M  \\
  B & 354.435305 & 0.291281 & 0.37 $\pm$ 0.16 & B  \\
  C & 354.430987 & 0.284249 & 1.67 $\pm$ 0.31 & - \\
  D & 354.415363 & 0.271211 & 4.44 $\pm$ 0.41 & M$^\bullet$ \\
  \bottomrule
 \end{tabular}
\end{table}

\section{Radio halo upper limits}
\label{sec:UL}

For the ten systems with no evidence of diffuse cluster emission, we estimate radio halo {flux densities} and power upper limits using two different methods. The predicted $P_{\rm 1.4 GHz}$ value for each cluster, based on the $P_{\rm 1.4 GHz}-M_{\rm 500,SZ}$ scaling relation, is given in the fifth column of Table \ref{tab:UL}. 

\subsection{Method 1: fixed halo radius}
\label{subsec:oldUL}

\begin{table*}
 \centering
 \caption{Radio halo power upper limits for clusters with non-detections. We use a generic spectral index of 1.3 to scale the 610 MHz upper limit to 1.4 GHz. We first use a maximum physical diameter of 1 Mpc for the simulated halo at the redshift of the cluster (method 1), and then use scaling relations to convert the $M^{\rm UPP}_{500}$ cluster masses from \citet{Hasselfield.2013.ACTE} to simulated radio halo sizes (method 2). See text for details. Columns: (1) ACT cluster name. (2) Redshift. (3) Integrated Compton-$y$ parameter. (4) Cluster mass. (5) Predicted radio power based on the $P_{\rm 1.4 GHz}-M_{500,SZ}$ scaling relation. (6,7) Upper limit {flux density} and radio power, respectively, based on a 1 Mpc halo. (8-10) Injected halo size, and the corresponding upper limit {flux density} and radio power, respectively.}
 \label{tab:UL}
 \begin{tabular}{lcuuc|cc|ccc}
  \toprule
   & & & & & \multicolumn{2}{c|}{\bf $R_H$ = 1 Mpc} & \multicolumn{3}{c}{\bf $R_H$ = variable}  \\
  Cluster Name & $z$ & \multicolumn{1}{c}{$Y_{\rm 500,SZ}$} & \multicolumn{1}{c}{$M_{\rm 500,SZ}$} & $P_{\rm 1.4 GHz}^{\rm th.}$ & $S_{\rm 610 MHz}$ & $P_{\rm 1.4 GHz}$  & $R_H$ & $S_{\rm 610 MHz}$ & $P_{\rm 1.4 GHz}$\\
  (ACT-CL...) & & \multicolumn{1}{c}{($10^{-4}$ arcmin$^2$)} & \multicolumn{1}{c}{($10^{14} M_\odot$)} & ($10^{24}$ W/Hz) & (mJy/b) & ($10^{24}$ W/Hz) & (Mpc) & (mJy/b) & ($10^{24}$ W/Hz)\\
  \midrule
  J0059.1$-$0049 & 0.786 & 3.5 , 0.6 & 5.2 , 0.9 & 0.7$^{+0.3}_{-0.3}$ & $<$8 & $<$9.7 & 0.602 & $<$3 & $<$3.6\\
  J0152.7+0100   & 0.230 & 13.0 , 2.3 & 5.7 , 1.1 & 0.9$^{+0.4}_{-0.3}$ & $<$8 & $<$0.5 & 0.653 & $<$5 & $<$0.3 \\
  J0239.8$-$0134 & 0.375 & 9.4 , 1.6 & 6.7 , 1.3 & 1.7$^{+0.5}_{-0.4}$  & $<$6 & $<$1.1 & 0.754 & $<$4 & $<$0.8 \\
  J2051.1+0215   & 0.321 & 7.6 , 2.5 & 5.3 , 1.4 & 0.7$^{+0.3}_{-0.2}$ & $<$7 & $<$0.9 & 0.613 & $<$4 & $<$0.5\\
  J2129.6+0005   & 0.234 & 11.4 , 2.4 & 5.3 , 1.1 & 0.7$^{+0.3}_{-0.2}$ & $<$10 & $<$0.6 & 0.612 & $<$4 & $<$0.2\\
  J2135.2+0125   & 0.231 & 14.2 , 2.4 & 6.3 , 1.2 & 1.4$^{+0.5}_{-0.4}$ & $<$9 & $<$0.5 & 0.714 & $<$6 & $<$0.4\\
  J2135.7+0009   & 0.118 & 10.8 , 6.3 & 2.6 , 1.1 & 0.1$^{+0.1}_{-0.1}$ & $<$25 & $<$0.3 & 0.326 & $<$9 & $<$0.1 \\
  J2154.5$-$0049 & 0.488 & 3.4 , 0.9 & 4.3 , 0.9 & 0.3$^{+0.3}_{-0.2}$ & $<$11 & $<$4.0 & 0.509 & $<$3 & $<$1.1\\
  J2327.4$-$0204 & 0.705 & 10.1 , 1.0 & 9.4 , 1.5 & 6.1$^{+2.5}_{-1.9}$ & $<$10 & $<$9.1 & 1.017 & $<$10 & $<$9.1\\
  J2337.6+0016   & 0.275 & 11.5 , 2.2 & 6.1 , 1.2 & 1.2$^{+0.5}_{-0.3}$ & $<$9 & $<$0.8 & 0.694 & $<$6 & $<$0.5\\
  \bottomrule
 \end{tabular}
 \vspace*{0.3cm}
\end{table*}

The first method we implemented is that described in \citet{Brunetti.2007.CRandGRH} and \citet{Kale.2013.EGRHSUL}. For each cluster, we assume a maximum halo diameter of 1 Mpc, typical of giant radio halos, and model the average brightness profile of well-studied halos \citep{Brunetti.2007.CRandGRH} with seven concentric, optically thin spheres with diameters ranging from 400 kpc to 1 Mpc. For each model, approximately 50\% of the flux is in the largest sphere. 

For each cluster we inject several simulated 1 Mpc halos with total {flux densities} in the range $S_{\rm inj} \sim 3 - 300$ mJy and image the altered $uv$-data at several resolutions. The angular size of the injected halo varies between 131{\arcsec} - 470{\arcsec} over the redshift range of the nine clusters. We find that extended emission is securely established for fluxes above a value of $S_{\rm sim} = 11$ mJy for all but the lowest redshift cluster, ACT-CL J2135.7+0009, which has an upper limit of 25 mJy owing to the large angular size of the injected halo in this case. For the other nine clusters, injected halo fluxes in the range 6{\textendash}11 mJy result in positive residuals in the full-resolution image with integrated {flux density} $\sim 4\sigma$ above the noise level with indications of extended emission in the low-resolution image. 

As in the literature, for each cluster the upper limit is taken to be the {flux density} for which the halo is just undetected in the low-resolution image. Table \ref{tab:UL} lists the 610 MHz radio halo upper limit {flux densities} for the clusters in which no evidence of central residual emission is found. Using a spectral index of $\alpha = 1.2$, we extrapolate the {flux densities} to 1.4 GHz to produce a k-corrected, 1.4 GHz radio power upper limit for each cluster using the following equation, where $D_{\rm L}$ is the luminosity distance at the cluster redshift $z$:
\begin{multline}
 \left(\frac{P_{\rm 1.4 GHz}}{\rm W Hz^{-1}}\right) = 4 \pi \left(\frac{D_{\rm L}}{\rm m}\right)^2 \left(\frac{S_{\rm 610 MHz}}{\rm m^{-2} W Hz^{-1}}\right) \\ 
 \times \left(\frac{1400 \rm MHz}{610 \rm MHz}\right)^{\alpha} (1 + z)^{-(1 + \alpha) },  \label{eqn:3:power}
\end{multline}

\begin{figure*}
 \centering
\includegraphics[width=0.48\textwidth, clip=true, trim=50 0 50 0]{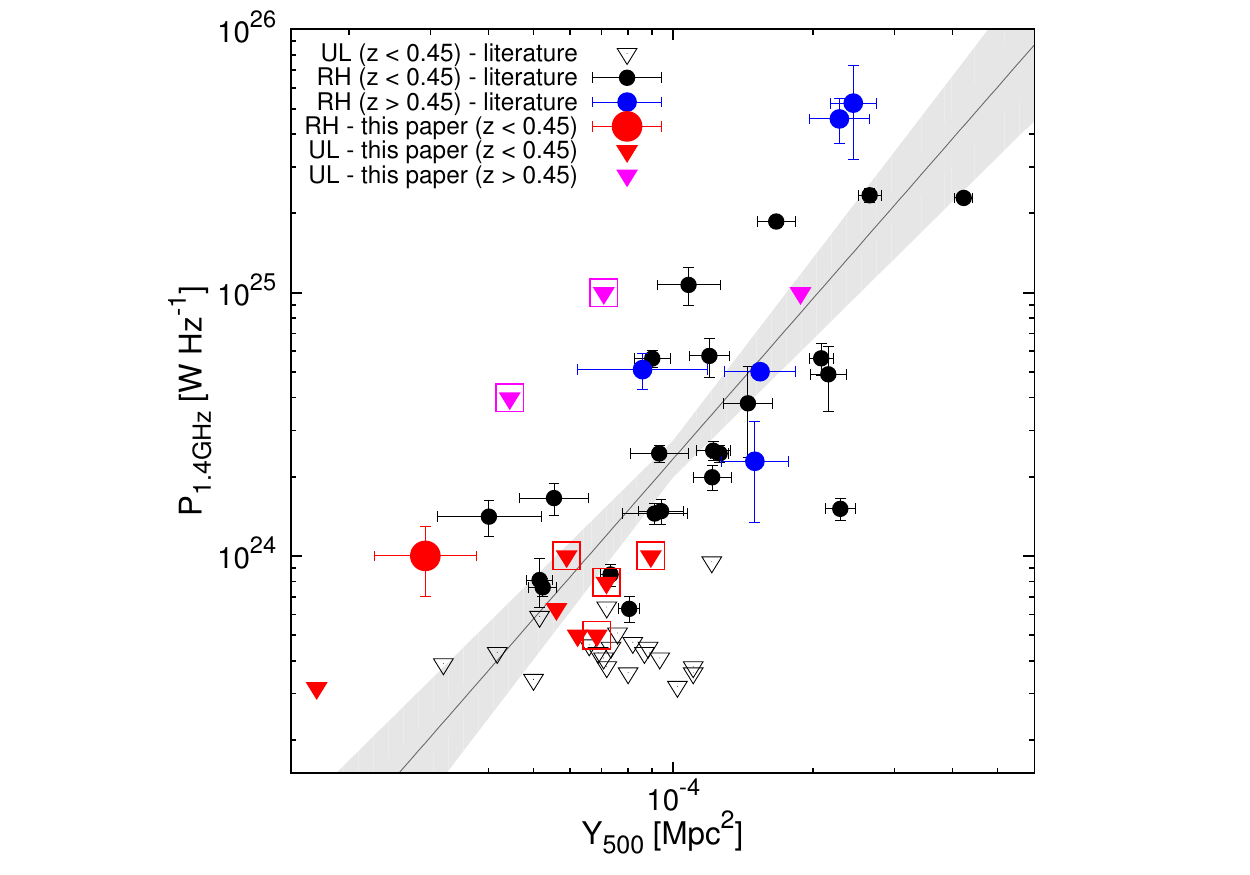}
\includegraphics[width=0.48\textwidth, clip=true, trim=50 0 50 0]{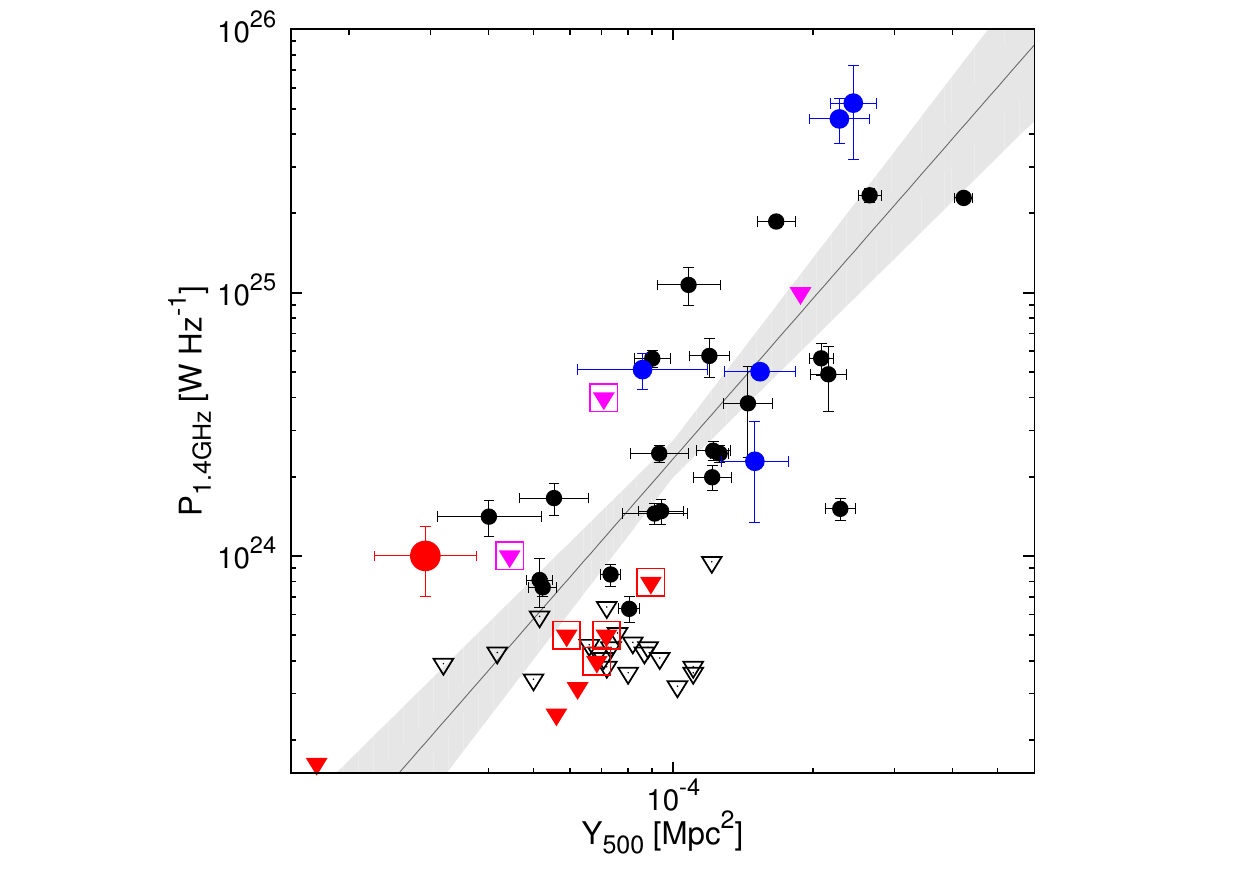}\\
\includegraphics[width=0.48\textwidth, clip=true, trim=50 0 50 0]{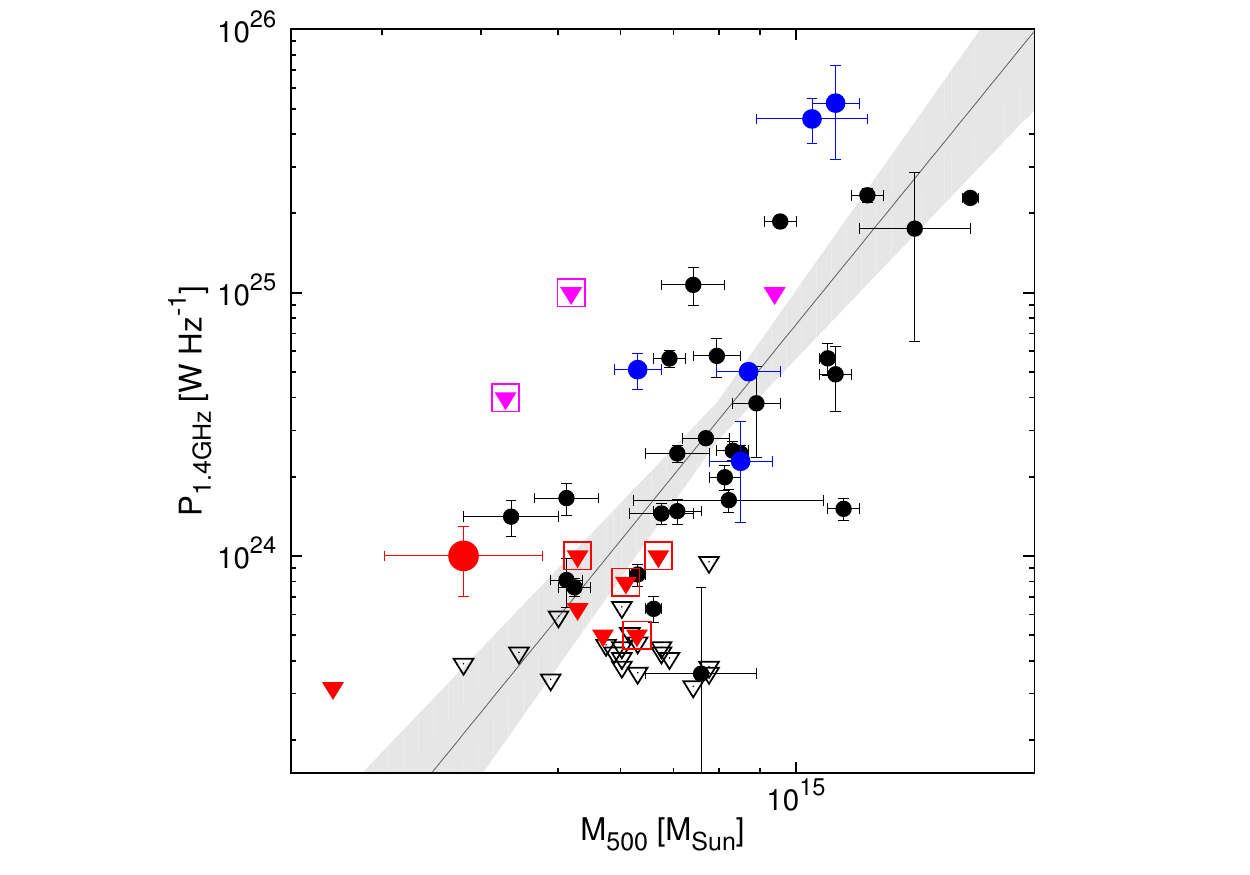}
\includegraphics[width=0.48\textwidth, clip=true, trim=50 0 50 0]{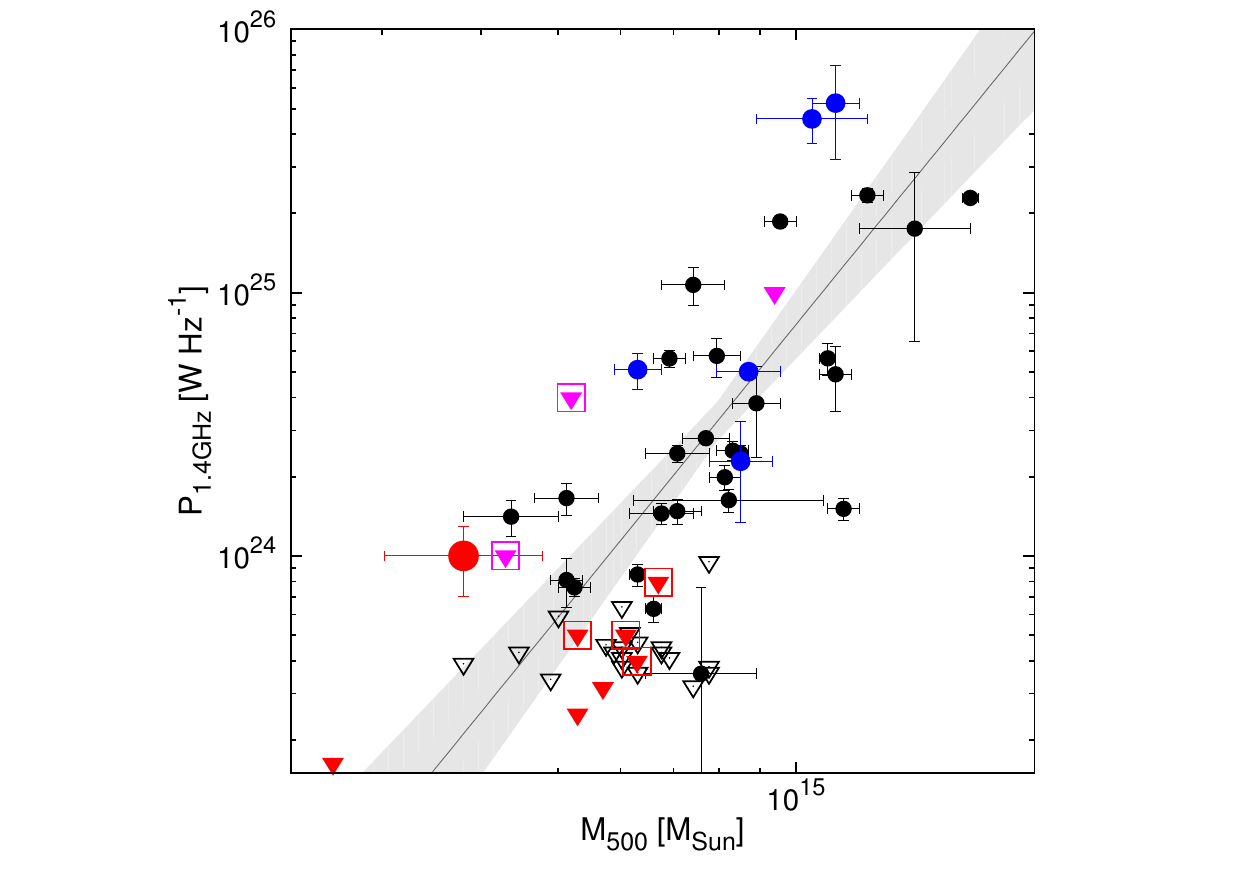}
 \caption{Scaling relations between 1.4 GHz radio halo power versus integrated SZ Compton-$y$ parameter (\textit{top row}) and versus SZ-derived mass (\textit{bottom row}). Radio halos and upper limits from the literature are shown as filled circles and empty triangles, respectively (black - $z < 0.45$, blue $z > 0.45$). The radio halo in ACT-CL J0256.5+0006 is indicated by a large red filled circle. Upper limits in this paper are shown by filled triangles (red - $z < 0.45$, magenta - $z > 0.45$); boxes indicate the disturbed systems with non-detections. The legend in the upper left panel is the same for all panels. \textit{Left panel:} Upper limits determined using method 1. \textit{Right panel:} Upper limits determined using method 2. See text for details.}
 \label{fig:P1.4corr}
\end{figure*}

\vspace{0.1cm}
These upper limits are shown as filled triangles in the $P_{\rm 1.4 GHz}-Y_{500}$ and $P_{\rm 1.4 GHz}-M_{500}$ planes shown in the top left and bottom left panels of Figure \ref{fig:P1.4corr}, respectively. Upper limits for disturbed clusters in our sample are indicated by a boxed triangle. The upper limits for the high redshift clusters (shown in magenta) are significantly higher than the existing upper limits from the literature, which are shown as empty triangles, and are above the scaling relation in all cases. In addition, some of the upper limits for the low redshift clusters are also above or consistent with the scaling relation. This indicates that we did not reach the sensitivity to measure the predicted halo { density} in many of these observations. There are several effects which contribute to this. Firstly, the noise in some of our final maps are up to a factor of two higher than the target noise, due to necessary RFI flagging of a significant percentage of data (30\% - 40\%). For ACT-
CL J2327.4$-$0204 and ACT-CL J2135.7+0009, this increases the observed upper limit relative to the observationally targeted value. Secondly, the preliminary cluster masses that we used in the pilot proposal turned out to be larger than the final published masses, which resulted in the exposure times being underestimated. A further factor is that in our prediction of halo {flux densities} we estimated a radio halo size using the \citet{Cassano.2007.NewScalRel} correlation between radio power and halo size. This means that some of the predicted radio halos differed, at times significantly, from the 1 Mpc size assumed for the upper limits computed here. We investigate the effect of this in the following section.

\subsection{Method 2: mass-based halo radius}
\label{subsec:newUL}

\begin{table*}
 \centering
 \caption{Results of the survival analysis for the $P_{\rm 1.4 GHz}$ vs $Y_{\rm 500,SZ}$ and $P_{\rm 1.4 GHz}$ vs $M_{\rm 500,SZ}$ relations, for three combinations of the data: 0 - literature only; 1 - literature combined with method 1 upper limits; 2 - literature combined with method 2 upper limits. The fit is given for the intercept $A$, slope $B$, and log-normal scatter $s$.}
 \begin{tabular}{c|ccc|ccc}
  \toprule
   & \multicolumn{3}{c|}{$P_{\rm 1.4 GHz}$ vs $Y_{\rm 500,SZ}$} & \multicolumn{3}{c}{$P_{\rm 1.4 GHz}$ vs $M_{\rm 500,SZ}$}\\
   & 0 & 1 & 2 & 0 & 1 & 2\\
  \midrule
  
  $A$ & 24.55$^{+0.07}_{-0.08}$ & 24.54$^{+0.07}_{-0.07}$ & 24.49$^{+0.07}_{-0.06}$  & 24.57$^{+0.08}_{-0.08}$ & 24.55$^{+0.08}_{-0.08}$ & 24.51$^{+0.08}_{-0.08}$\\
  $B$ & 1.66$^{+0.34}_{-0.32}$ & 1.74$^{+0.30}_{-0.28}$ & 2.00$^{+0.28}_{-0.27}$ & 2.88$^{+0.64}_{-0.62}$ & 2.99$^{+0.59}_{-0.60}$ & 3.39$^{+0.60}_{-0.54}$\\
  $s$ & 0.37$^{+0.07}_{-0.05}$ & 0.38$^{+0.06}_{-0.05}$ & 0.37$^{+0.06}_{-0.05}$ & 0.39$^{+0.07}_{-0.06}$ & 0.41$^{+0.06}_{-0.05}$ & 0.41$^{+0.07}_{-0.06}$\\

  \bottomrule
 \end{tabular}
 \label{tab:survival}
\end{table*}

{
The second method for computing upper limits takes into account the scaling of radio halo size with cluster mass, as found by \citet{Cassano.2007.NewScalRel}. A similar method has been employed by \citet{Bonafede.2017} in which they adopt realistic halo profiles instead of optical spheres. Here we maintain the use of optically thin spheres in order to compare the results with the method in the previous section and investigate the effect of the change in predicted radio halo size. }

We re-compute upper limits for our clusters, injecting simulated halos with sizes determined by extrapolating a 1.4 GHz radio power from the cluster SZ mass, and then using the \citet{Cassano.2007.NewScalRel} scaling relation to obtain the halo size. The halo size, new upper limit {flux densities}, and associated radio powers are given in Table \ref{tab:UL}. These revised upper limits are shown in the right top and bottom panels of Figure \ref{fig:P1.4corr}. All upper limits have decreased compared to method 1, except for ACT-CL J2327.4$-$0204 which remained the same due to the predicted halo size being very close to 1 Mpc. The other two high redshift upper limits are still well above the region populated by literature upper limits, but are now well within the scatter of the existing scaling relations. These higher upper limits may be explained by the similarity in angular size of the injected halos compared to the low resolution synthesised beam of the images ($\sim$ 28\arcsec). In this case the 
map noise has a greater effect on the detection limit than for clusters with larger, and therefore more resolved, injected halos. The higher upper limits for high redshift systems may indicate that the lack of flux sensitivity is a major problem when observing high-redshift systems of intermediate SZ-signal and mass, and that correcting the size of the simulated halos is insufficient on its own. The upper limits for the low redshift systems are now all within the region of literature upper limits, however ACT-CL J2135.7+0009, being at a substantially lower redshift than the other clusters, is still above the correlation. 

\subsection{Survival analysis}
\label{subsec:survanaly}

In order to check whether or not our upper limits could belong to the population of detections, we perform a survival analysis of our data by measuring scaling relations and comparing between the different sets of data. We use the Bayesian \citet{Kelly.2007} method to perform the linear regression as it takes into account measurement {uncertainties}. We used three sets of data: (0) the previous detections from the literature only, (1) the literature detections with our upper limits from method 1, and (2) the literature detections with our upper limits from method 2. For each dataset, we fit for $P_{\rm 1.4GHz}-Y_{\rm 500,SZ}$ and $P_{\rm 1.4GHz}-M_{\rm 500,SZ}$ scaling relations of the following forms:
\begin{equation}
 \log P_{\rm 1.4GHz} = A + B\;\log\left(\frac{M_{\rm 500,SZ}}{8.1 \times 10^{14}\;M_\odot}\right),
\end{equation}
\vspace{-0.2cm}
\begin{equation} 
 \log P_{\rm 1.4GHz} = A + B\;\log\left(\frac{Y_{\rm 500,SZ}}{1.2 \times 10^{-4}\;\rm Mpc^2}\right).
\end{equation}

\noindent In addition to fitting for $A$ and $B$ in each case, we also fit for the log-normal intrinsic scatter, $s$. The results of the analysis are shown in Table \ref{tab:survival}. There is marginally more scatter in $P_{\rm 1.4 GHz}$--$M_{\rm 500,SZ}$ relation, however, given the current measurement uncertainties, in the case of both scaling relation there is no significant difference in the fit parameters between the three datasets. Adding the upper limits therefore does not significantly modify the scaling relations. Moreover, although the method 2 upper limits are generally lower than their method 1 counterparts, there is no significant difference in the fit results for either scaling relation when considering the different upper limit methods. This suggests that the upper limits, including those for the disturbed systems, are consistent with the scaling relations.

{ To compare our scaling relation fits with the literature, we use the RH+USS BCES bisector results from \citet[][see their Table 3]{Cassano.2013.GRHScalRel} and convert the values such that the fitting forms match those used here. Their results convert to $B_M = 2.02 \pm 0.28$ and $A_M = 24.369 \pm 0.070$ for the $P_{\rm 1.4 GHz}$--$M_{\rm 500,SZ}$ fit, and $B_Y = 3.70 \pm 0.56$ and $A_Y = 24.55 \pm0.56$ for the $P_{\rm 1.4GHz}-Y_{\rm 500,SZ}$ relation. Our $P_{\rm 1.4GHz}-M_{\rm 500,SZ}$ fit produces a steeper slope for all three samples, although the results are consistent with the literature within two sigma. Conversely, we obtain a significantly flatter slope for the $P_{\rm 1.4GHz}-Y_{\rm 500,SZ}$ correlation, with our literature-only sample result consistent with the literature value at three sigma.}

\section{Radio sources in the full fields}
\label{sec:fovsrcs}

Due to the sensitivity of our 610 MHz maps, several extended and diffuse sources can be seen in the full field-of-view maps. Appendix \ref{app:fovsrcs} provides full details on three prominent sources with diffuse emisison in our maps. We also comment on the numerous FR-I and FR-II sources in the fields and provide source catalogs for all pointings. Here we summarize our results on the three discrete sources.

{In the ACT-CL J0045.2$-$0152 map, NE of the cluster, lies a face-on spiral galaxy NGC 0245 at $z=0.0136$. NCG0245 is a metal-rich starforming galaxy, with star forming hot spots in the nucleus and spiral arms \citep{PerezGonzalez.2003}. Our GMRT data reveals diffuse emission throughout the optical galactic disk, providing radio morphology for this full region for the first time. Our GMRT emission and spectral index map for the source are shown in Figure \ref{fig:NGC0245}.}

{The second source of interest is the active radio galaxy PKS 2324-02 ($z = 0.188$), the brightest source in the ACT-CL J2327.4$-$0204 field of view. FIRST data on this source shows a dual jet structure with large synchrotron lobes. Using our GMRT data, which has comparable resolution to FIRST, we have investigated the spatial spectral index distribution of PKS 2324-02 for the first time. The spectral index map shown in Figure \ref{fig:PKS2324-02} reveals multiple epochs of synchrotron activity, with the outer edges of the lobes consisting of very steep ($\alpha < -2.5$) spectrum emission.}

{The final discrete source discussed is that of the merging galaxy pair Arp 118, SW of the ACT-CL J0256.5+0006 position. We detect diffuse emission over the full optical source, extending beyond the optical emission. Interestingly, the radio peak is offset from the cores of both pair galaxies and does not spatially correlate with visible star formation in the merger. The GMRT emission for Arp 118 is shown in Figure \ref{fig:ARP118}.}

\section{Summary and conclusions}
\label{sec:summary}
We present here the GMRT 610 MHz continuum imaging results of a subset of the ACT Equatorial cluster sample, with a completeness above $M_{500} > 5 \times 10^{14} M_\odot$ of $\sim$ 73\%. We include data from our dedicated ACT follow-up as well as archival GMRT data. A summary of our results are as follows:
\begin{itemize}
 \item We detect a new radio mini-halo in ACT-CL J0022$-$0036, the highest redshift mini-halo detection to date. The {estimated} 610 MHz mini-halo flux density is {$S_{610} = 6.4 \pm 1.8$} mJy. X-ray and optical substructure measurements indicate this cluster is a cool-core system undergoing minor merger activity. The origin of the mini-halo could be from hadronic process in the cool-core, or turbulence driven due to gas sloshing in the cluster centre.
 
 \item Through reprocessing of data with updated software, we are able to spatially resolve the new radio halo detection of ACT-CL J0256.5+0006 presented in \citet{Knowles.2016.J0256}. We measure a more constrained 610 MHz halo {flux density} of 6.9 $\pm$ 0.7 mJy, which relates to a 1.4 GHz log radio power (W Hz\per) of $24.0 \pm 0.1$.
 
 \item We detect low-resolution diffuse emission in ACT-CL J0045.9$-$0152 and ACT-CL J0014.9$-$0057. Further radio follow-up is required to confirm the detections and assist in accurate classification.
 
 \item We cannot confirm the mini-halo in ACT-CL J2129.6+0005 reported by \citet{Kale.2015.EGRHS} due to contamination of the BCG region by a sidelobe from a bright $\sim$ 1 Jy source to the South.
 
 \item We determine radio halo upper limits for clusters with no diffuse emission detection in two ways: using the standard method \citep[see e.g.,][]{Kale.2013.EGRHSUL}, and using an adapted method where the size of the injected halo depends on the virial radius of the cluster. A survival analysis of the $P_{\rm 1.4GHz}-Y_{\rm 500,SZ}$ and $P_{\rm 1.4 GHz}-M_{\rm 500,SZ}$ scaling relations and both sets of upper limits indicates no statistically significant difference between either method of determining upper limits. 
 
 \item As 43\% of our cluster sample lies at high redshift ($z \gtrsim 0.5$), we can for the first time comment on the occurrence of diffuse emission in this regime. We note that above our mass cut of $M_{\rm 500,SZ} > 5 \times 10^{14}\;M_\odot$, the radio follow-up above a redshift of $z \gtrsim 0.5$ is 100\% complete. Of the six high redshift clusters, we have one detection and two clusters with unclassified diffuse emission. Although we have small number statistics, if the presence of the unclassified emission is confirmed to be diffuse halo or relic emission, we would have a diffuse emission occurrence rate higher than theoretical models predict for high redshift \citep{Brunetti.2007.CRandGRH}. However, deeper observations of our high redshift cluster sub-sample are required to confirm the statistics.
 
 \item In Appendix \ref{app:fovsrcs} we present new 610 MHz imaging and spectral index maps for three bright, extended sources in fields of three different cluster pointings: a face-on, star-forming spiral galaxy NCG0245, the active radio galaxy PKS2324-02, and the merging galaxy pair Arp 118. We also comment on other extended sources in the 610 MHz maps, most of which are FR-I and FR-II sources.
 
 \item {Finally, we} provide a catalog of the 610 MHz radio sources in our full resolution cluster fields{, available online}. The density of sources varies significantly in some images due to a wide variety in image quality and depth, particularly in the case of archival data.
\end{itemize}

The occurrence of diffuse radio emission, above a mass limit of $M_{500} > 5 \times 10^{14} M_\odot$, is $27^{+20}_{-14}\%$ in our cluster sample\footnote{Uncertainties determined from \citet{Gehrels.1986}.}, when taking into account the unclassified emission. This is similar to what has been found in previous X-ray and SZ samples \citep{Cassano.2013.GRHScalRel}. Given the current understanding of, and open questions underlying, the mechanisms powering diffuse cluster emission, and in particular radio halos, increasing the number of detections while focusing on statistically complete samples is critical in order to improve constraints on the theoretical models. We note that non-detections can be equally enlightening; in particular, the lack of diffuse emission in two high mass merging clusters, A370 and ACT-CL J2135.2+0125, presented here warrants further investigation. A larger follow-up programme with the upgraded GMRT, focusing on a statistical sample of 40 ACTPol clusters, is currently underway.

\section*{Acknowledgements}

KK acknowledges funding support from NRF/SKA South Africa Project grants 82761 and 96665, and the Claude Leon Foundation. AJB acknowledges support from NSF grant AST-0955810. JPH acknowledges financial support for this work from the National Aeronautics and Space Administration through Chandra Award Number GO4-15125X issued by the Chandra X-ray Center, which is operated by the Smithsonian Astrophysical Observatory for and on behalf of the National Aeronautics Space Administration under contract NAS8-03060. CL thanks CONICYT for grant Anillo ACT-1417. KM acknowledges support from National Research Foundation of South Africa under grants 85467 and 93565. 

We thank the staff of the GMRT who made these observations possible. GMRT is run by the National Centre for Radio Astrophysics of the Tata Institute of Fundamental Research. 

This research has made use of the NASA/IPAC Extragalactic Database (NED) which is operated by the Jet Propulsion Laboratory, California Institute of Technology, under contract with the National Aeronautics and Space Administration. It also made use of the Ned Wright online cosmology calculator \citep{Wright.2006}. Funding for SDSS-III has been provided by the Alfred P. Sloan Foundation, the Participating Institutions, the National Science Foundation, and the U.S. Department of Energy Office of Science.




\bibliographystyle{mnras}
\bibliography{references} 



\clearpage
\appendix

\section{Relaxed clusters with diffuse emission non-detections}
\label{app:nondets}
Here we discuss the results from the four clusters in our sample which show no evidence of diffuse radio emission at the level of the noise of our images, and have no indication of disturbed morphology given current data. 

\subsection{ACT-CL J0152.7+0100}

\begin{figure}
 \centering
 \includegraphics[height=0.48\textwidth,clip=True,trim=40 10 10 35]{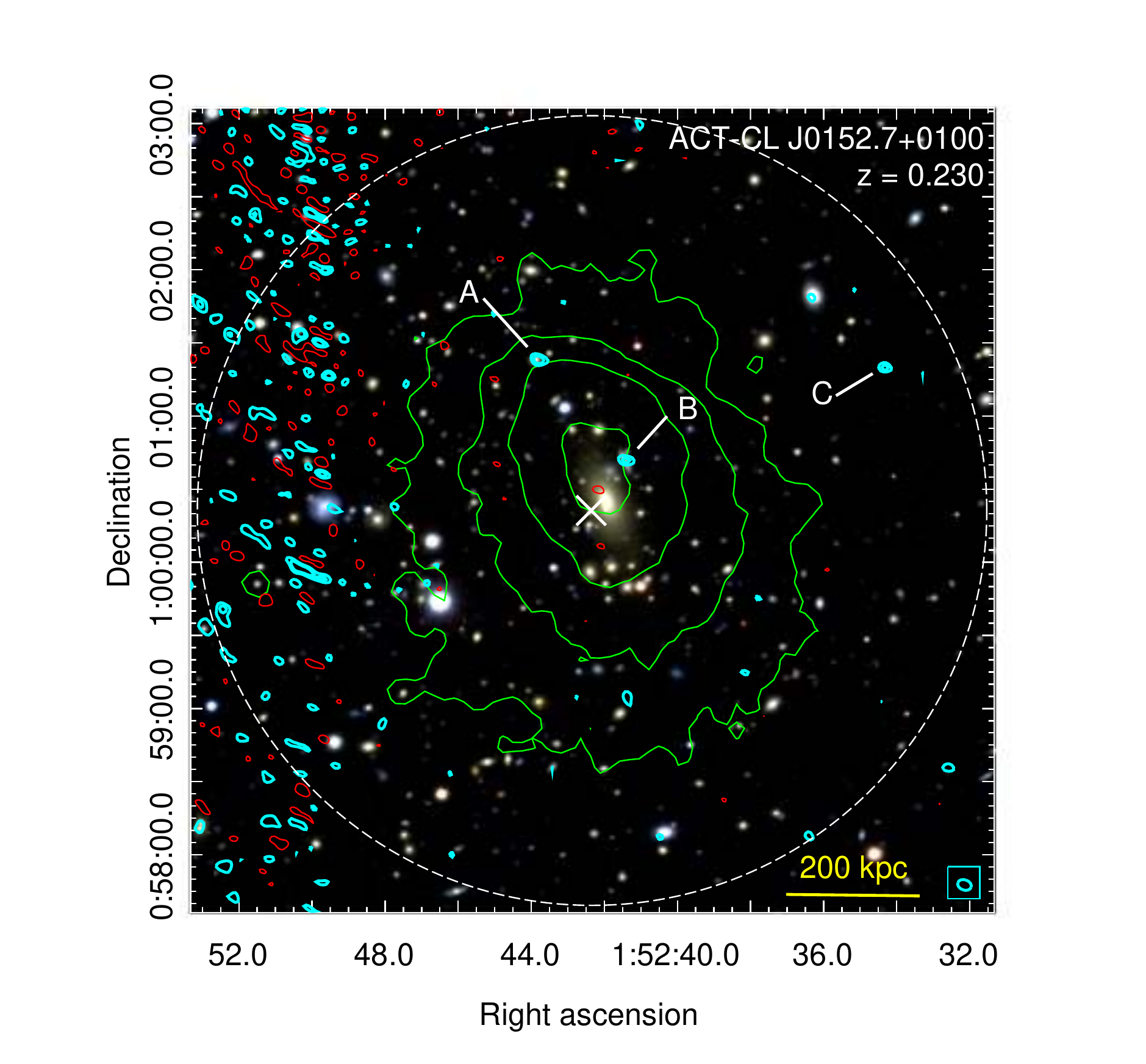}
 \vspace{-0.5cm}
 \caption{SDSS $gri$-image of ACT-CL J0152.7+0100 with 610 MHz GMRT (+ve: thick, cyan, -ve: thin, red) and \textit{XMM-Newton} (green{, [10,20,40,80] counts}) contours overlaid.  GMRT contour levels are $[\pm3,5]\sigma$, where $1\sigma = 98\; \mu$Jy beam\per, with the beam shown by the boxed ellipse. {Flux densities} for sources A-C are given in Table \ref{tab:j0152}. The X and dashed circle indicate the cluster SZ peak and $R_{500}$ region, respectively, and the physical scale at the cluster redshift is shown by the bar in the bottom right corner.}
 \label{fig:j0152}
\end{figure}

ACT-CL J0152.7+0100 ($z = 0.230$) is an Abell cluster \citep[A267, hereafter;][]{Abell.1958} and has been studied at multiple wavelengths from optical through to X-ray \citep[see e.g.,][]{Dahle.2002, Bauer.2005, Cavagnolo.2009}. This non-cool core cluster \citep{Cavagnolo.2009, Mahdavi.2013} was defined as a ``fossil system'' by \citet{EigenthalerZeilinger.2009} and \citet{Zarattini.2014} based on the magnitude separation in $r$-band between the brightest and second brightest cluster galaxies, as defined by \citet{Jones.2003} and later revised by \citet{Voevodkin.2010}. Simulations of fossil groups indicate that these systems only continue to grow through minor mergers, and therefore should not show evidence of strong substructure \citep{DOnghia.2005.fossilsims}. The dynamical state of A267 is somewhat unclear, with X-ray morphological parameter results $w = 0.008$ \citep{Pratt.2016} and $P_3/P_0 = (0.27 \pm 0.22) \times 10^{-7}$ \citep{Mahdavi.2013} indicating strong evidence for a relaxed cluster\footnote{
Based on citerion defined by \cite{Cassano.2010.GRHMergerConn}}, whereas DS test results based on spectroscopic measurements of 226 cluster members provide strong evidence for the presence of substructure \citep[$S_\Delta < 0.01$;][]{Pratt.2016}. The high significance of substructure from the DS test may be related to the clumpy nature of the outskirts of the cluster, whereas the morphological parameters in the X-ray are determined within the central $R_{500}$ region of the cluster. The SDSS 3-colour $gri$ image of the A267 cluster region is shown in the left panel of Figure \ref{fig:j0152}, with arbitrary \textit{XMM-Newton} contours (ObsID: 0084230401; 30 ks) overlaid.

A267 was observed with the GMRT at 610 MHz as part of the Extended GMRT Radio Halo Survey \citep[EGRHS;][]{Kale.2013.EGRHSUL}. Our full resolution contours from the reprocessing of their data are shown in cyan in Figure \ref{fig:j0152}. The Eastern part of the cluster region is contaminated by sidelobes of a bright source North of the cluster, however we confirm the EGRHS finding of no diffuse emission in the cluster region and detect three faint radio sources above $5\sigma$ within $R_{500}$. Not surprisingly, none of these sources are detected in FIRST. Our 610 MHz {flux densities} for these sources, labelled A-C, are provided in Table \ref{tab:j0152}. There is no radio emission from the cluster BCG marked by the cross in Figure \ref{fig:j0152}. Only source B is spectroscopically confirmed as a cluster member at $z = 0.2248$. From a visual inspection of the SDSS $gri$-image, source A is likely a cluster member based on colour comparison. There is no optical counterpart for source C.

\begin{table}
 \centering
 \caption{610 MHz {flux densities} for the sources in the J0152.7+0100 and J2129.6+0005 $R_{500}$ cluster regions. Source labels are indicated in Figure \ref{fig:j0152}. Notes are as in Table \ref{tab:j0022}.}
 \label{tab:j0152}
 \begin{tabular}{cccuc}
  \toprule
  ID & R.A. & Dec. & \multicolumn{1}{c}{$S_{\rm 610 MHz}$} & Notes \\
   & (deg) & (deg) & \multicolumn{1}{c}{\rm (mJy)} & \\
  \midrule
  A & 28.182501 & 1.023076 & 0.77 , 0.19 & M$^*$  \\
  B & 28.172523 & 1.011619 & 0.69 , 0.18 & M \\
  C & 28.142997 & 1.022179 & 0.53 , 0.16 & -  \\
  \bottomrule
 \end{tabular}
\end{table}

\subsection{ACT-CL J2129.6+0005}

\begin{figure}
 \centering
 \includegraphics[width=0.48\textwidth, clip=True, trim=30 20 30 40]{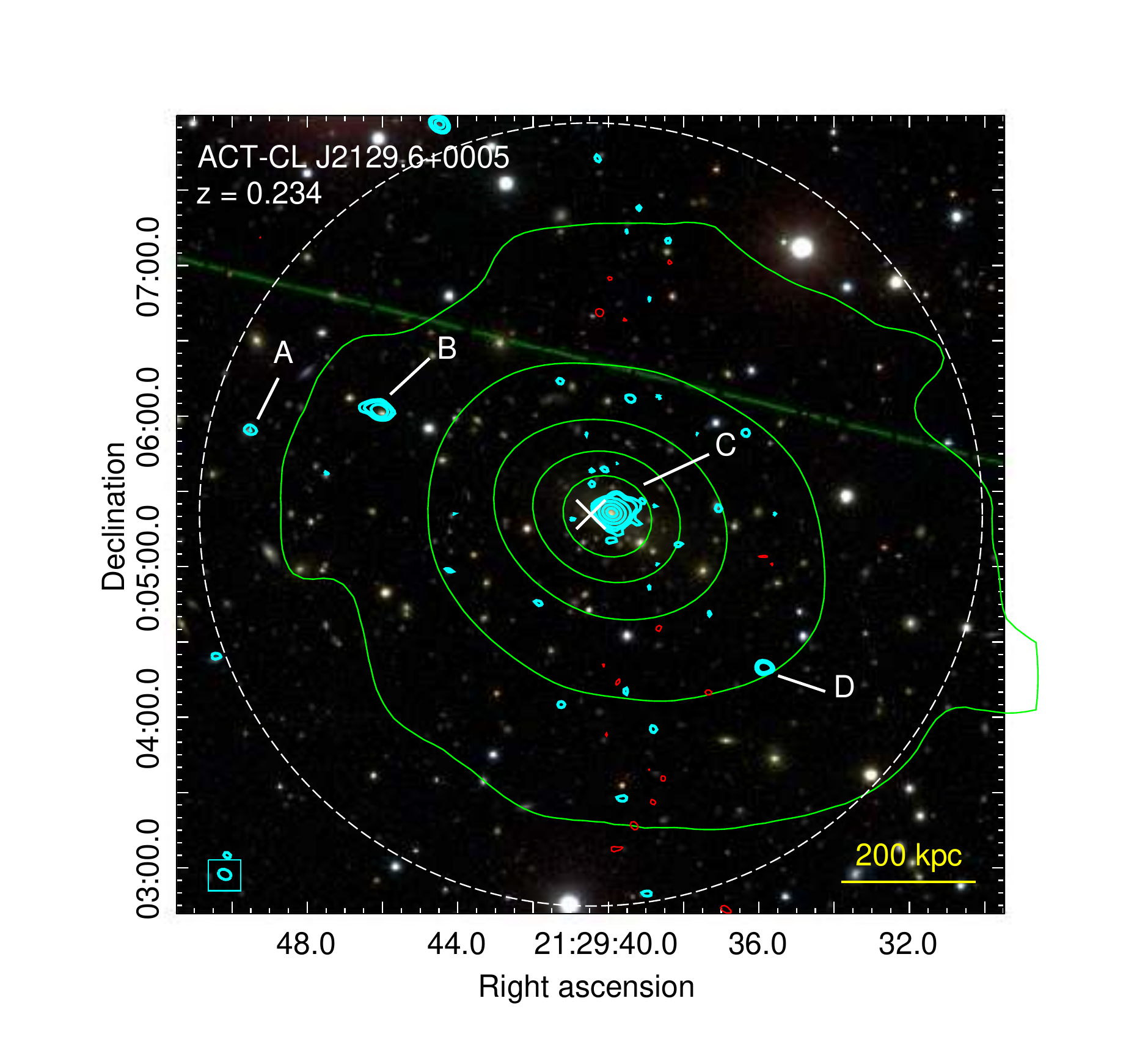}
 \vspace{-0.5cm}
 \caption{SDSS DR12 3-colour \textit{gri}-image of ACT-CL J2129.6+0005 with 610 MHz GMRT contours overlaid (+ve: thick, cyan, -ve: thin, red). Contour levels are $[\pm3,5,10,50,200,500]\sigma$, where $1\sigma = 50\; \mu$Jy beam\per, with the beam shown by the boxed ellipse. The cross and dashed circle are the SZ peak and $R_{500}$ region for the cluster, respectively. \textit{XMM-Newton} contours are overlaid in green { with levels [10, 34.4, 107.5, 229.4, 400] counts}. The physical scale at the cluster redshift is shown by the bar in the bottom right corner. {Flux densities} for sources A-D are provided in Table \ref{tab:j2129}. }
 \label{fig:j2129}
\end{figure}

A REFLEX cluster \citep[RXC J2129+0005;][]{Bohringer.2004}, ACT-CL J2129.6+0005 is a relaxed system at a redshift of $z = 0.234$, with a central BCG. \citet{Kale.2015.EGRHS} studied it as part of the EGRHS at both 235 MHz and 610 MHz. They reported a detection of a $\sim$ 225 kpc radio mini-halo at 610 MHz after removing unresolved radio emission from the BCG. After reprocessing their 610 MHz data, we produce an image with slightly higher resolution and lower noise. Our full resolution, primary beam-corrected contours (+ve: thick, cyan, -ve: thin, red) are overlaid on the SDSS $gri$-image shown in the right panel of Figure \ref{fig:j2129}, with \textit{XMM-Newton} X-ray contours (ObsID: 0093030201) overlaid in green. We detect four sources, labelled A-D,  above 5$\sigma$ of the central noise level of 0.05 mJy beam\per. Our 610 MHz {flux densities} of these sources are given in Table \ref{tab:j2129}.

Source C is the BCG, detected in FIRST with a 1.4 GHz {flux density} of 24.3 $\pm$ 1.2 mJy, giving a spectral index of $\alpha_{610}^{1435} = -0.8 \;\pm\; 0.2$. There is some indication of extended mini-halo emission around this source, as well as some unresolved 3$\sigma$ emission. However, there is a bright 0.9 Jy source 12.6{\arcmin} directly south of the BCG. This source has a North-South sidelobe which could be contributing to the higher signal around the BCG. Indeed, after subtracting the full-resolution BCG emission, a low resolution image reveals a signal in the centre of the cluster, but due to the sidelobe contamination we cannot reliably recover or confirm the extended emission reported by \citet{Kale.2015.EGRHS}. Sources A and B are spectroscopically confirmed as a foreground galaxy ($z = 0.2886$) and cluster member ($z = 0.2379$), respectively. Source D has no discernible optical counterpart. \\

\begin{table}
 \centering
 \caption{610 MHz {flux densities} for the sources in the J0152.7+0100 and J2129.6+0005 $R_{500}$ cluster regions. Source labels are indicated in Figure \ref{fig:j2129}. Notes are as in Table \ref{tab:j0022}.}
 \label{tab:j2129}
 \begin{tabular}{cccuc}
  \toprule
  ID & R.A. & Dec. & \multicolumn{1}{c}{$S_{\rm 610 MHz}$} & Notes \\
   & (deg) & (deg) & \multicolumn{1}{c}{rm (mJy)} & \\
  \midrule
  A & 322.456321 & 0.098469 & 0.29 , 0.07 & F  \\
  B & 322.442018 & 0.100606 & 1.66 , 0.11 & M  \\
  C & 322.416229 & 0.089324 & 48.86 , 2.45 & M$^\bullet$ \\
  D & 322.399400 & 0.072164 & 0.57 , 0.08 & - \\
  \bottomrule
 \end{tabular}
\end{table}

\subsection{ACT-CL J2135.7+0009}
ACT-CL J2135.7+0009 is a low mass, low redshift ($z = 0.118$) cluster observed as part of the initial pilot study. Although the preliminary mass estimate placed this system within the selection criteria of our study, the published mass is 50\% lower than the defined cut off of $M_{500}^{\rm UPP} = 5 \times 10^{14} M_\odot$. We include the results here for completeness. There is no available dynamical state information on this system.

\begin{figure}
 \centering
 \includegraphics[width=0.47\textwidth, clip=True, trim=20 0 10 40]{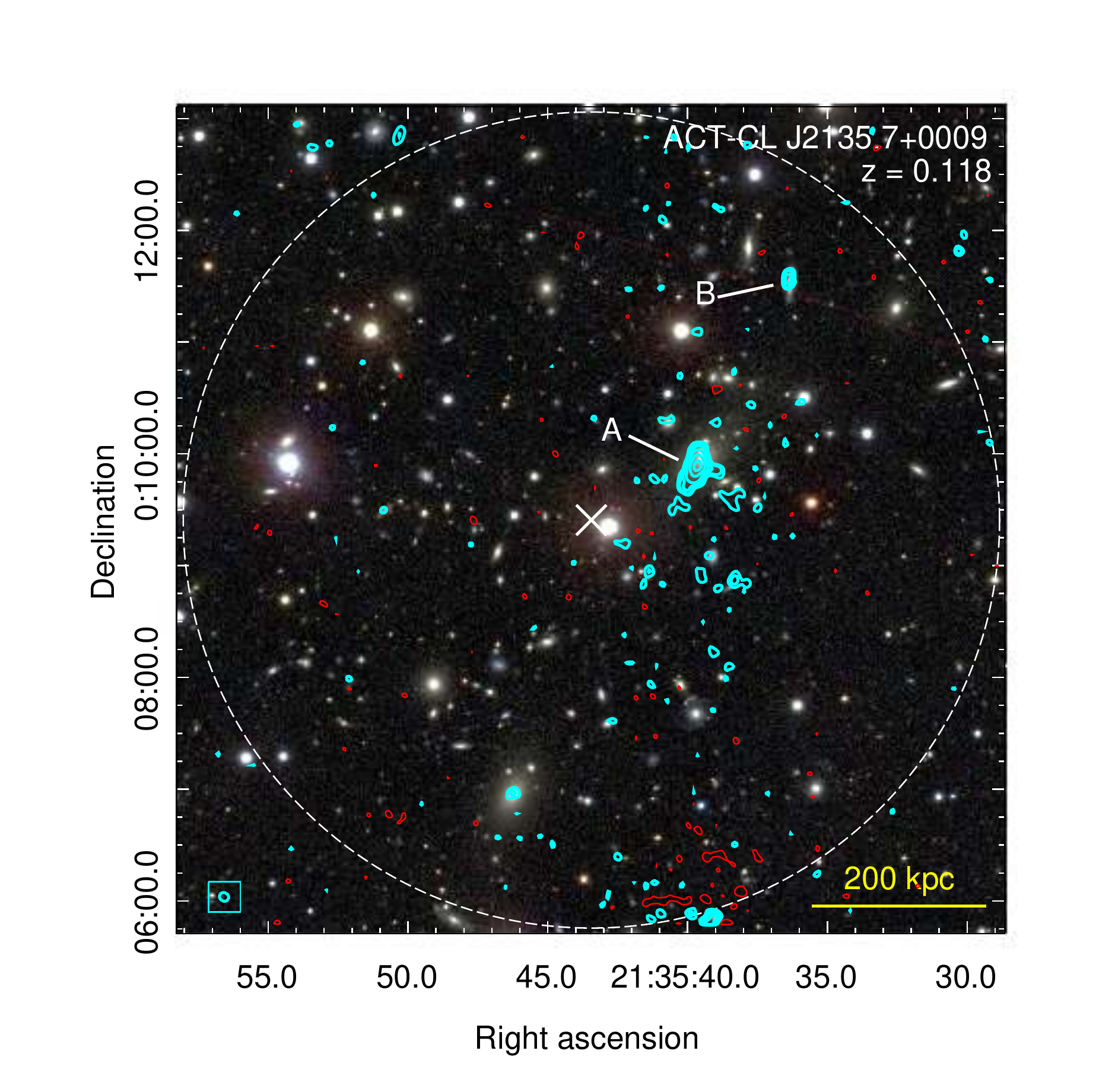}
 \vspace{-0.4cm}
 \caption{SDSS DR12 3-colour \textit{gri}-image of ACT-CL J2135.7+0009 with 610 MHz GMRT contours overlaid (+ve: thick, cyan, -ve: thin, red). Contour levels are $[\pm3,5,10,20,50,100,200] \times 1\sigma = 90\; \mu$Jy beam\per, with the beam shown by the boxed ellipse. The X and dashed circle are the SZ peak and $R_{500}$ region for the cluster, respectively. {Flux densities} for sources A and B are provided in Table \ref{tab:j2135}. The physical scale at the cluster redshift is shown by the bar in the bottom right corner.}
 \label{fig:j2135}
\end{figure}

Our 610 MHz full resolution radio contours are overlaid on the SDSS $gri$-band image of the cluster $R_{500}$ region shown in Figure \ref{fig:j2135}. Two compact sources, labelled A and B, are detected above 10$\sigma$, where $\sigma = 95 \mu$Jy beam{\per} is the rms noise in the central region of the map. This is slightly higher than the map noises for the other clusters observed in the pilot program due to contamination of the field of view by a sidelobe of a bright source South of the cluster region. Our 610 MHz {flux densities} for these sources are given in Table \ref{tab:j2135}. Source A is detected in FIRST data with a {flux density} of 14.32 mJy, giving a spectral index of $\alpha^{1400}_{610} = -1.19 \pm 0.17$.\\

\begin{table}
 \centering
 \caption{610 MHz {flux densities} for the sources in the ACT-CL J2135.7+0009 $R_{500}$ cluster region. Source labels are indicated in Figure \ref{fig:j2135}. Notes are as in Table \ref{tab:j0022}.}
 \label{tab:j2135}
 \begin{tabular}{cccuc}
  \toprule
  ID & R.A. & Dec. & \multicolumn{1}{c}{$S_{\rm 610 MHz}$} & Notes \\
   & (deg) & (deg) & \multicolumn{1}{c}{\rm (mJy)} & \\
  \midrule
  A & 322.417554 & 0.087071 & 40.01 , 2.12 & M$^\bullet$ \\
  B & 323.901868 & 0.191379 & 2.12 , 0.19 & - \\
  \bottomrule
 \end{tabular}
\end{table}

\subsection{ACT-CL J2327.4$-$0204}
\begin{figure}
 \centering
 \includegraphics[width=0.47\textwidth, clip=True, trim=20 20 10 40]{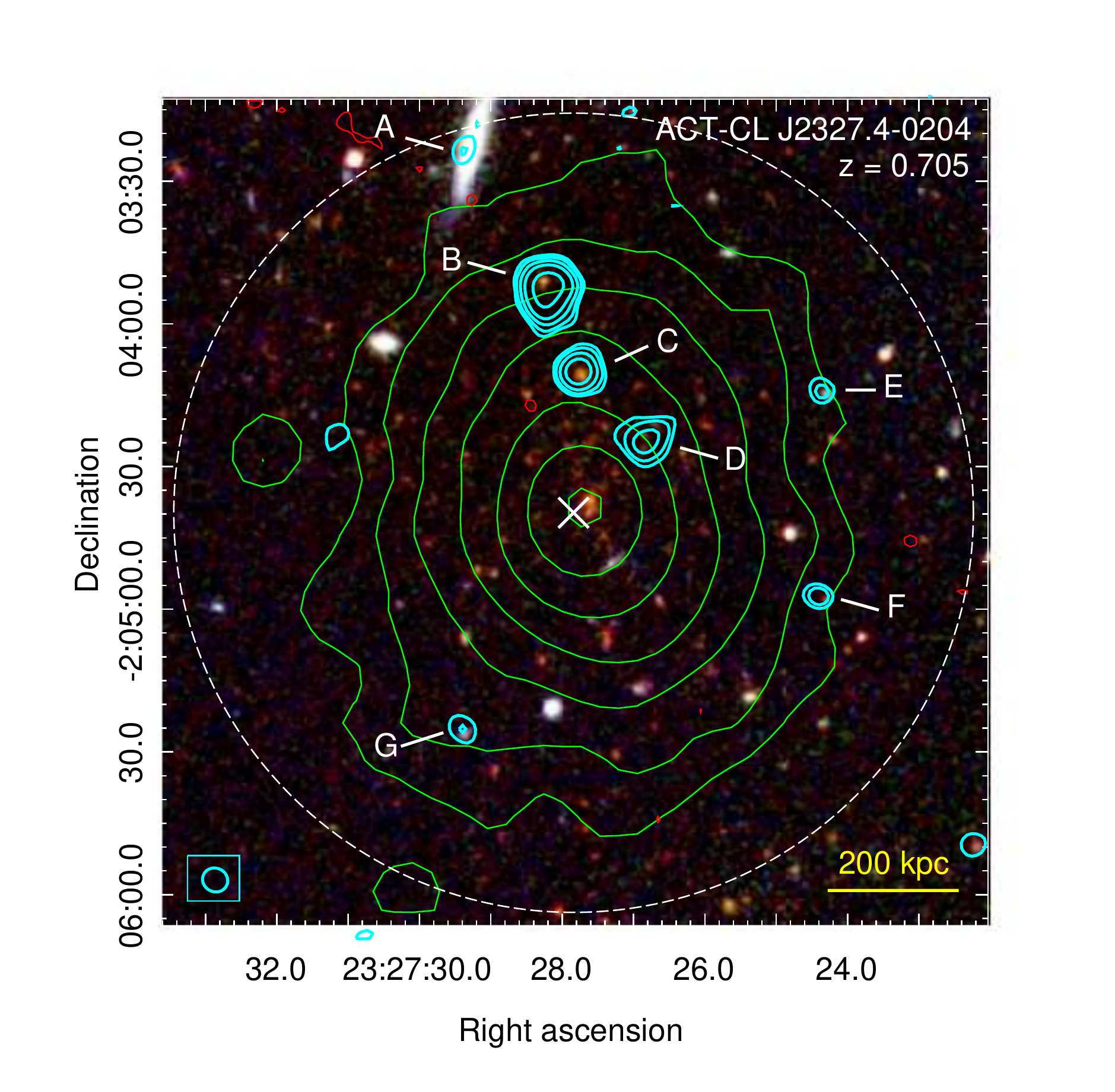}
 \caption{SDSS DR12 3-colour \textit{gri}-image of ACT-CL J2327.2$-$0204 with 610 MHz GMRT contours overlaid (+ve: thick, cyan, -ve: thin, red). Contour levels are $[\pm3,5,10,20,50,100,200] \times 1\sigma = 58\; \mu$Jy beam\per, with the beam shown by the boxed ellipse. \textit{Chandra} {counts} contours are overlaid in green {with levels of 6, 8, 12, 20, 36, 68, 125}. The X and dashed circle are the SZ peak and $R_{500}$ region for the cluster, respectively. The physical scale at the cluster redshift is shown by the bar in the bottom right corner.}
 \label{fig:j2327}
\end{figure}

The richest galaxy cluster in the Second Red-Sequence Cluster Survey \citep[RCS2;][]{Gilbank.2011.RCS2} and the most significant ACTPol detection over $\sim$ 1000 deg$^2$ \citep{Hilton.2017.ACTPol}, ACT-CL J2327.4$-$0204 (hereafter J2327) is a well studied, massive cluster at $z = 0.705$. A comprehensive multiwavelength study of this cluster was carried out by \citet{Sharon.2015.J2327}; they confirmed the X-ray findings of \citet{Menanteau.2013.ACTE} that J2327 is dynamically relaxed, with no evidence for significant substructure using X-ray, gravitational lensing, SZ, and optical measurements. The SDSS DR12 composite $gri$-image of the cluster is given in Figure \ref{fig:j2327}, with \textit{Chandra} contours (arbitrary levels) overlaid in green. The \textit{Chandra} data is taken from project 14025 (PI: Hicks).

Our full resolution, primary beam-corrected 610 MHz image of J2327 is show by cyan contours overlaid on Figure \ref{fig:j2327}. We detect seven compact sources, labelled A-G, above $5\sigma$ of the rms noise. Their 610 MHz {flux densities} are provided in Table \ref{tab:j2327}. All seven sources have a likely or potentiual optical counterpart. Source C is the only one with a spectroscopic redshift, identifying it as a cluster member at $z_C = 0.7073$. Sources A, E, and G are likely foreground sources, based on colour matching in the SDSS image. Similarly, sources B, D, and F are likely emanating from cluster members. We note however that there is a spatial offset between the radio peaks of sources B and D and their potential optical counterparts; it is possible that these sources are in fact AGN jets associated with source C.

\begin{table}
 \centering
 \caption{610 MHz {flux densities} for the sources in the ACT-CL J2327.2$-$0204 $R_{500}$ cluster region. Source labels are indicated in Figure \ref{fig:j2327}. Notes are as in Table \ref{tab:j0022}. $^\dagger$ Possible optical counterparts, but could also be AGN jets associated with source C.}
 \label{tab:j2327}
 \begin{tabular}{cccuc}
  \toprule
  ID & R.A. & Dec. & \multicolumn{1}{c}{$S_{\rm 610 MHz}$} & Notes\\
   & (deg) & (deg) & \multicolumn{1}{c}{\rm (mJy)} & \\
  \midrule
  A & 351.872422 & -2.056597 & 0.41 , 0.16 & F  \\
  B & 351.867499 & -2.064711 & 11.88 , 0.62 & M$^\dagger$  \\
  C & 351.865698 & -2.069451 & 3.11 , 0.21 & M \\
  D & 351.861740 & -2.073479 & 1.92 , 0.16 & M$^\dagger$\\
  E & 351.851487 & -2.070650 & 0.48 , 0.17 & F \\
  F & 351.851689 & -2.082577 & 0.49 , 0.15 & M\\
  G & 351.872487 & -2.090330 & 0.34 , 0.11 & F\\

  \bottomrule
 \end{tabular}
\end{table}

\section{Radio sources in the full fields}
\label{app:fovsrcs}

Due to the sensitivity of our 610 MHz maps, several extended and diffuse sources can be seen in the full field-of-view maps. Here we discuss three prominent sources, and comment on the numerous FR-I and FR-II sources. We use FIRST data to measure spectral indices for the sources, and where spectral index maps have been made, the GMRT and FIRST images are smoothed to the same resolution and regridded before masking out pixels below 3$\sigma$ in each image. The spectral index $\alpha^{1400}_{610}$ is then computed, with the spectral index uncertainty $\Delta\alpha^{1400}_{610}$ calculated as 
\begin{equation}
 \Delta\alpha^{1400}_{610} = \left[\log\left(\frac{\nu_1}{\nu_2}\right)\right]^{-1} \sqrt{\frac{\epsilon_1^2 + \sigma_1^2}{S_1^2} + \frac{\epsilon_2^2 + \sigma_2^2}{S_2^2}},
\end{equation}
where $\epsilon$ is the {flux density} uncertainty \citep[$\sim$ 5\% for both FIRST and 610 MHz GMRT][respectively]{Helfand.2015, Chandra.2004.GMRTfluxerr}, $\sigma$ is the local rms noise, $S$ is the {flux density}, $\nu$ is the image frequency, and subscripts 1 and 2 denote properties of the FIRST and GMRT images, respectively.

\subsection{NGC0245}
\label{subsubsec:ngc0245}

\begin{figure*}
 \centering
  \includegraphics[width=0.32\textwidth,clip=True,trim=32 0 70 55]{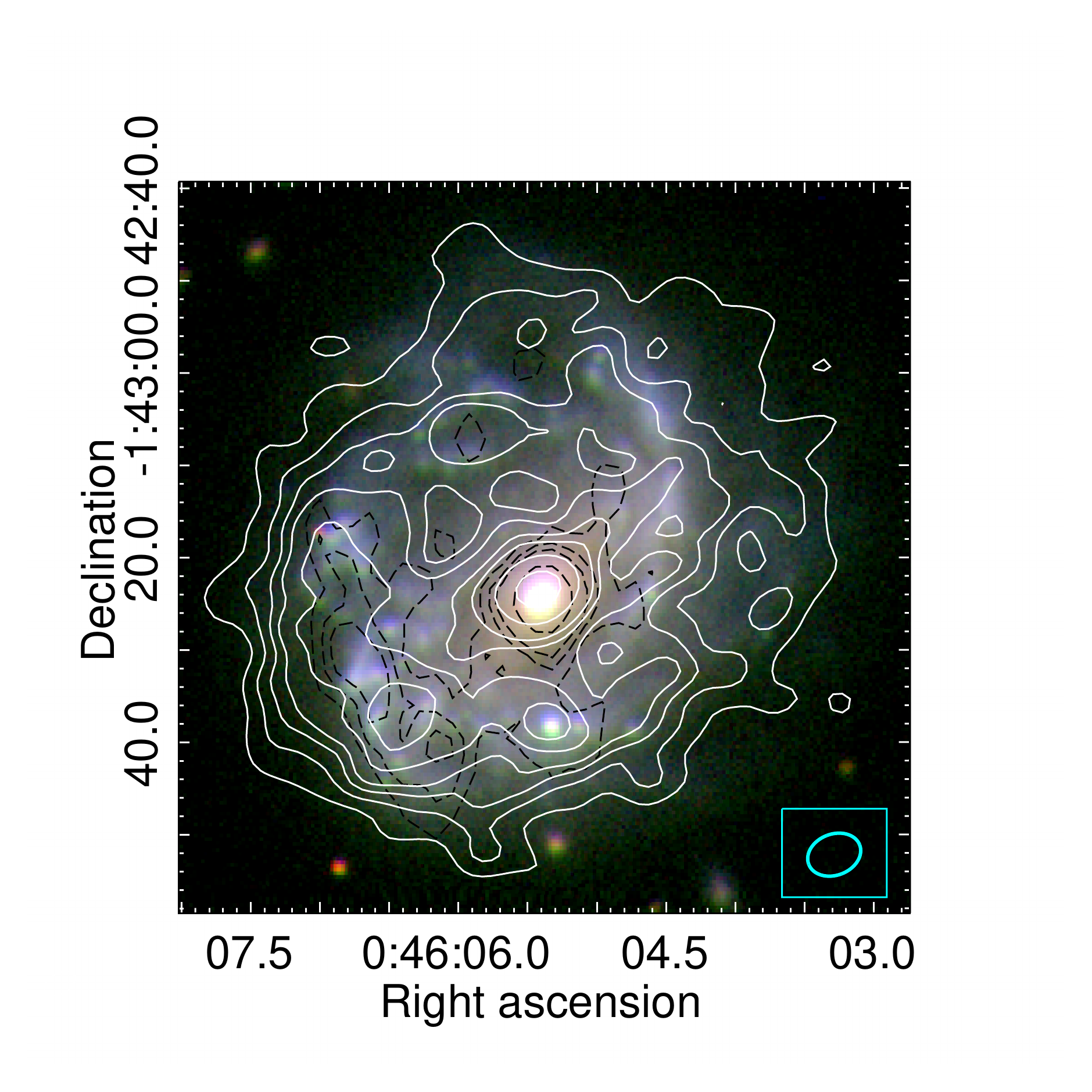}
  \includegraphics[width=0.31\textwidth,clip=True,trim=5 15 10 30]{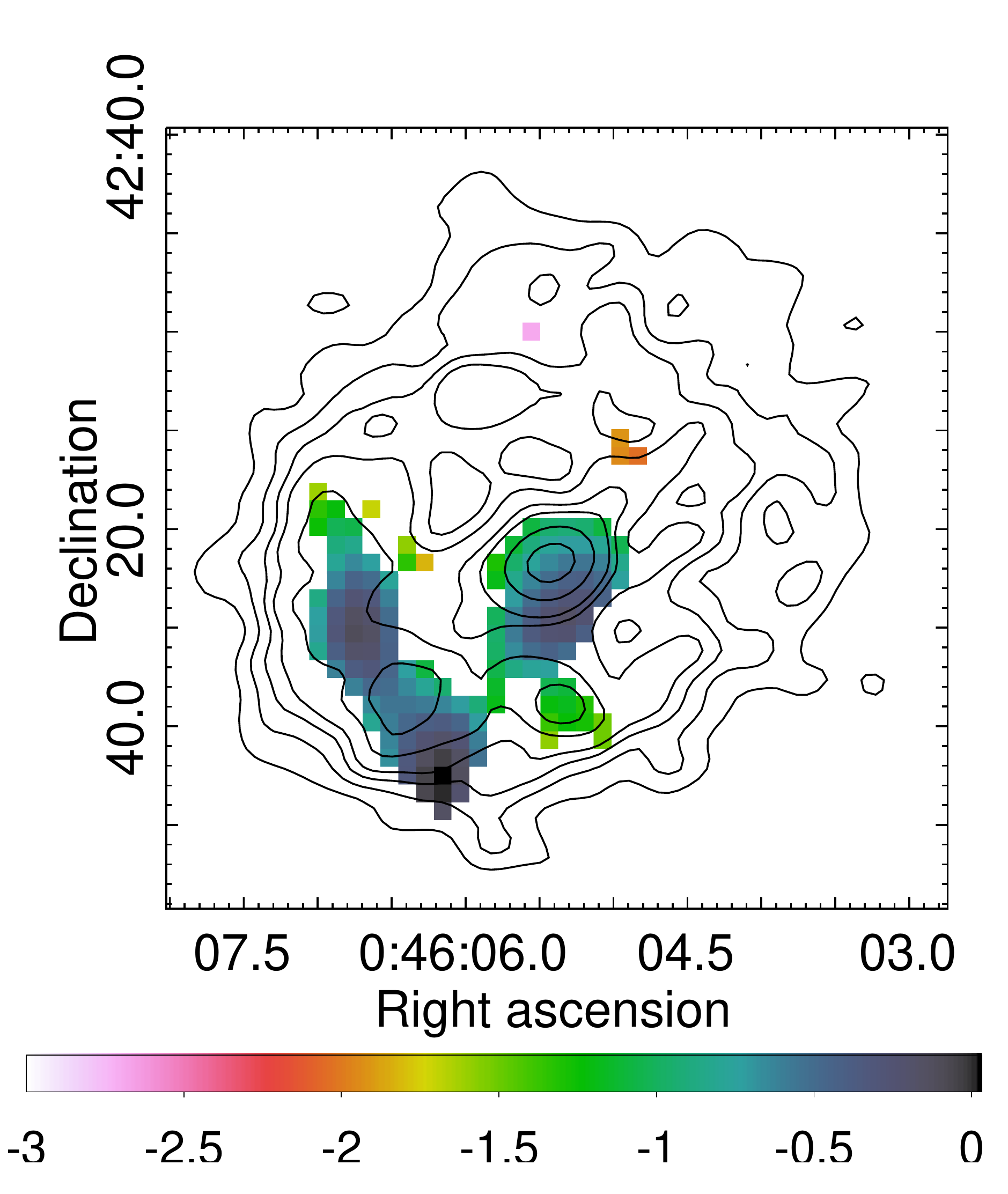}
  \includegraphics[width=0.31\textwidth,clip=True,trim=5 15 10 30]{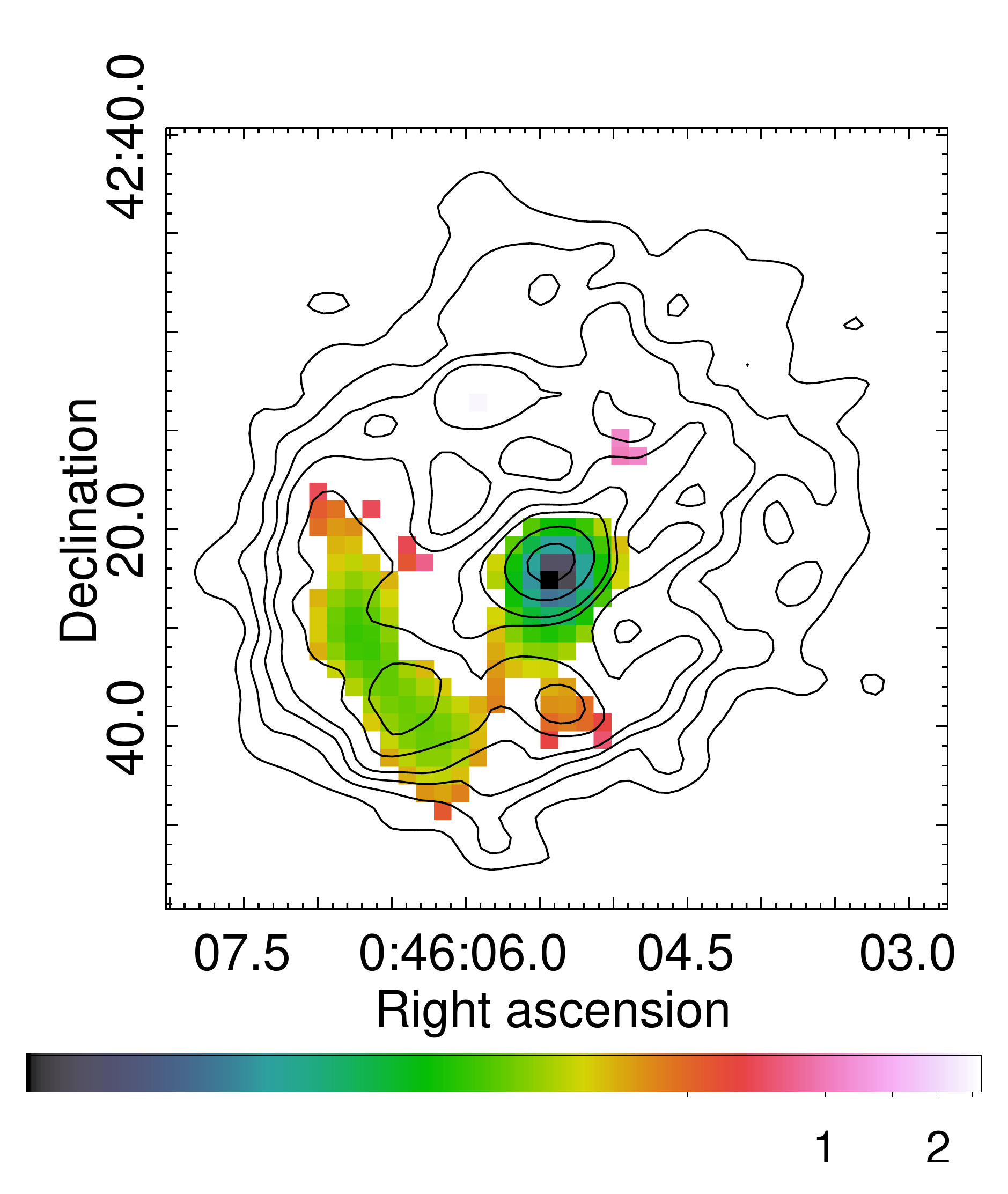}
  \caption{NGC0245. \textit{Left:} SDSS DR12 $gri$ image with 610 MHz GMRT (white, solid) and FIRST (black, dashed) contours overlaid. The GMRT contour levels are [3,5,10,20,50,100] $\times$ 1$\sigma = 42 \mu$Jy beam\per, with the beam indicated by the boxed ellipse. FIRST contour levels are [3,10,15,20,40] $\times$ 1$\sigma = 50 \mu$Jy beam\per. \textit{Middle:} Spectral index map of $\alpha^{1400}_{608}$ with the 610 MHz GMRT contours overlaid. \textit{Right:} Spectral index uncertainty map for $\alpha^{1400}_{608}$, with the 610 MHz GMRT contours overlaid.}
  \label{fig:NGC0245}
\end{figure*}

\begin{figure*}
 \centering
  \includegraphics[width=0.32\textwidth]{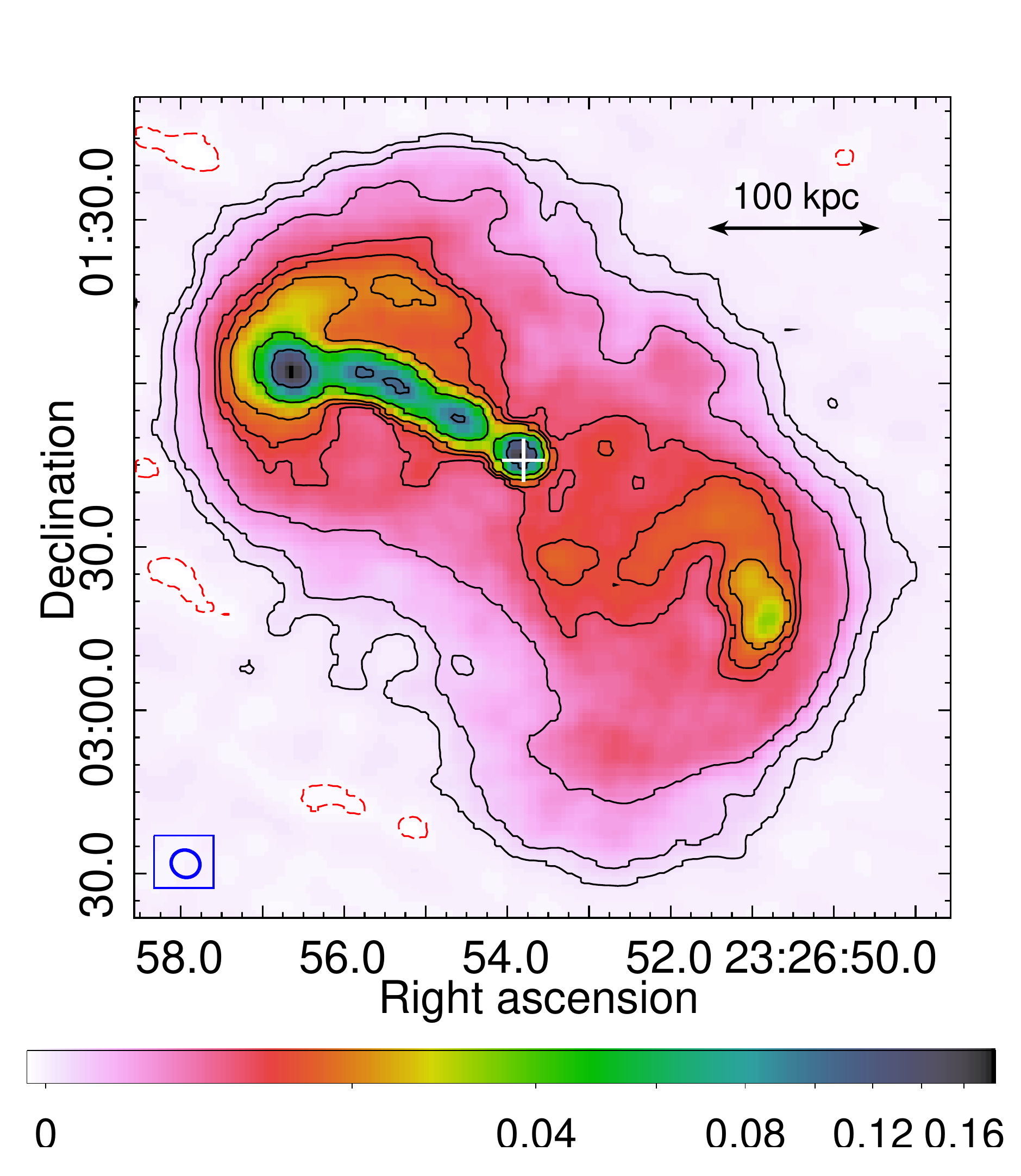}
  \includegraphics[width=0.32\textwidth]{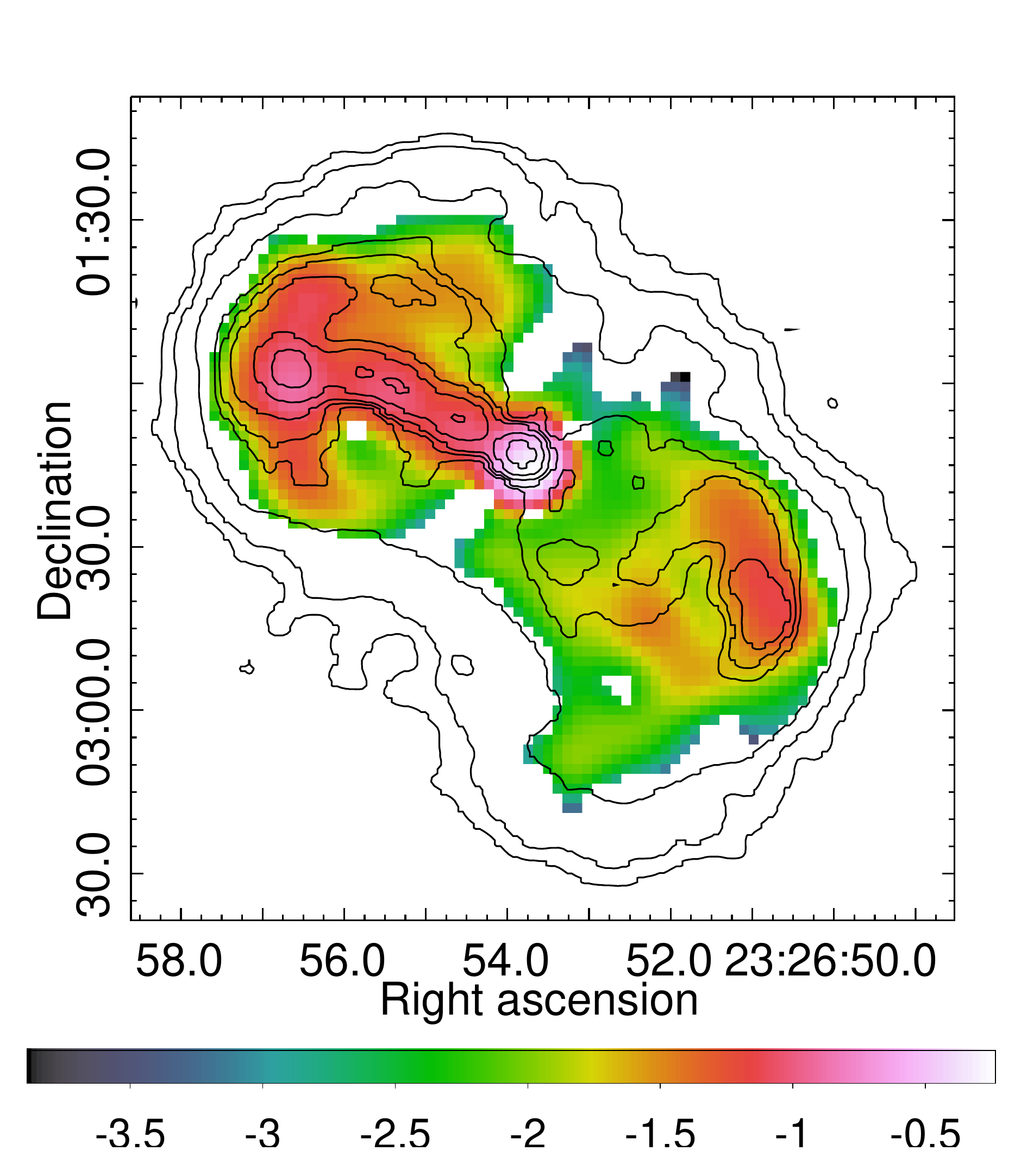}
  \includegraphics[width=0.32\textwidth]{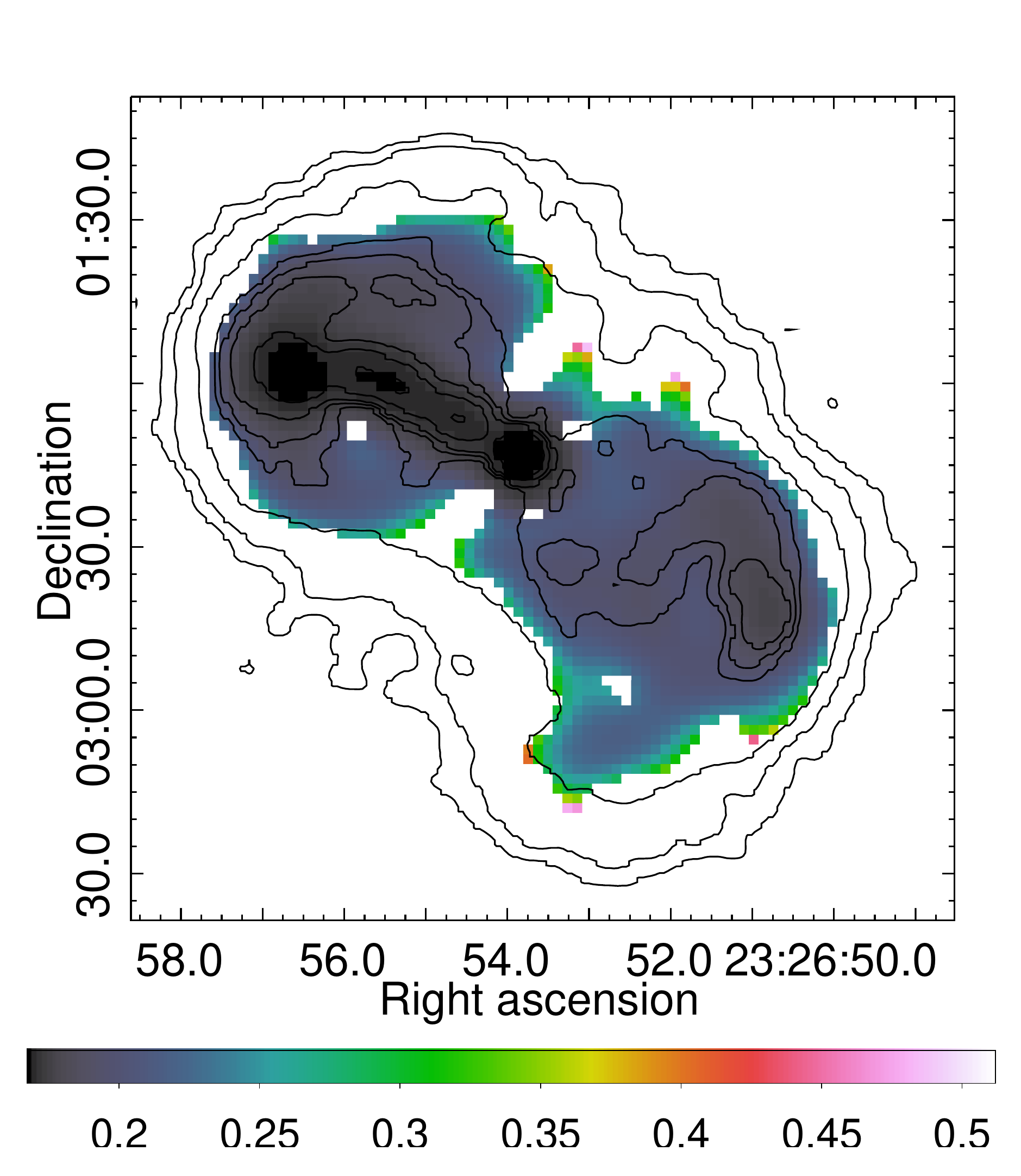}
  \caption{Active radio galaxy PKS 2324-02. \textit{Left:} 610 MHz GMRT image with [3,10,{30,60,75},100,200,500] $\times$ 1$\sigma = 0.2$ mJy beam\per contours overlaid in black. The $-3\sigma$ level is shown by the {dashed,} red contours. The beam is indicated by the boxed ellipse. The position of PKS2324-02 is indicated by the white cross. \textit{Middle:} Spectral index map of $\alpha^{1400}_{608}$, using our 610 MHz GMRT and FIRST data{; contours as per the left panel}. \textit{Right:} Spectral index {uncertainty} map for $\alpha^{1400}_{608}${; contours as per the left panel}.}
  \label{fig:PKS2324-02}
\end{figure*}

The radio galaxy NGC 0245 (R.A. = 00h46m05.35s, Dec. = $-$01d43m24.35s, J2000) lies within the GMRT field-of-view of ACT-CL J0045.2$-$0152. NGC 0245 is a nearly face-on spiral galaxy at $z$ = 0.0136 and has been identified as a metal-rich starburst galaxy, with recent star formation occurring in the galaxy nucleus as well as in regions of the spiral arms \citep{PerezGonzalez.2003}. It is part of several galaxy samples studied in the infrared and near and far ultraviolet wave bands \citep{Moshir.1990, Rego.1993, MoustakasKennicutt.2006, Hao.2011}.

In our 610 MHz map, we detect the bright core of NGC0245, $S_{\rm GMRT,core} = 9.63 \pm 0.08$ mJy, surrounded by a halo of faint diffuse emission with a largest angular diameter of 1.14{\arcmin} ($\sim$19 kpc at the galaxy redshift). The integrated {flux density} for the diffuse disk emission is $S_{\rm GMRT,disk} = 49.17 \pm 0.58$ mJy. The left panel of Figure \ref{fig:NGC0245} shows our 610 MHz radio contours overlaid in white on the 3-colour \textit{gri} image of NGC0245 from SDSS Data Release 12 \citep{Alam.2015}. The GMRT emission extends over the entire galactic disk, with radio hot spots roughly coinciding with the regions of star forming activity. These radio hotspots are detected in FIRST (dashed black contours in the SDSS image), however our GMRT data provides, for the first time, radio morphology beyond these star forming regions. NGC0245 is detected in NVSS data beyond the optical disk, out to a radius of $\sim$ 15 kpc at the 5$\sigma$ level, but structure within the disk is unresolved 
due to the larger 
NVSS beam.

The $\alpha^{1400}_{610}$ spectral index and spectral index {uncertainty} maps for this source are shown in the middle and right panels of Figure \ref{fig:NGC0245}, respectively. As FIRST does not detect the full galactic disk, we can only provide a spectral index map for the galactic core and the two radio hotspots in the star forming regions of the disk. The spectral index is flattest at the radio peaks, gradually steepening further, however the spectral index uncertainties indicate consistency with a flat spectrum for the full region covered by the maps. The galactic core of NGC0245 has an integrated spectral index of $\alpha_{\rm core} = -0.7 \pm 0.3$, based on the FIRST core {flux density} of $S_{\rm FIRST,core} = 5.15 \pm 0.3$ mJy. As we cannot disentangle the core emission from that of the disk in the NVSS data, we determine an integrated spectral index of $\alpha_{\rm NGC0245} = -0.59 \pm 0.42$ for the entire radio galaxy.

\subsection{PKS-0204}
\label{subsubsec:pks0204}

The dominating source in the field-of-view of the ACT-CL J2327$-$0204 observation is a bright, extended source identified as the active radio galaxy PKS 2324-02 at $z = 0.188$\footnote{Using the NASA/IPAC Extragalactic Database: \url{https://ned.ipac.caltech.edu/}}. Figure \ref{fig:PKS2324-02} shows our 610 MHz radio map of PKS 2324-02 (top). The galaxy core is situated at a J2000 position of R.A = 23h26m53.843s, Dec. = $-$02d02m13.09s and has a {flux density} of $S_{\rm 610,core}$ = 180.0 $\pm$ 0.1 mJy in our 610 MHz GMRT, primary beam corrected map. Diffuse structures surrounding the galaxy core are oriented roughly NE to SW and have a largest angular extent of 2.4\arcmin, corresponding to 454.5 kpc at the redshift of this source. The source's physical extent places it below the size threshold of the rare giant radio galaxies \citep[$> 700$ kpc, see e.g.][]{Molina.2014}. The 610 MHz {flux density} of the entire object is $S_{610} = 4.439 \pm 0.003$ Jy. There is bright compact emission 
within the NE lobe with a bridge of compact emission joining it to the galaxy core. This bridge-like feature is also detected in the 1.4 GHz FIRST radio image of this galaxy, shown in the bottom left panel of Figure \ref{fig:PKS2324-02}. The FIRST data shows similar structures within the diffuse emission as found in our 610 MHz map. The FIRST core {flux density} is $S_{\rm core, FIRST} = 144.3 \pm 0.3$ mJy, giving a flat spectral index of $\alpha_{610}^{1400} = -0.27$.

The middle and right panels of Figure \ref{fig:PKS2324-02} show the spectral index map and spectral index {uncertainty} map for PKS2324-02, respectively. As indicated by our integrated spectral index above, the core of the radio galaxy has a flat spectrum. The NE jet emanating from the core has a spectral index of around -1, with regions of flatter spectra corresponding to the radio hotspots within the jet. Further away from the jet, the spectrum steepens in line with expectations from synchrotron ageing. Most of the SW lobe has a steep synchrotron spectrum, with spectral indices less than -1.5. Due to the brightness of the source and the quality of the data, the uncertainties on the spectral index are mostly below $\sim$ 0.25.

\subsection{Galaxy pair Arp 118}
\label{subsec:galpair}

\begin{figure}
 \centering
 \includegraphics[width=0.45\textwidth, clip=True, trim=40 70 30 35]{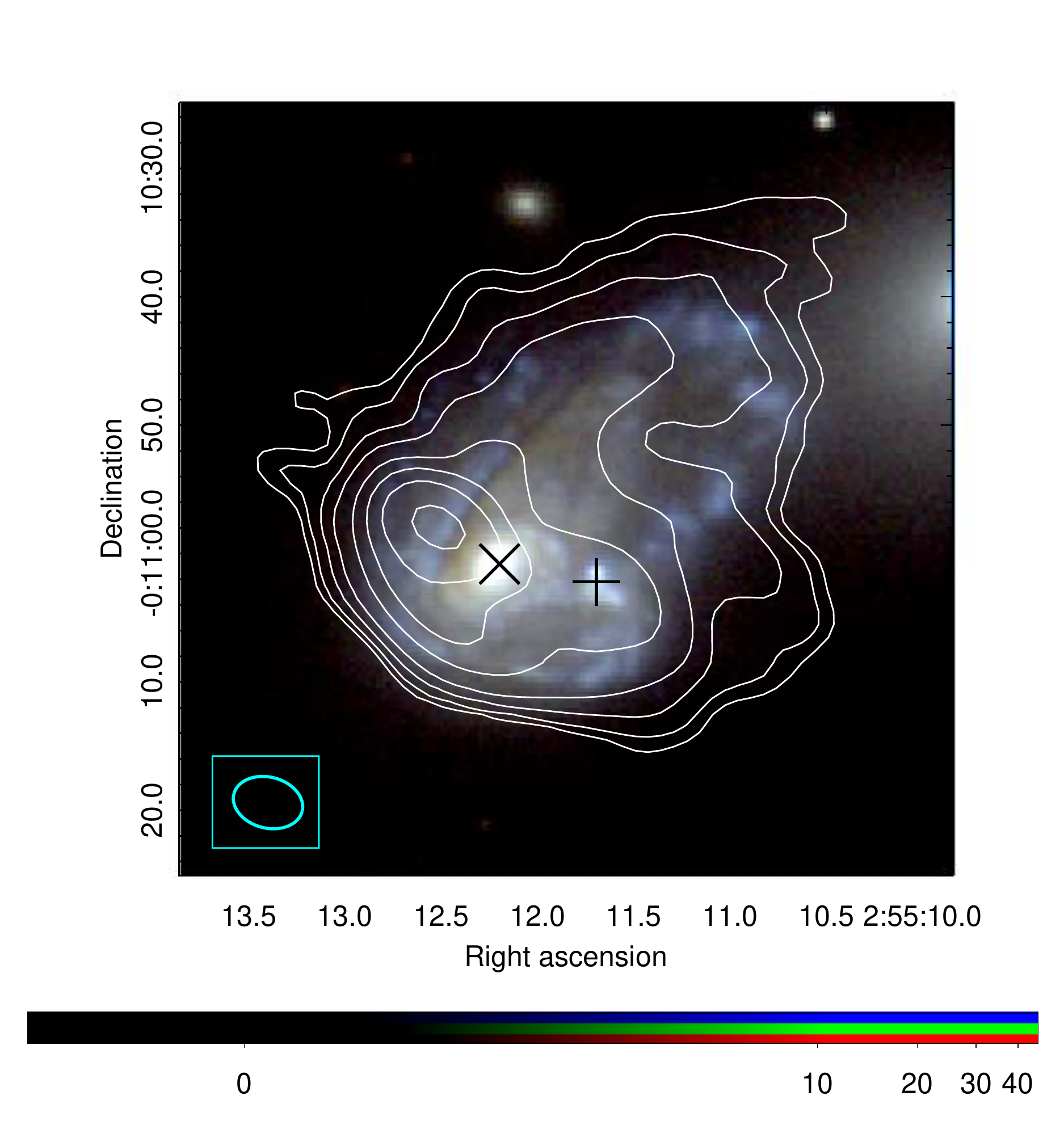}
 \caption{SDSS 3-colour $gri$-image of the galaxy pair Arp 118. Full resolution 610 MHz GMRT contours are overlaid with levels of $[3,5,10,20,50,100,200]\sigma$, where $1\sigma = 120\; \mu$Jy beam\per. The beam is shown by the boxed ellipse. The two galaxies are indicated by the X and $+$.}
 \label{fig:ARP118}
\end{figure}

The merging galaxy pair Arp 118 lies within the primary beam of our ACT-CL J0256.5+0006 pointing. Diffuse emission extending beyond the optical Arp 118 emission is detected in the full resolution 610 MHz image. We measure an integrated 610 MHz {flux density} for Arp 118 of 279.5 $\pm$ 14.0 mJy. A large region of this emission is also detected in FIRST, with a total integrated {flux density} at 1.435 GHz of 141.0 $\pm$ 7.1 mJy for the galaxy pair. The global spectral index for the source is therefore -0.8 $\pm$ 0.2. Figure \ref{fig:ARP118} shows the composite $gri$-image of Arp 118 with our 610 MHz full resolution contours overlaid in white. The radio peak is offset from both galaxies involved in the merger, namely the main and in-falling galaxies denoted by the X and $+$, respectively. The offset\footnote{Given a maximum FIRST astrometric uncertainty of 0.5{\arcsec} \citep{Helfand.2015} and an astrometric match between the GMRT and FIRST data.}, 5.5 $\pm$ 1.0 arcseconds and 13.0 $\pm$ 1.
0 arcseconds from the main and in-falling 
galaxies, respectively, does not appear to correlate with any star forming region in the galaxy merger. 


\subsection{Other sources of interest}
\label{subsubsec:soi}

In each of our cluster pointings there are several extended sources within the field of view, many of which are FR-I and FR-II sources. Primary beam-corrected images of the full field of view for each cluster pointing are given in Appendix \ref{app:fovimgs}. Figure \ref{fig:soi} shows a gallery of some of these sources of interest from the J0059$-$0049 field. The top left panel is a potential galaxy-galaxy lensing source. Including the three sources discussed above, there are a total of 218 extended sources in our cluster sample which we identify as interesting for potential follow-up. These sources are given the code type `M' or `C' in the source catalog discussed in the next section.

\begin{figure*}
 \centering
 \includegraphics[width=0.32\textwidth, clip=True, trim=20 0 20 0]{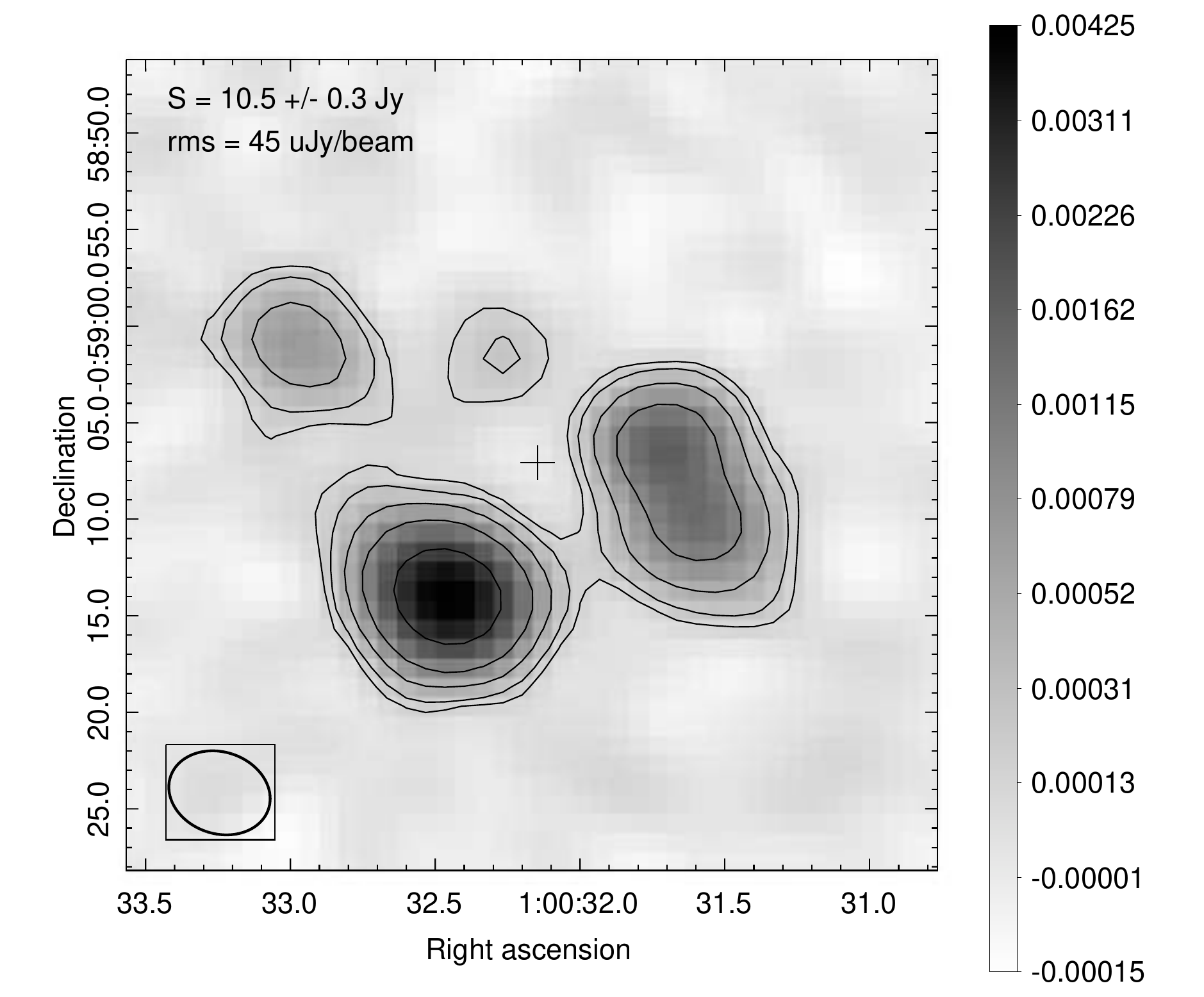}
 \includegraphics[width=0.32\textwidth, clip=True, trim=20 0 20 0]{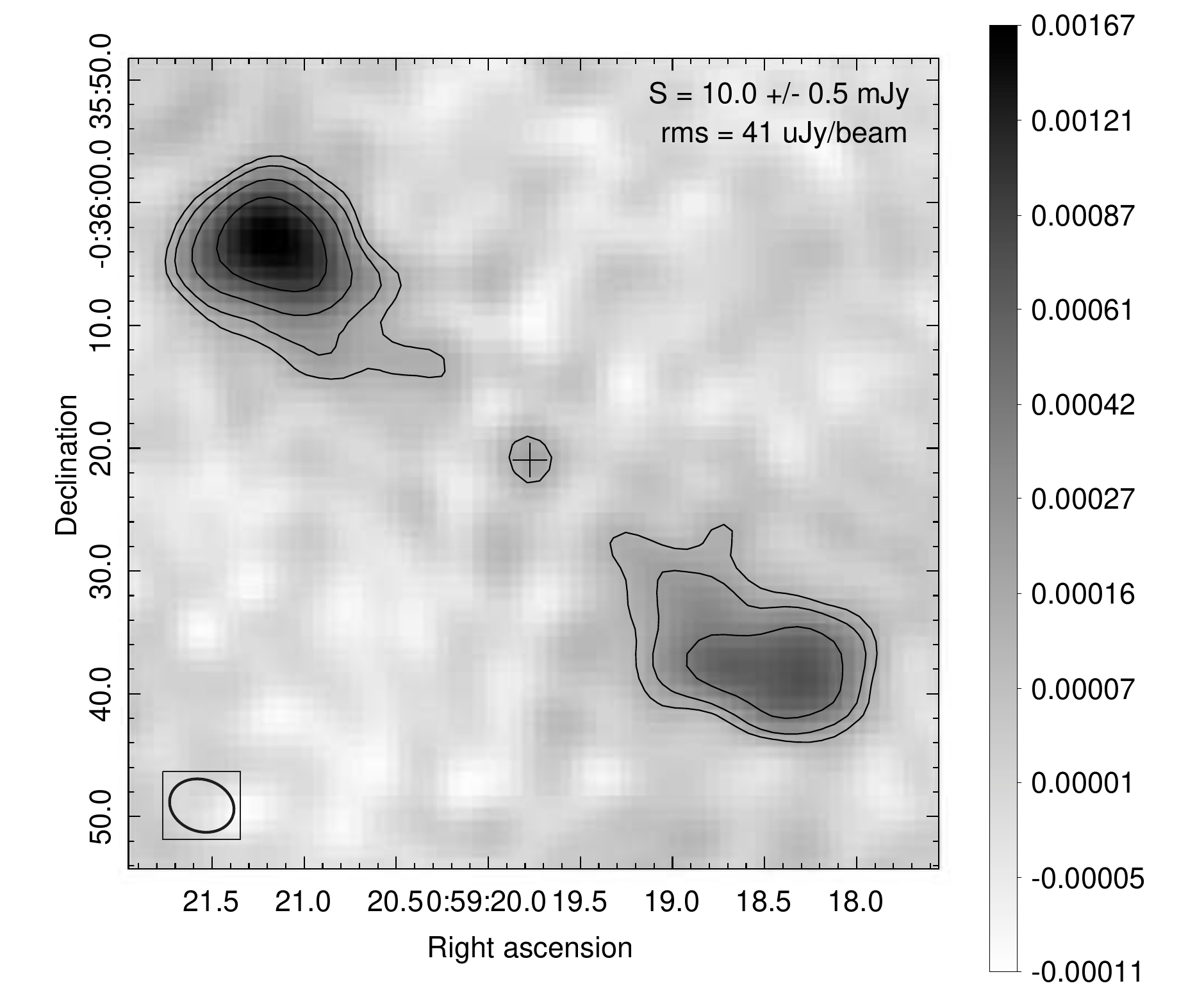}
 \includegraphics[width=0.32\textwidth, clip=True, trim=20 0 20 0]{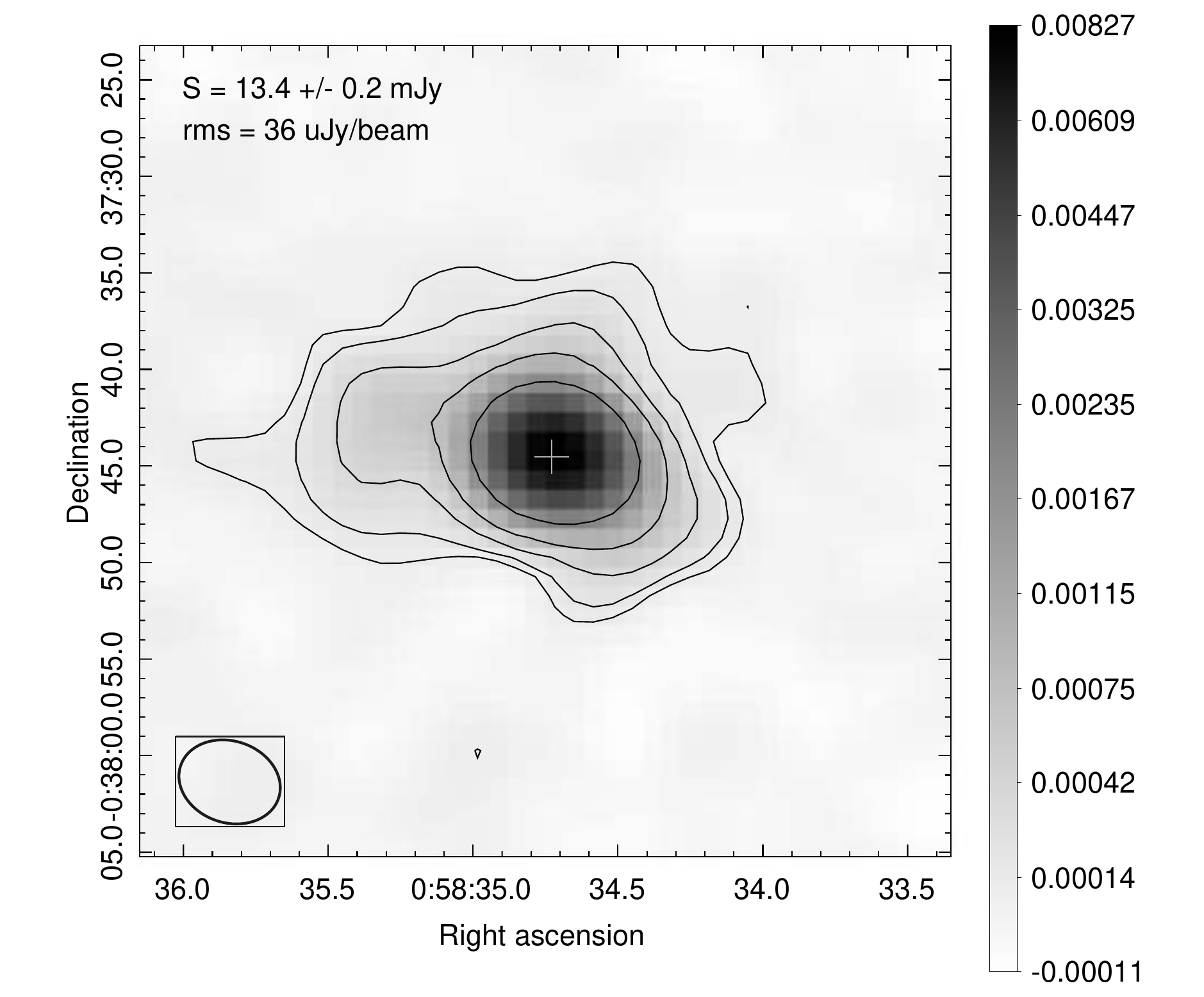}\\
 \includegraphics[width=0.32\textwidth, clip=True, trim=20 -20 20 -20]{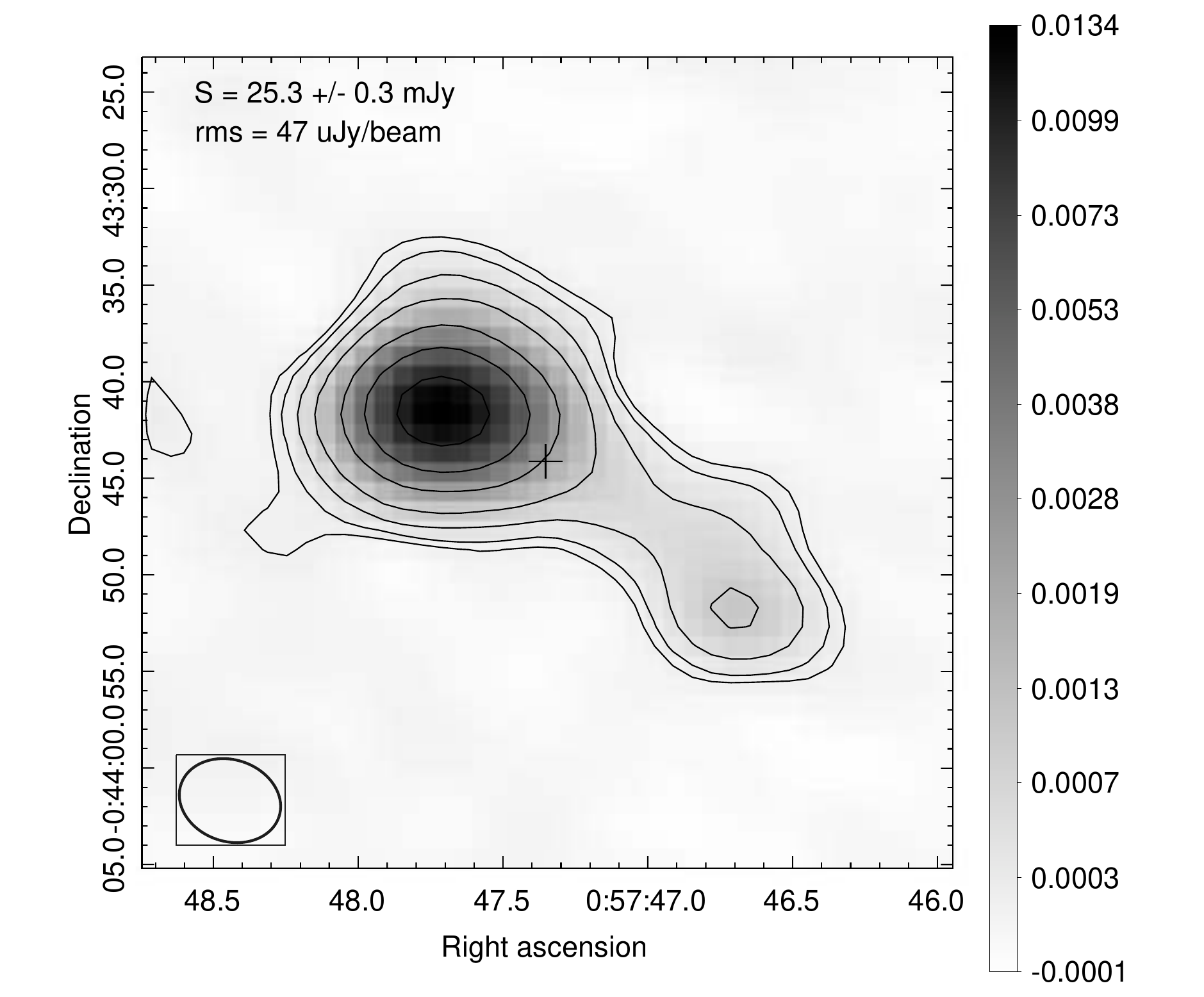}
 \includegraphics[width=0.32\textwidth, clip=True, trim=20 -20 20 -20]{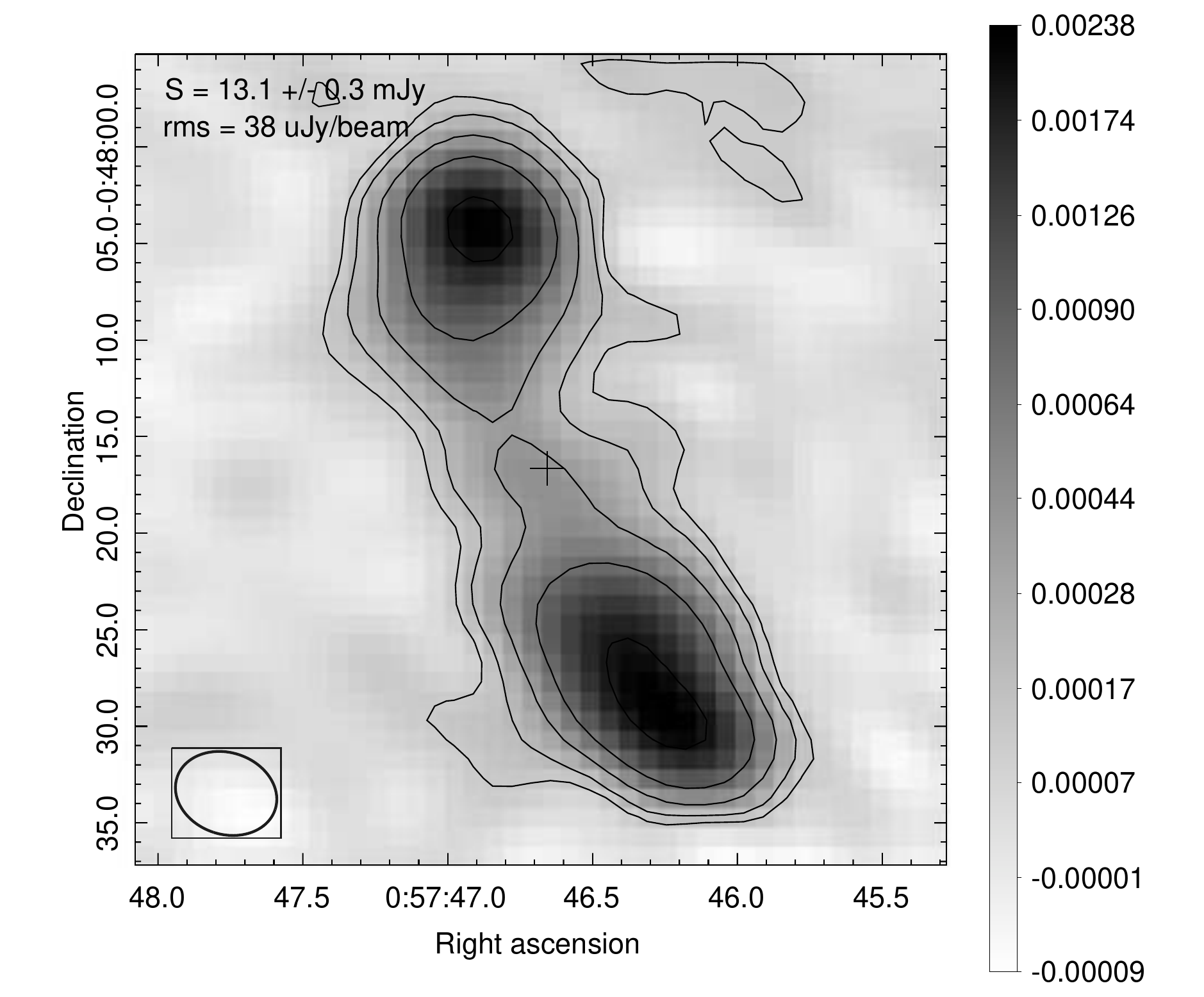}
 \includegraphics[width=0.32\textwidth, clip=True, trim=20 -20 20 -20]{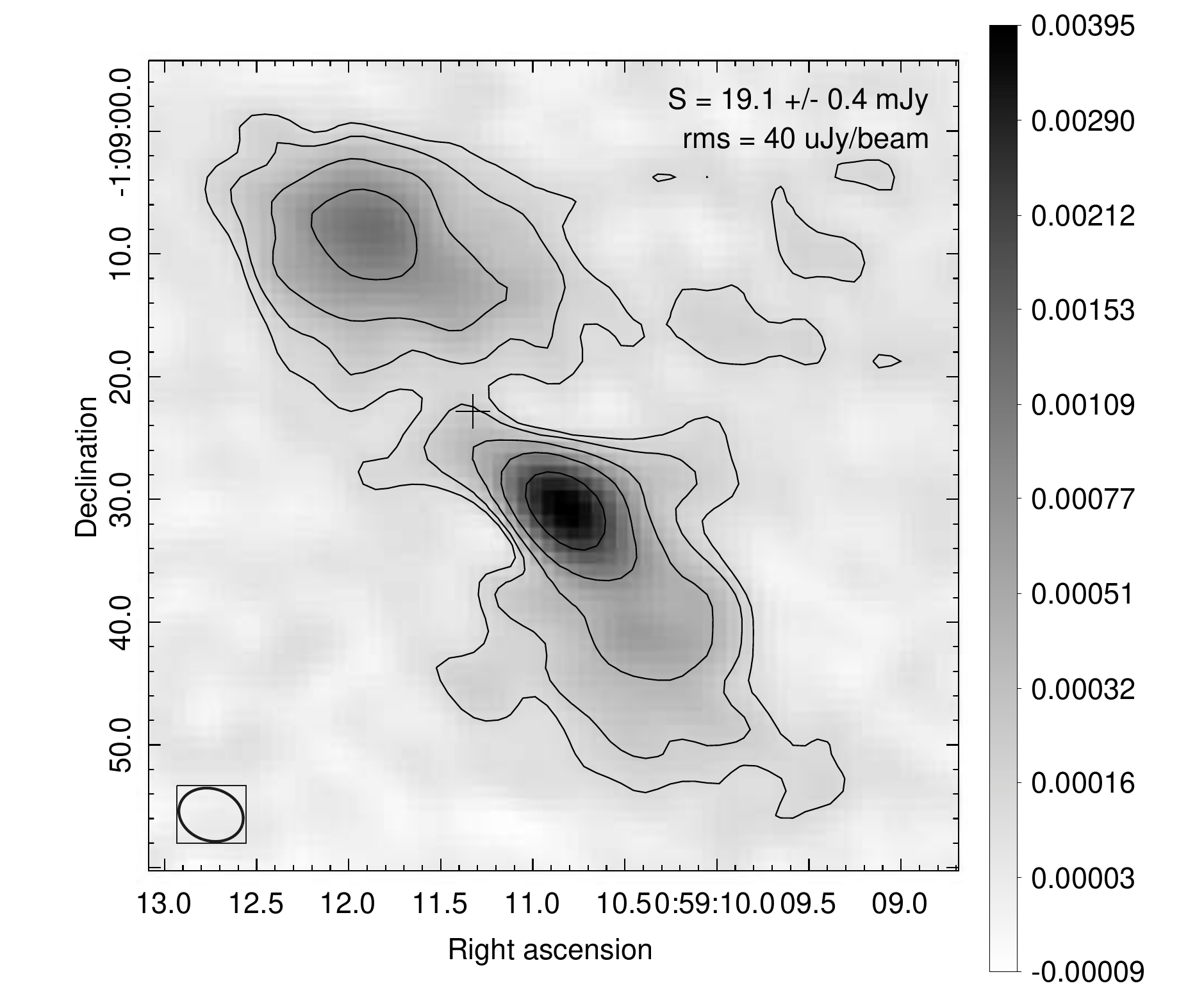}\\
 \includegraphics[width=0.32\textwidth, clip=True, trim=20 0 20 0]{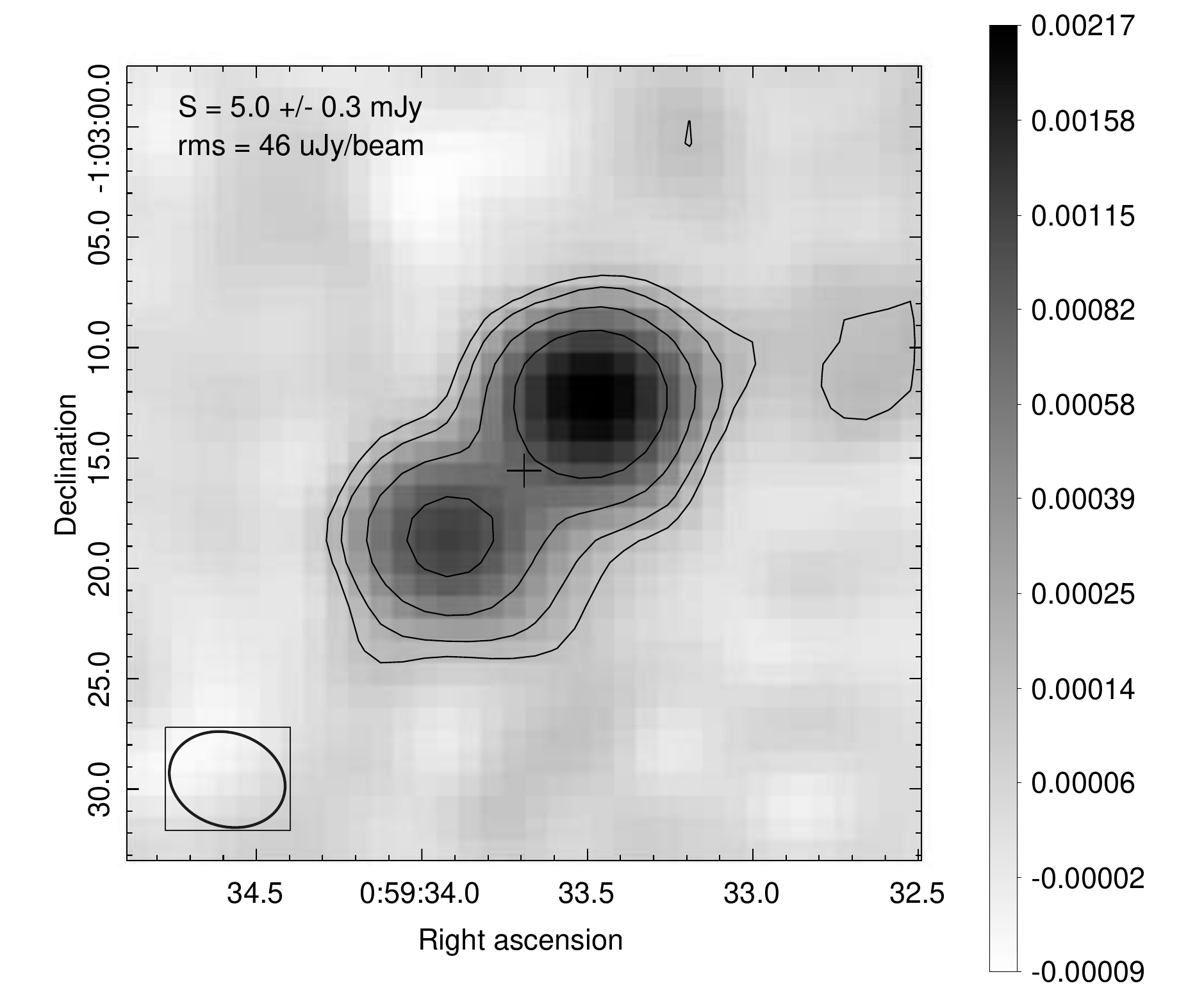}
 \includegraphics[width=0.32\textwidth, clip=True, trim=20 0 20 0]{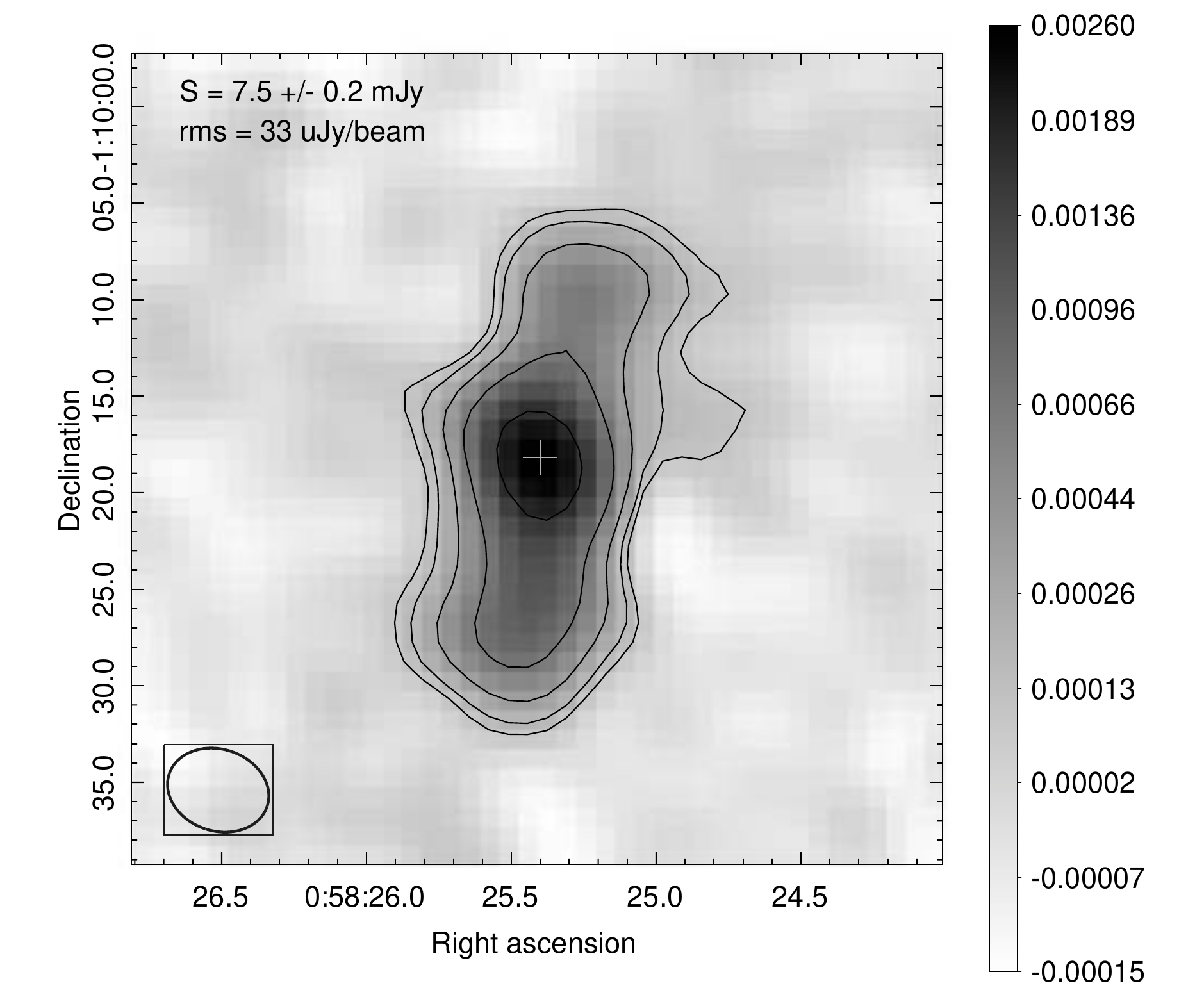}
 \includegraphics[width=0.32\textwidth, clip=True, trim=20 0 20 0]{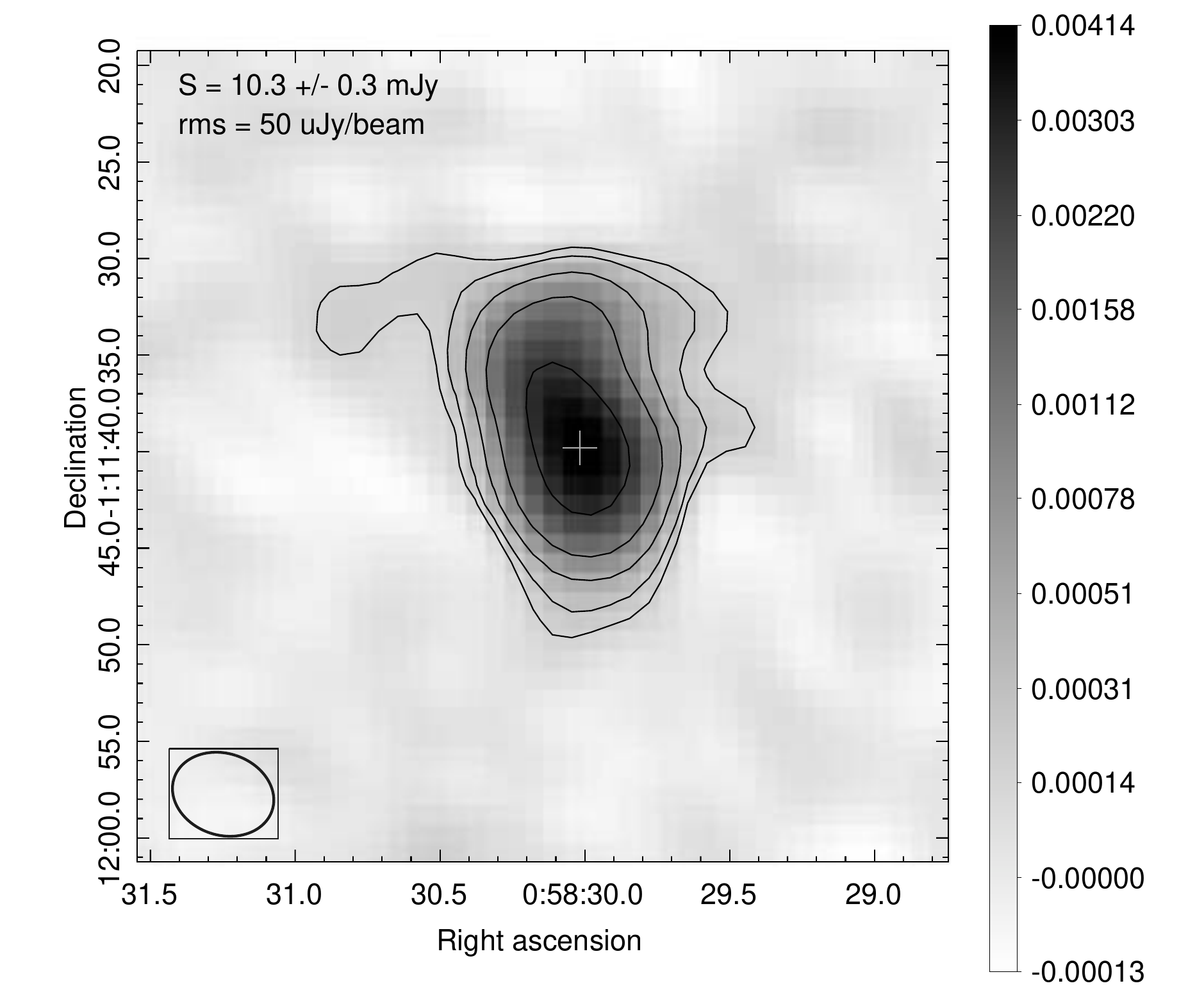}
 \caption{Example gallery of some of the extended and/or double lobed sources in the field of view of the J0059 cluster pointing. Contours are [3,5,10,20,50]$\sigma$, where $\sigma$ is the local rms noise at the position of the source. The synthesised beam is shown by the boxed ellipse in the bottom left corner of each panel.}
 \label{fig:soi}
\end{figure*}


\subsection{Source catalogs}
\label{subsec:srccats}

For each of our cluster pointings, we determine source catalogs using \textsc{pybdsf} \citep{Mohan.2015} with an island and source cut of 3$\sigma$ and 5$\sigma$, respectively. As some of our images are affected by bright source sidelobes, we visually check the \textsc{pybdsf} results, and flag those sources which lie within the sidelobes. In the worst cases, up to 25\% of the initial \textsc{pybdsf} source detections are removed. Over the 14 cluster pointings, we have a total of 3960 sources, with a minimum detection of 79 and a maximum of 563 sources per pointing. A full catalog of these sources, with 610 MHz positions, {flux densities}, and shape parameters is available online. Table \ref{tab:srcs} shows an example excerpt from this catalog. As there is a significant range in image sensitivity across our sample, we do not determine global source counts due to the numerous selection effects which would bias the numbers.

\begin{table*}
 \centering
 \caption{Excerpt of the full catalog of sources detected at 610 MHz. The full table is available online. (1) ACT-GMRT identifier. (2) J2000 Right Ascension. (3) J2000 Declination. (4) Total 610 MHz {flux density}. (5) {Uncertainty} on the total 610 MHz {flux density}. (6) Major axis. (7) Minor axis. (8) Parallactic angle. (9) Source type code: M and C indicate sources of interest. }
 \label{tab:srcs}
 \begin{tabular}{lcc.....c}
  \toprule
Source ID & RA$_{J2000}$ & Dec$_{J2000}$ & \multicolumn{1}{c}{$S_{610}$} & \multicolumn{1}{c}{$\Delta S_{610}$} & \multicolumn{1}{c}{$a$} & \multicolumn{1}{c}{$b$} & \multicolumn{1}{c}{\rm p.a} & Notes\\
 & (deg) & (deg) & \multicolumn{1}{c}{\rm (mJy)} & \multicolumn{1}{c}{\rm (mJy)} & \multicolumn{1}{c}{\rm (arcsec)} & \multicolumn{1}{c}{\rm (arcsec)} & \multicolumn{1}{c}{\rm (deg)} & \\
\midrule
J010058.1-005547.8 &  15.242272 &  -0.929958 &  27.0 &  0.2 &   6.1 &   4.5 &  79.7 & M \\
J010058.1-004348.7 &  15.242223 &  -0.730200 &   0.6 &  0.1 &   7.7 &   5.9 & 107.6 & S \\
J010056.0-004426.3 &  15.233488 &  -0.740643 &   0.5 &  0.1 &   6.8 &   5.1 &   6.7 & S \\
J010047.9-004239.8 &  15.199752 &  -0.711063 &   0.6 &  0.1 &   6.1 &   4.3 &  73.5 & S \\
J010047.7-003752.5 &  15.198552 &  -0.631260 &   1.1 &  0.1 &   6.8 &   4.4 &  67.9 & S \\
\bottomrule
  
 \end{tabular}

\end{table*}

\clearpage
\onecolumn

\section{Full field of view cluster images}
\label{app:fovimgs}

Here we present the primary beam-corrected images of the full field of view for each cluster pointing. In each image, the $R_{500}$ cluster region, centred of the SZ peak, is indicated by the dashed circle - zoomed in images of these regions are provided in the main body of this paper, in Sections \ref{sec:clusterDE} and \ref{sec:nondets}. The positions of sources of interest, discussed in Section \ref{sec:fovsrcs}, are indicated by the boxes. 

\noindent%
\begin{minipage}{\linewidth}
\makebox[\linewidth]{
  \includegraphics[width=\textwidth]{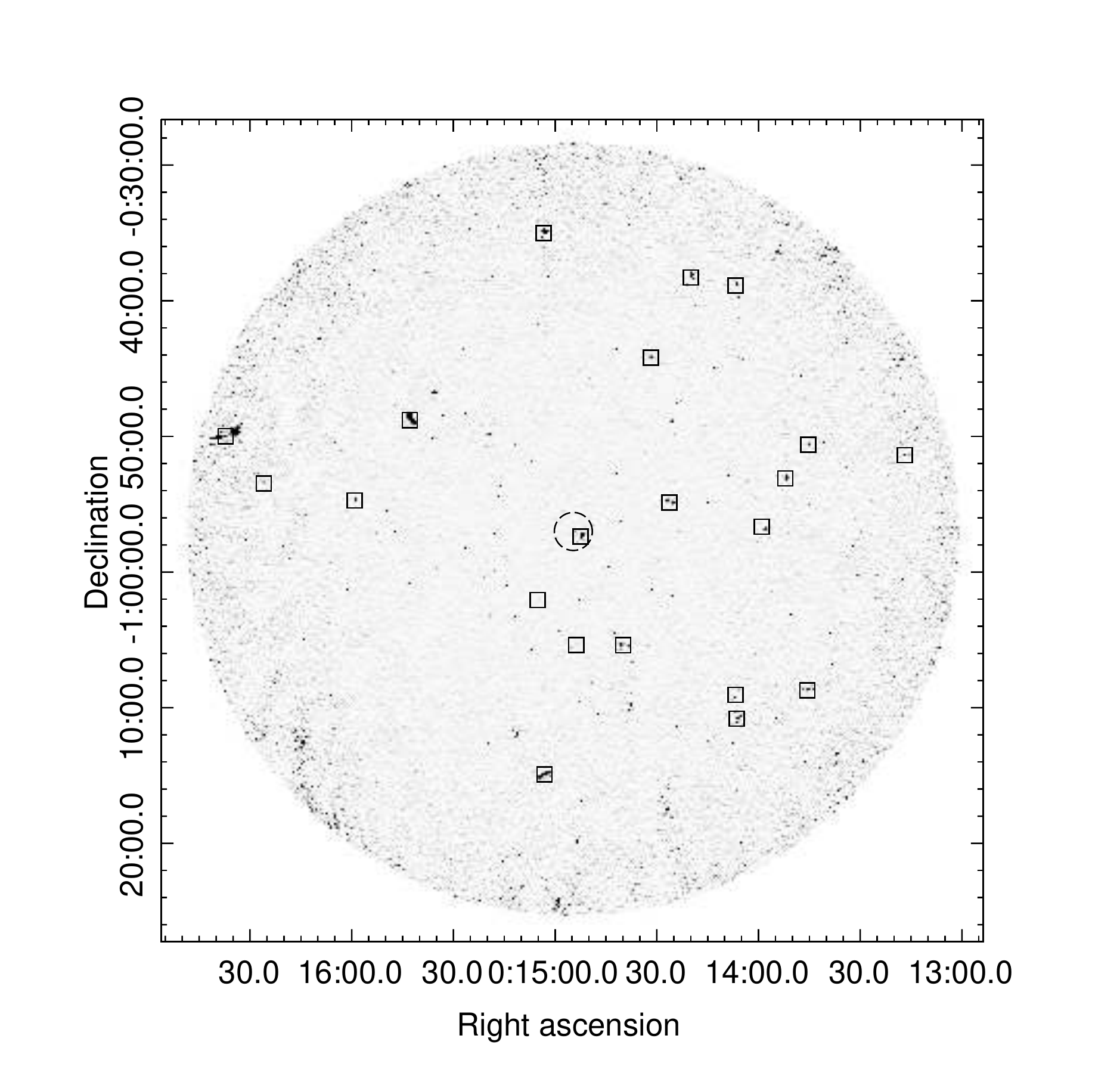}}
\captionof{figure}{ACT-CL J0014.9$-$0057. The dashed circle denotes the $R_{500}$ region centred on the cluster SZ peak. Boxes indicate the postion of sources of interest, discussed in Section \ref{sec:fovsrcs}.}\label{fig:j0014_FOV}
\end{minipage}

\begin{figure} 
 \centering
 \includegraphics[width=\textwidth,angle=90,origin=c]{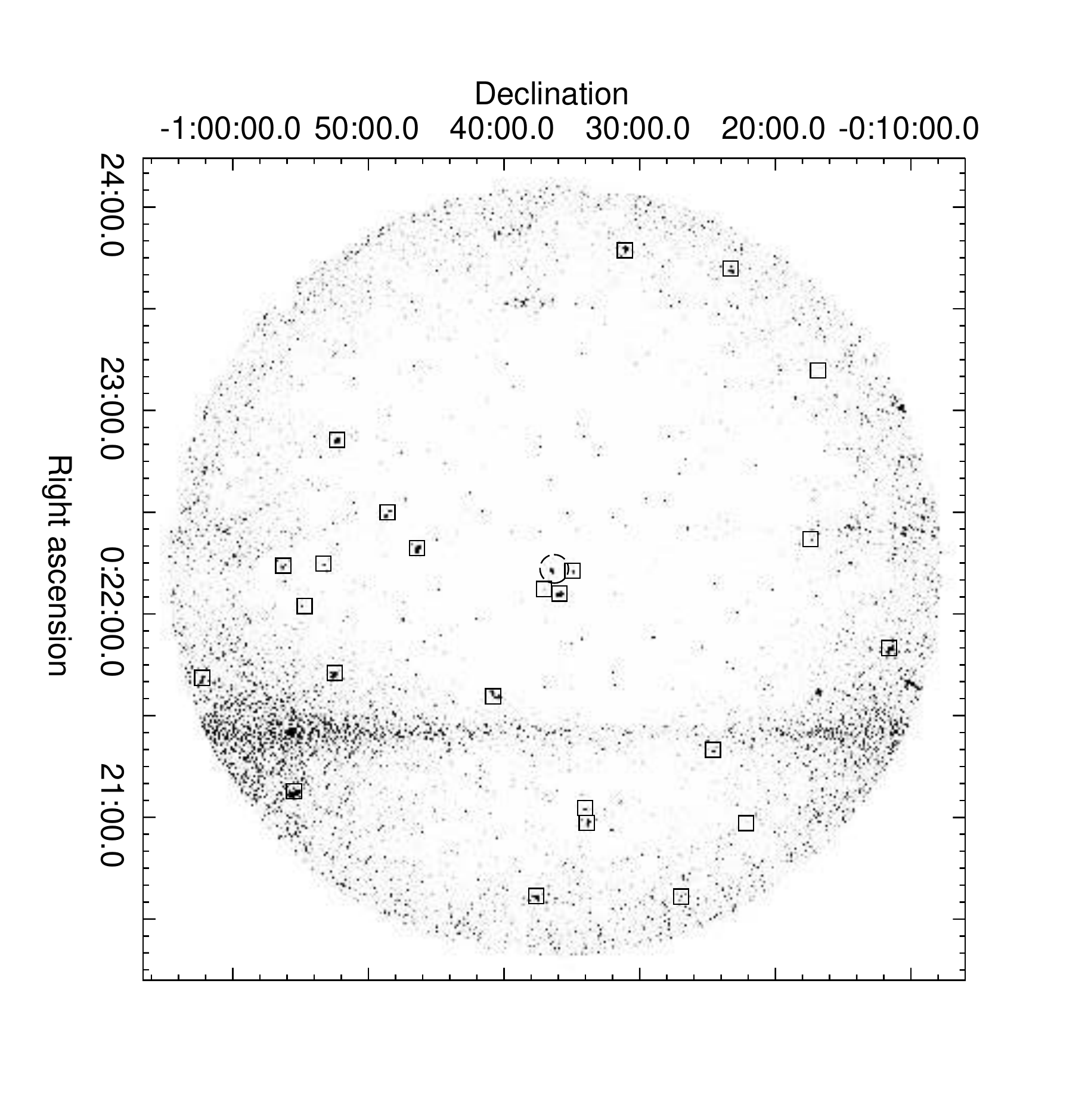}
 \caption{ACT-CL J0022.2$-$0036. The dashed circle denotes the $R_500$ region centred on the cluster SZ peak. Boxes indicate the postion of sources of interest, discussed in Section \ref{sec:fovsrcs}.}
 \label{fig:j0022_FOV}
\end{figure}

\begin{figure}
 \centering
 \includegraphics[width=\textwidth,angle=90,origin=c]{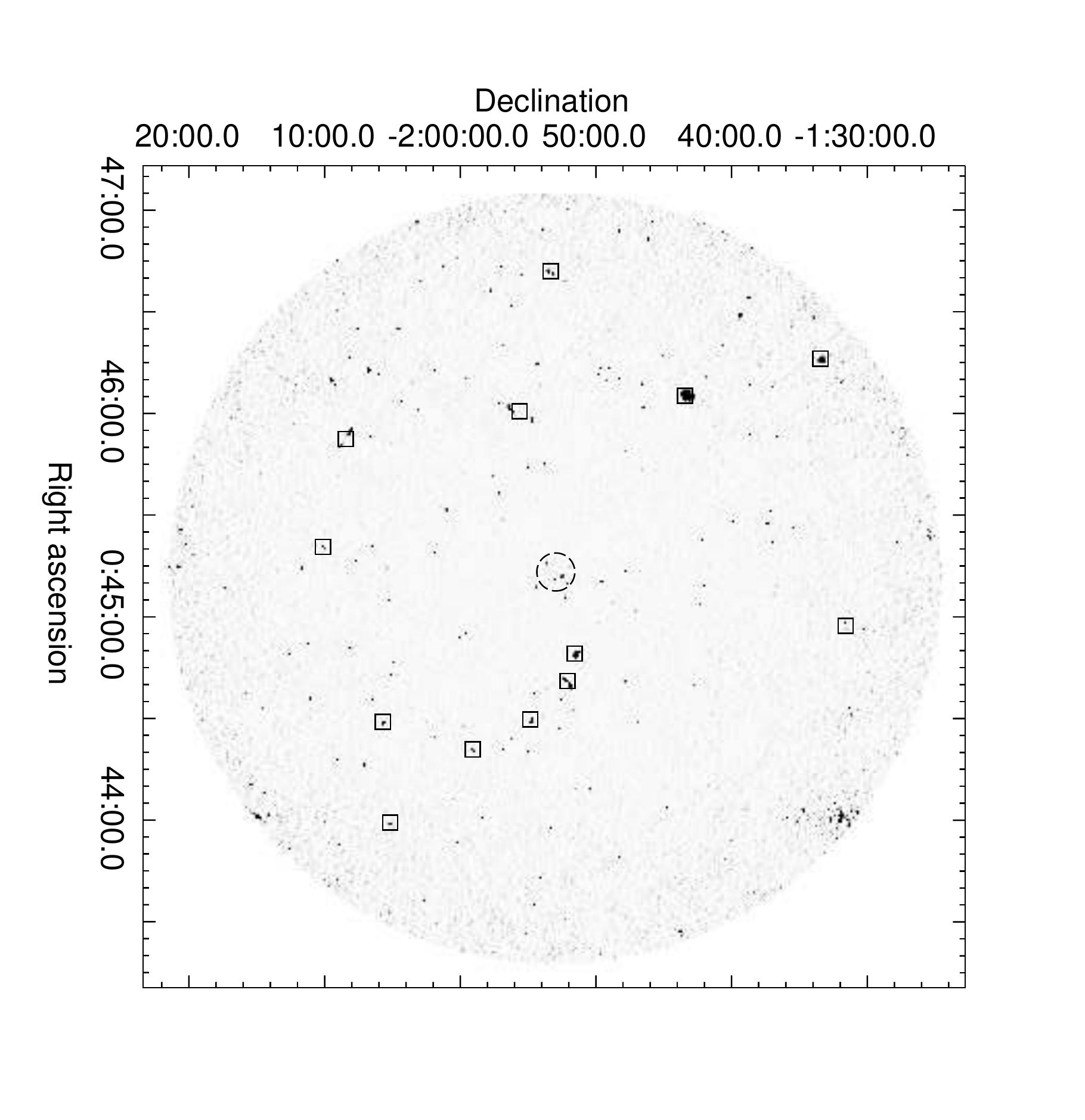}
 \caption{ACT-CL J0045.2$-$0152.  The dashed circle denotes the $R_{500}$ region centred on the cluster SZ peak. Boxes indicate the postion of sources of interest, discussed in Section \ref{sec:fovsrcs}.}
 \label{fig:j0045_FOV}
\end{figure}

\begin{figure}
 \centering
 \includegraphics[width=\textwidth,angle=90,origin=c]{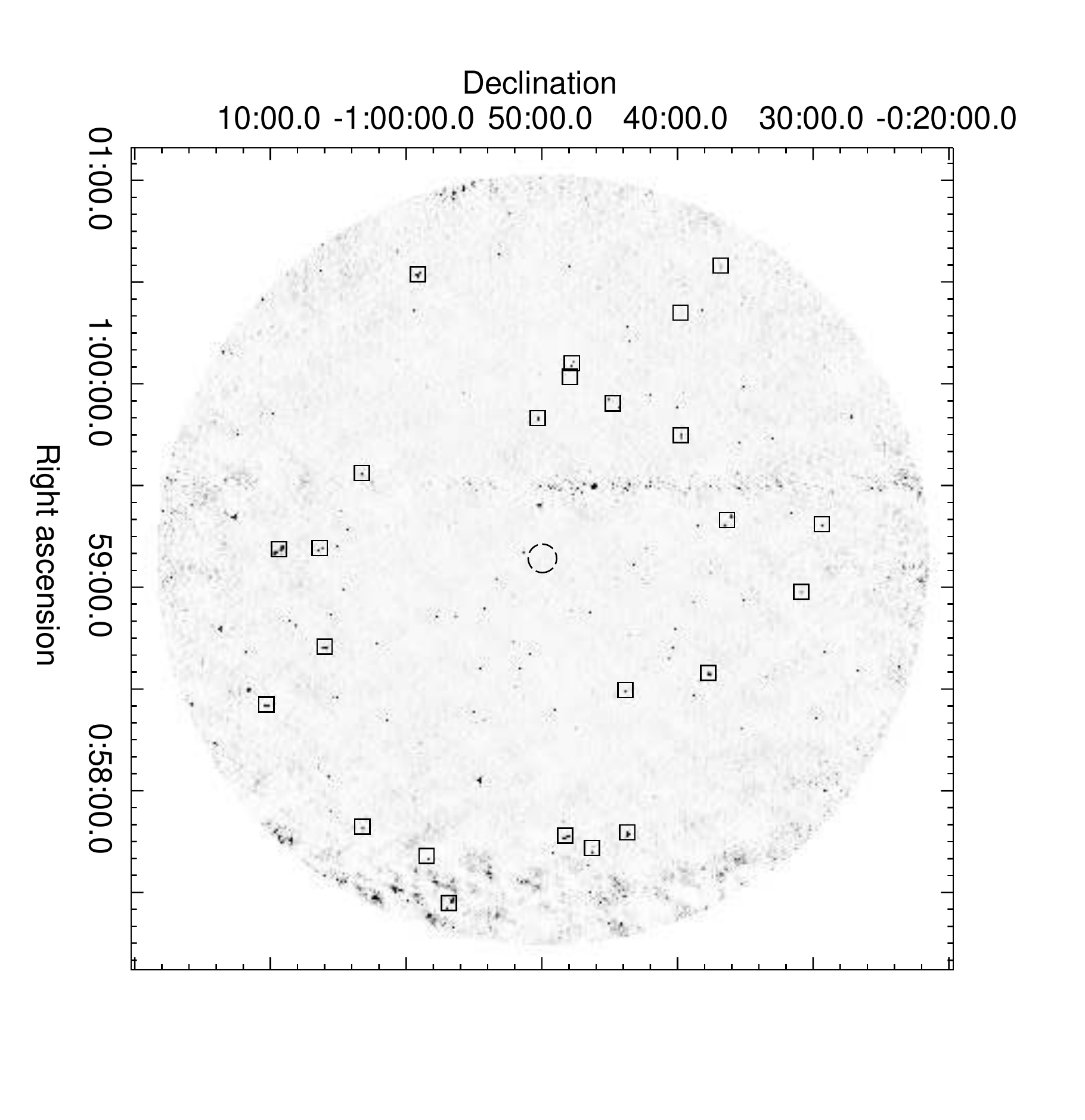}
 \caption{ACT-CL J0059.1$-$0049. The dashed circle denotes the $R_{500}$ region centred on the cluster SZ peak. Boxes indicate the postion of sources of interest, discussed in Section \ref{sec:fovsrcs}.}
 \label{fig:j0059_FOV}
\end{figure}

\begin{figure}
 \centering
 \includegraphics[width=\textwidth]{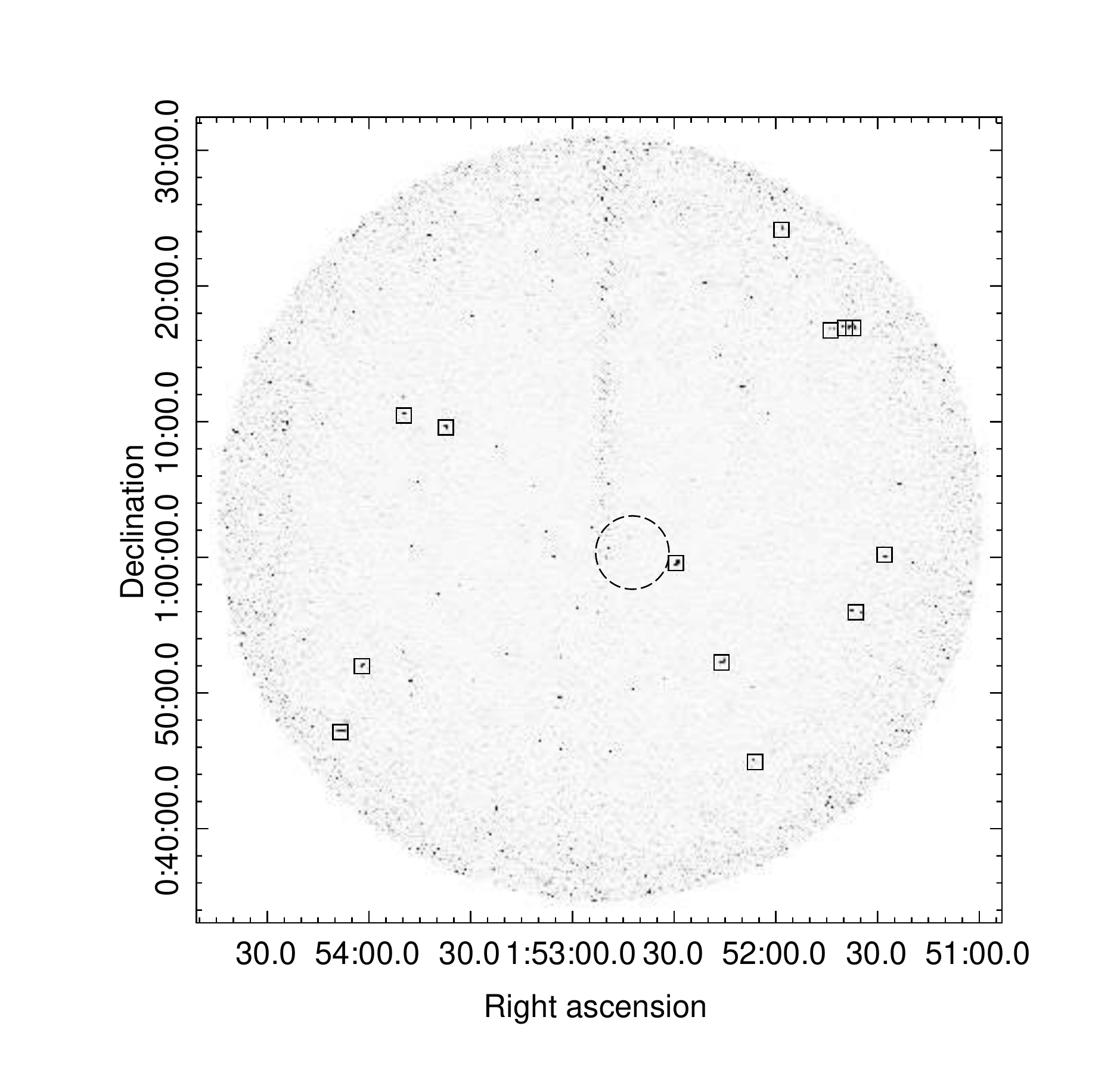}
 \caption{\textit{Left:} ACT-CL J0152.1+0100.  The dashed circle denotes the $R_{500}$ region centred on the cluster SZ peak. Boxes indicate the postion of sources of interest, discussed in Section \ref{sec:fovsrcs}.}
 \label{fig:j0152_FOV}
\end{figure}

\begin{figure}
 \centering
 \includegraphics[width=\textwidth]{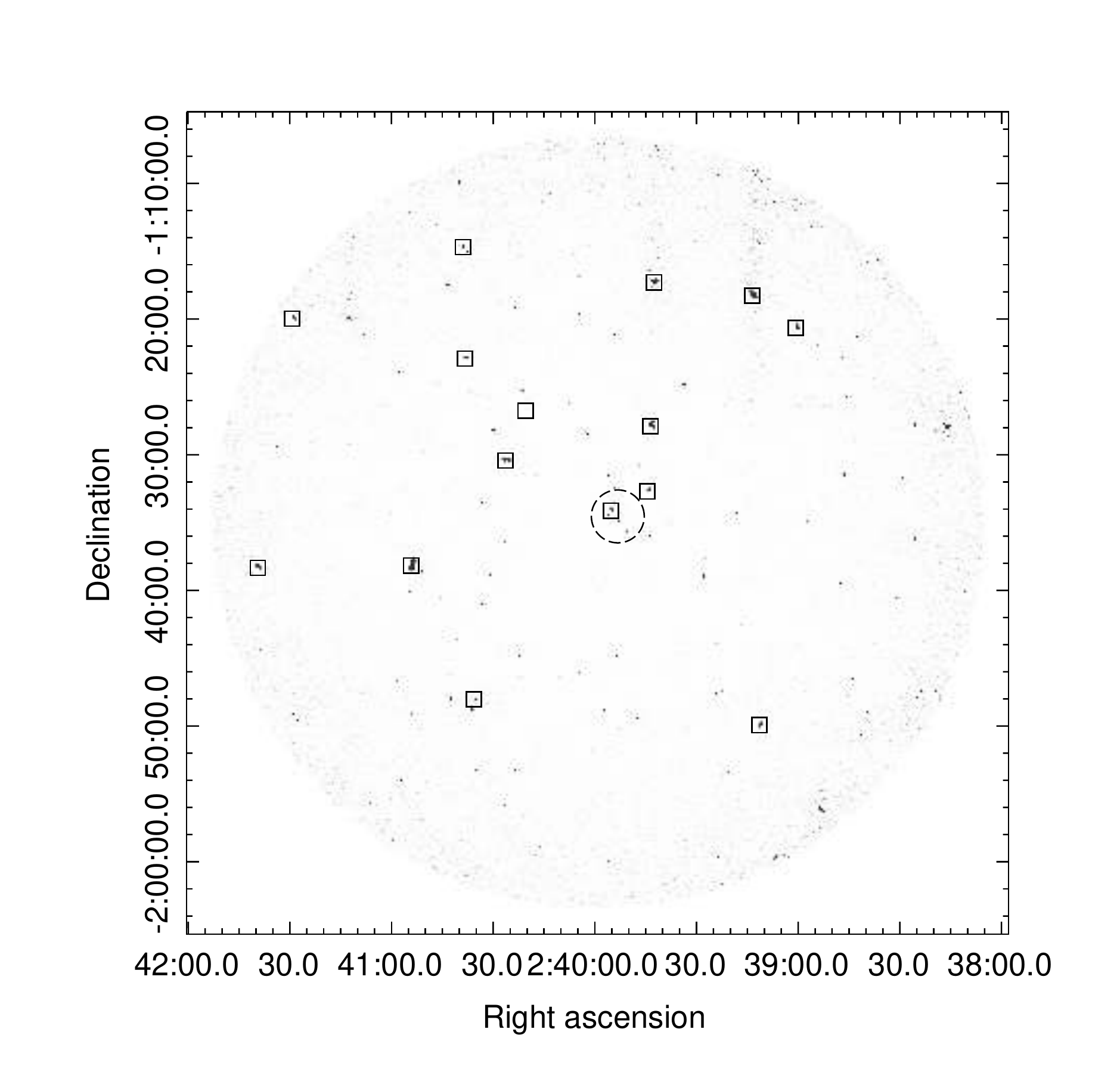}
 \caption{ACT-CL J0239.8$-$0134.  The dashed circle denotes the $R_{500}$ region centred on the cluster SZ peak. Boxes indicate the postion of sources of interest, discussed in Section \ref{sec:fovsrcs}.}
 \label{fig:j0239_FOV}
\end{figure}

\begin{figure}
 \centering
 \includegraphics[width=\textwidth]{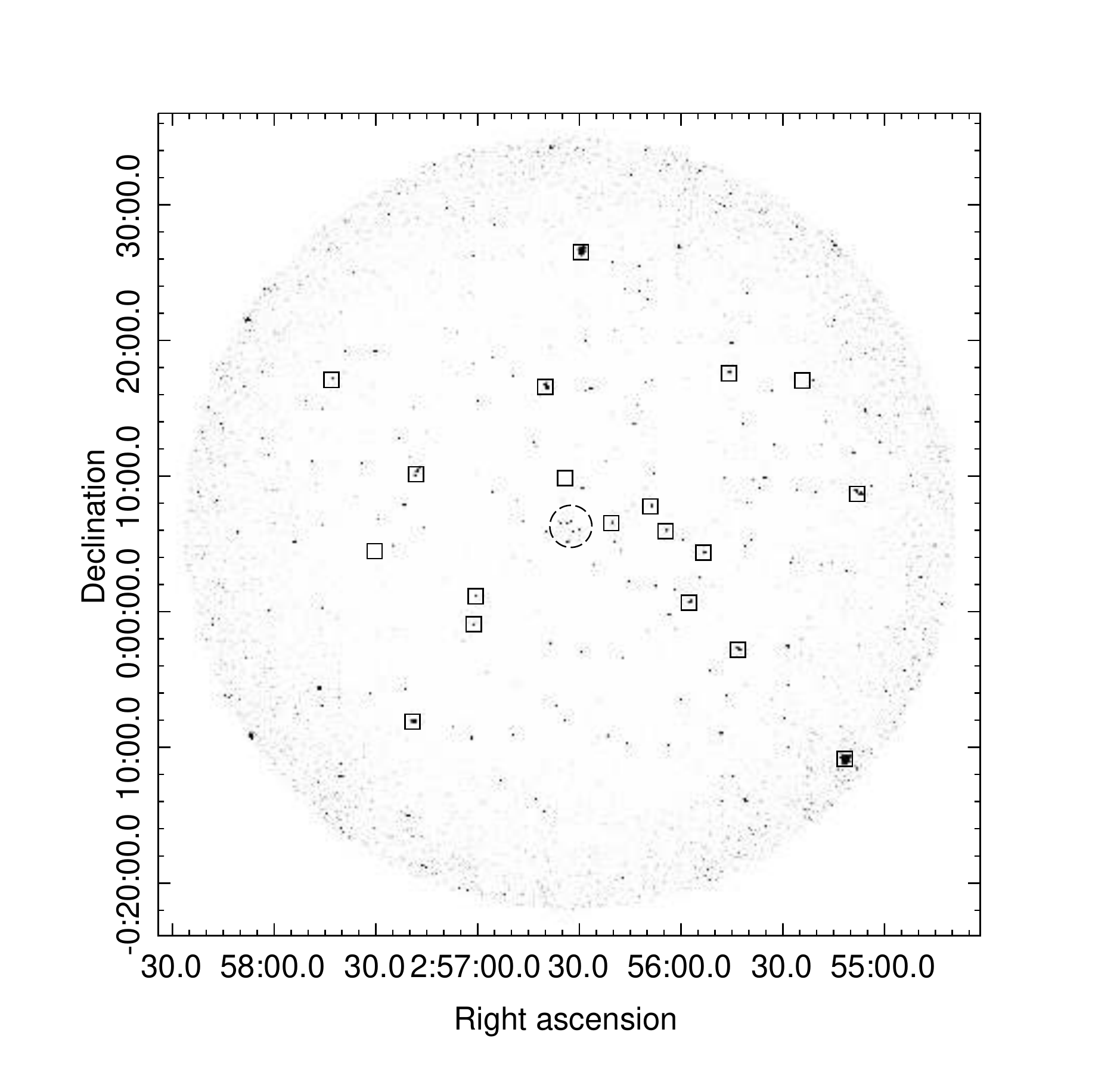}
 \caption{ACT-CL J0256.5+0006. The dashed circle denotes the $R_{500}$ region centred on the cluster SZ peak. Boxes indicate the postion of sources of interest, discussed in Section \ref{sec:fovsrcs}.}
 \label{fig:j0256_FOV}
\end{figure}

\begin{figure}
 \centering
 \includegraphics[width=\textwidth,angle=90,origin=c]{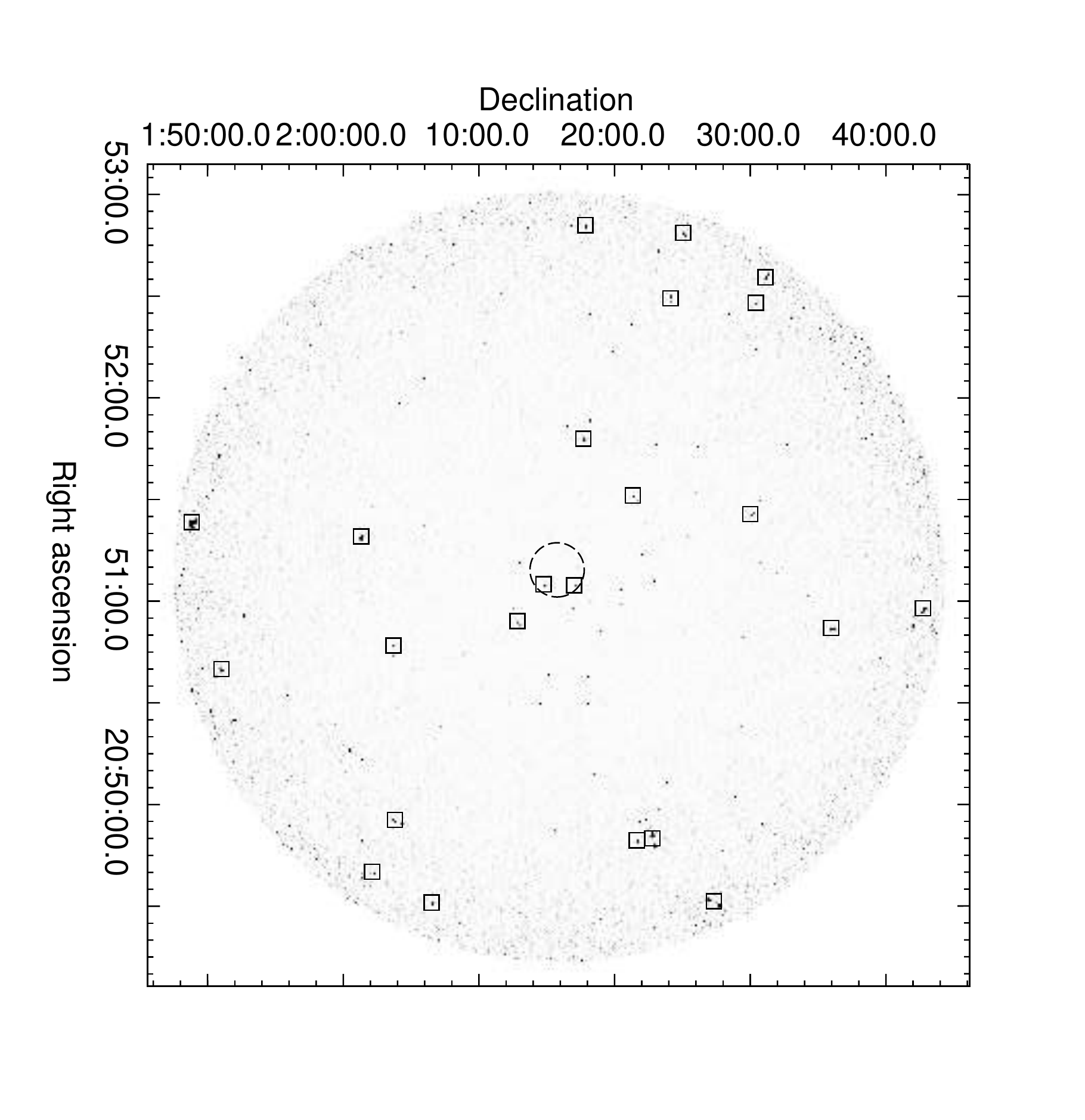}
 \caption{ACT-CL J2051.1+0215. The dashed circle denotes the $R_{500}$ region centred on the cluster SZ peak. Boxes indicate the postion of sources of interest, discussed in Section \ref{sec:fovsrcs}.}
 \label{fig:j2051_FOV}
\end{figure}

\begin{figure}
 \centering
 \includegraphics[width=\textwidth,clip=True,trim=10 100 10 100,angle=90,origin=c]{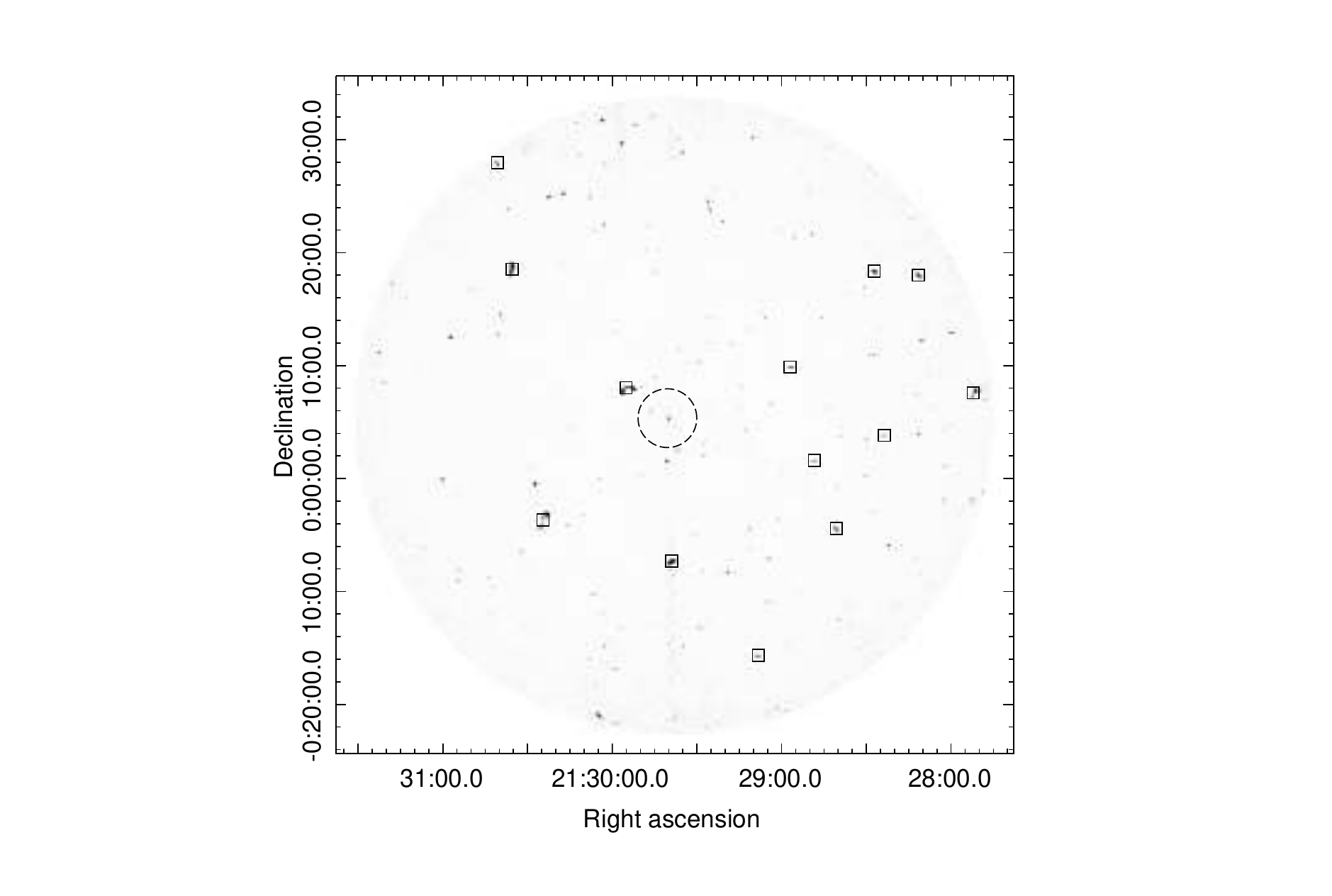}
 \caption{ACT-CL J2129.6+0005. The dashed circle denotes the $R_{500}$ region centred on the cluster SZ peak. Boxes indicate the postion of sources of interest, discussed in Section \ref{sec:fovsrcs}.}
 \label{fig:j2129_FOV}
\end{figure}

\begin{figure}
 \centering
 \includegraphics[width=\textwidth,angle=90,origin=c]{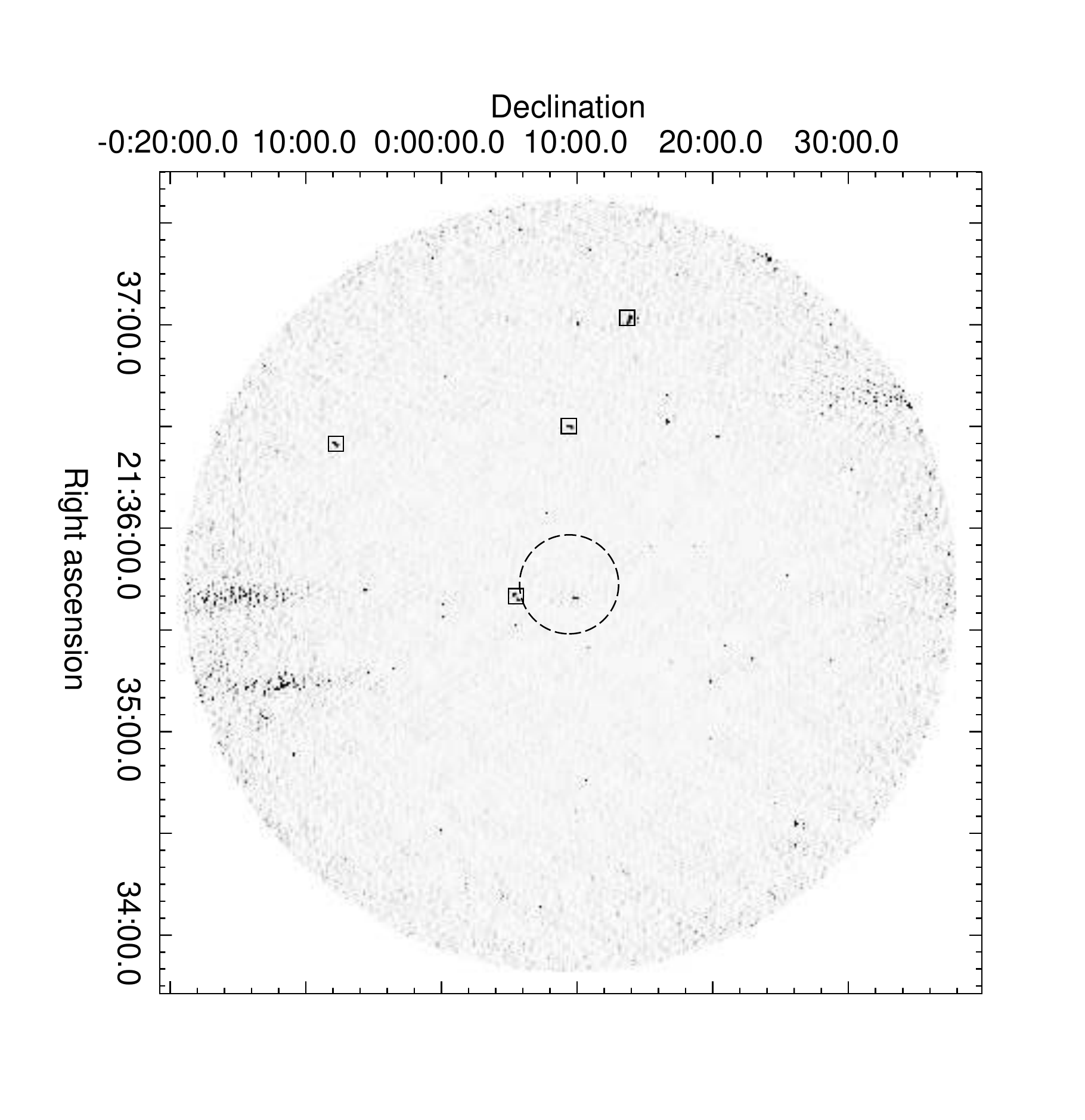}
 \caption{ACT-CL J2135.7+0009. The dashed circle denotes the $R_{500}$ region centred on the cluster SZ peak. Boxes indicate the postion of sources of interest, discussed in Section \ref{sec:fovsrcs}}
 \label{fig:j2135p0009_FOV}
\end{figure}

\begin{figure}
 \centering
 \includegraphics[width=\textwidth,angle=90,origin=c]{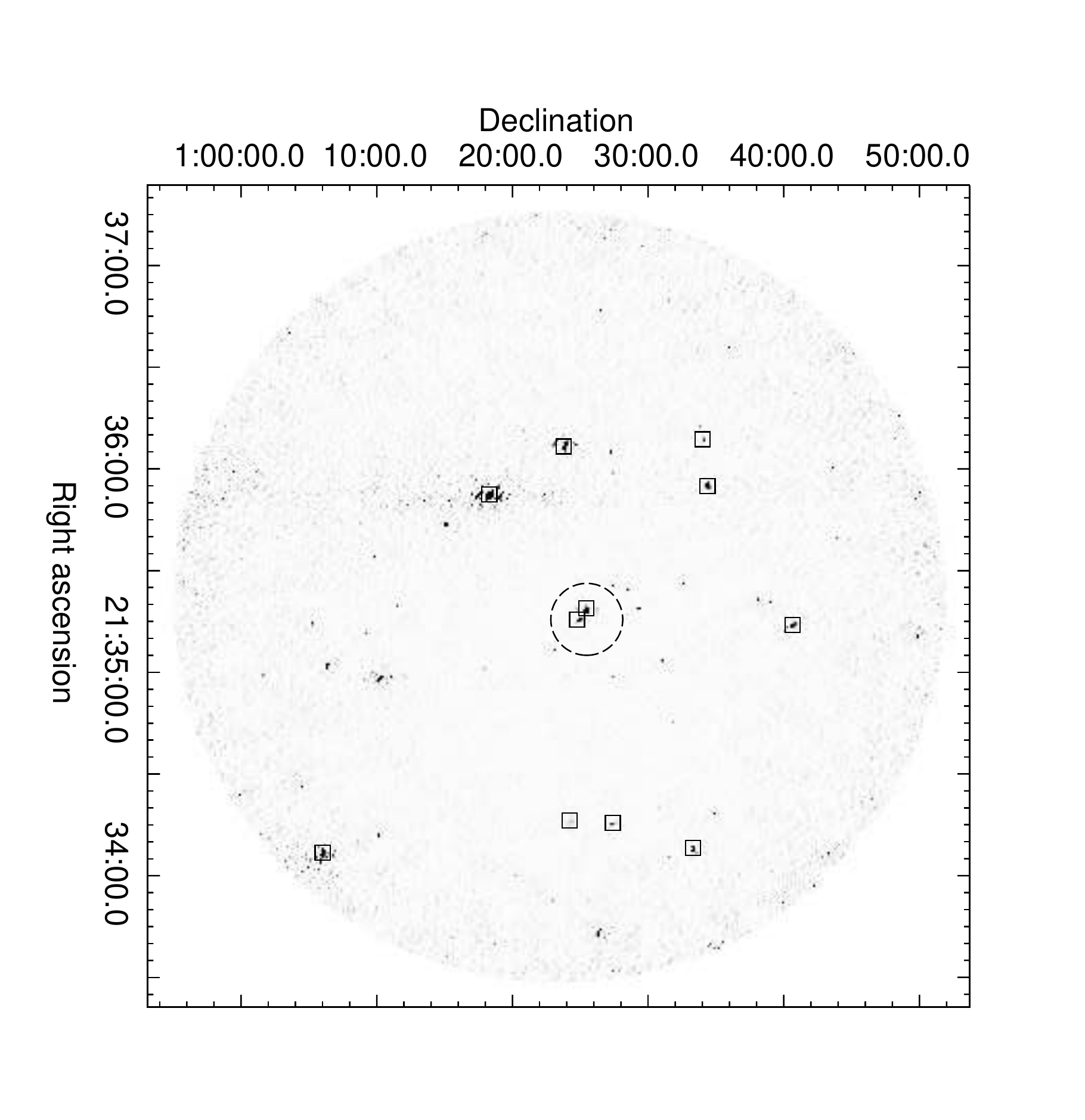}
 \caption{ACT-CL J2135.2+0125. The dashed circle denotes the $R_{500}$ region centred on the cluster SZ peak. Boxes indicate the postion of sources of interest, discussed in Section \ref{sec:fovsrcs}.}
 \label{fig:j2135p0125_FOV}
\end{figure}

\begin{figure}
 \centering
 \includegraphics[width=\textwidth,angle=90,origin=c]{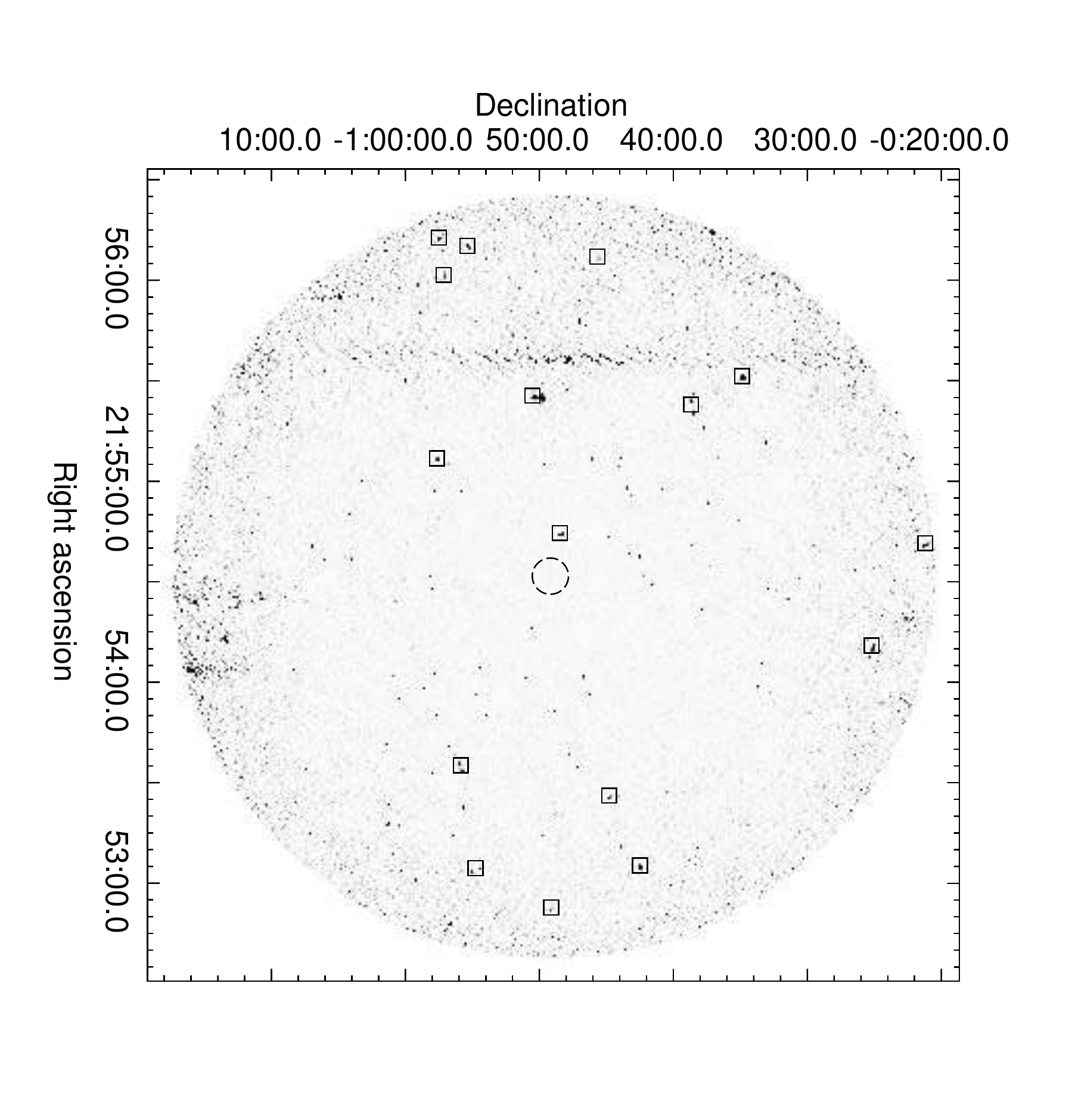}
 \caption{ACT-CL J2154.5$-$0049. The dashed circle denotes the $R_{500}$ region centred on the cluster SZ peak. Boxes indicate the postion of sources of interest, discussed in Section \ref{sec:fovsrcs}.}
 \label{fig:j2154_FOV}
\end{figure}

\begin{figure}
 \centering
 \includegraphics[width=\textwidth,angle=90,origin=c]{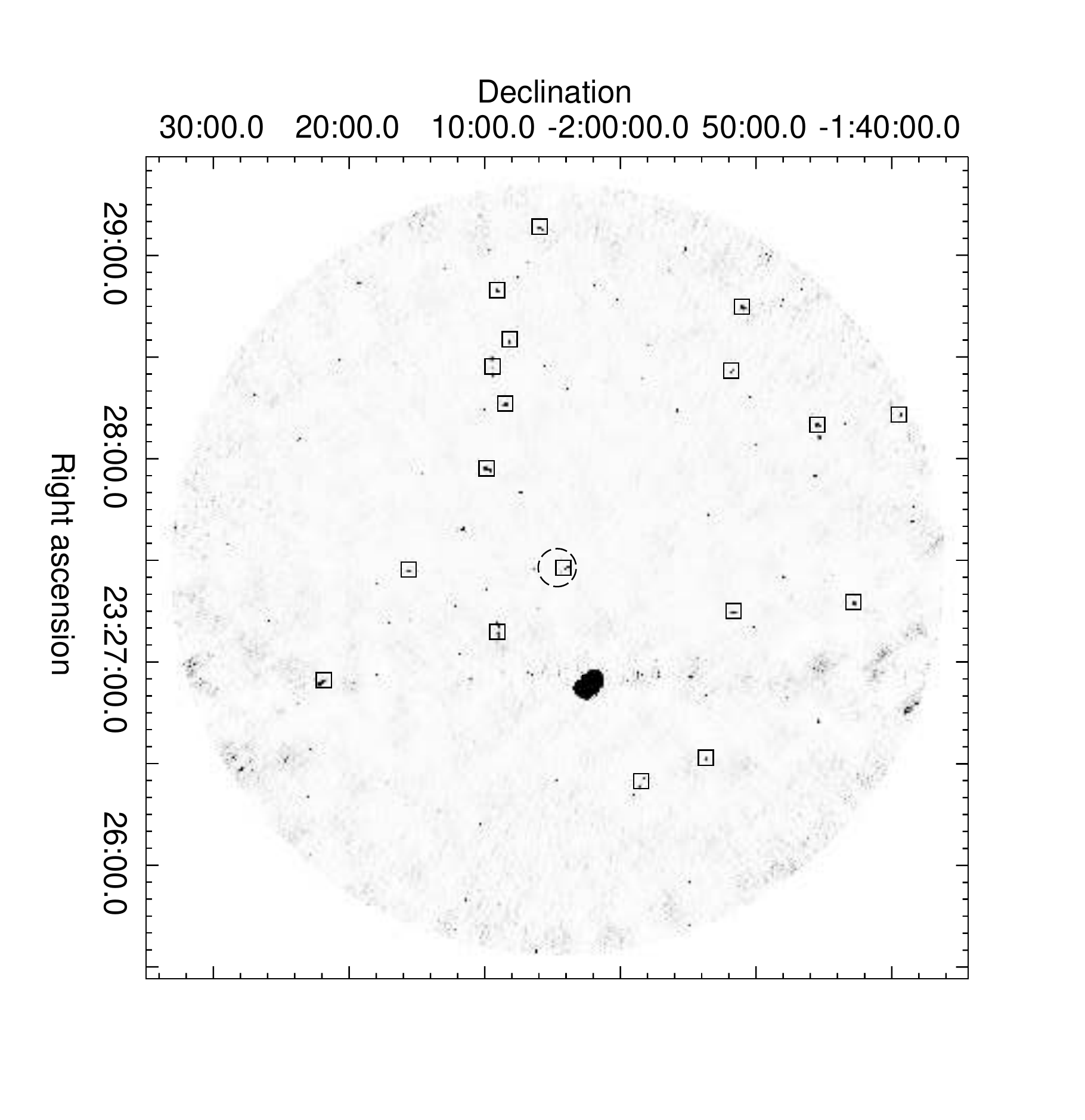}
 \caption{ACT-CL J2327.4$-$0204. The dashed circle denotes the $R_{500}$ region centred on the cluster SZ peak. Boxes indicate the postion of sources of interest, discussed in Section \ref{sec:fovsrcs}.}
 \label{fig:j2327_FOV}
\end{figure}

\begin{figure}
 \centering
 \includegraphics[width=\textwidth,angle=90,origin=c]{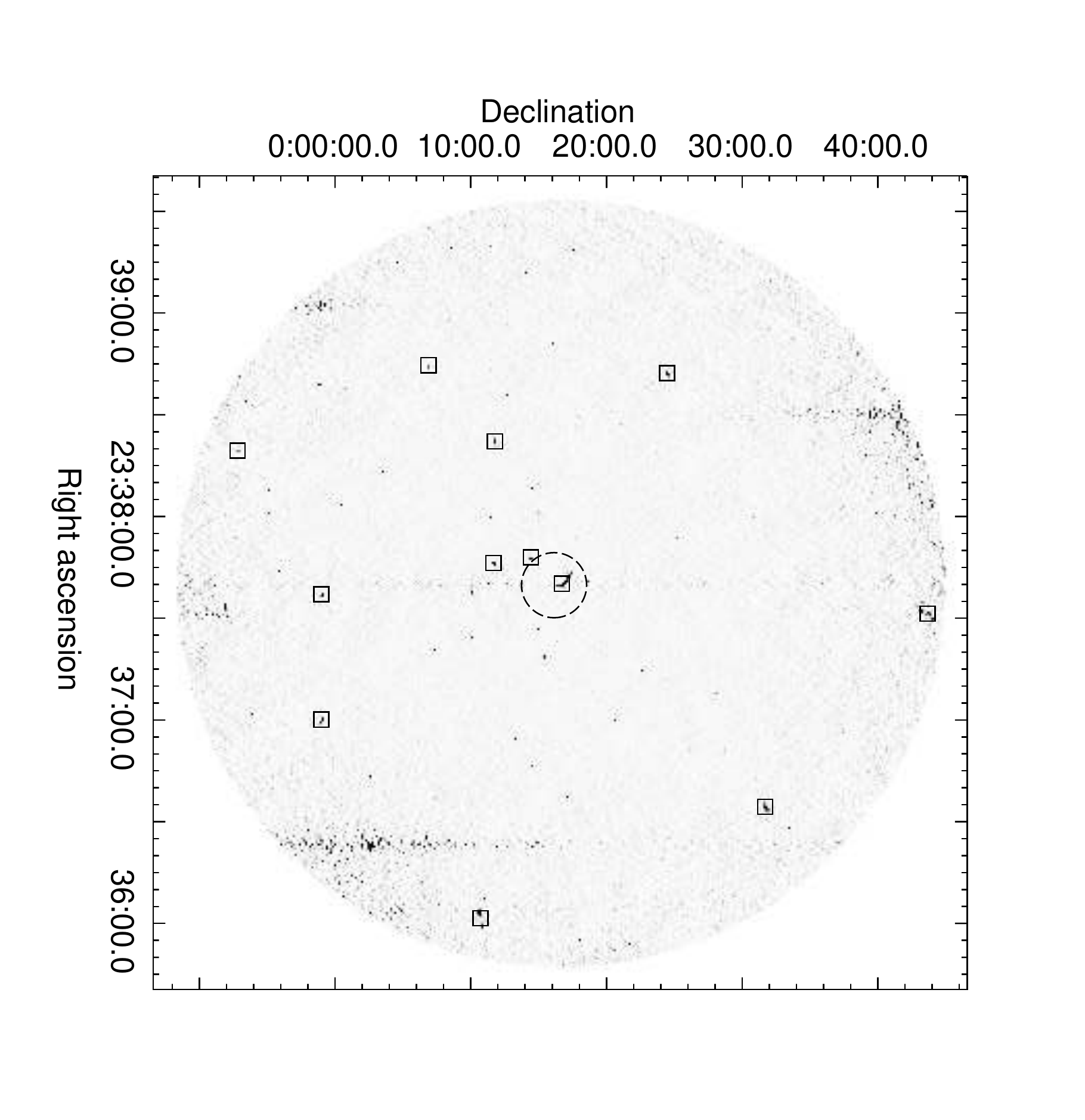}
 \caption{ACT-CL J2337.6+0016. The dashed circle denotes the $R_{500}$ region centred on the cluster SZ peak. Boxes indicate the postion of sources of interest, discussed in Section \ref{sec:fovsrcs}.}
 \label{fig:j2337_FOV}
\end{figure}


\bsp	
\label{lastpage}
\end{document}